\documentclass[aps,prb,amsmath,twocolumn,superscriptaddress]{revtex4-2}
\usepackage{graphicx}% Include figure files
\usepackage{dcolumn}% Align table columns on decimal point
\usepackage{bm}% bold math
\usepackage{xcolor}% text colors
\usepackage{hyperref}

\usepackage{lineno}
\begin{document}

\title{Classical periodic trajectories and quantum scars in many-spin systems}

\author{Igor Ermakov}
\email{ermakov1054@yandex.ru}
\affiliation{Department of Mathematical Methods for Quantum Technologies, Steklov Mathematical Institute of Russian Academy of Sciences,
8 Gubkina St., Moscow 119991, Russia. }
\affiliation{Laboratory for the Physics of Complex Quantum Systems, Moscow Institute of Physics and Technology, Institutsky per. 9, Dolgoprudny, Moscow  region,  141700, Russia.
}
\affiliation{
Skolkovo Institute of Science and Technology, Bolshoy Bulvar, 30, bld. 1, Moscow, 121205, Russia.}  
\author{Oleg Lychkovskiy}
 \email{lychkovskiy@gmail.com}
 \affiliation{
Skolkovo Institute of Science and Technology, Bolshoy Bulvar, 30, bld. 1, Moscow, 121205, Russia.} 
\affiliation{Laboratory for the Physics of Complex Quantum Systems, Moscow Institute of Physics and Technology, Institutsky per. 9, Dolgoprudny, Moscow  region,  141700, Russia.
}
\affiliation{Department of Mathematical Methods for Quantum Technologies, Steklov Mathematical Institute of Russian Academy of Sciences,
8 Gubkina St., Moscow 119991, Russia. } 

\author{Boris V. Fine}
 \email{boris.fine@uni-leipzig.de  \\ 
 (B.V.F. had affiliation $^2$ until 2022.)}
\affiliation{
Institute for Theoretical Physics,
University of Leipzig, Brüderstrasse 16, 04103 Leipzig, Germany. 
}
\affiliation{Laboratory for the Physics of Complex Quantum Systems, Moscow Institute of Physics and Technology, Institutsky per. 9, Dolgoprudny, Moscow  region,  141700, Russia.
}

\date{\today}% It is always \today, today,

\begin{abstract} 
To probe the limits of dynamic thermalization, we numerically investigate the stability of exceptional periodic classical trajectories in  chaotic many-spin systems and explore a possible connection between these trajectories  and  exceptional nonthermal quantum eigenstates known as ``quantum many-body scars".  The systems considered are chaotic spin chains with short-range interactions, both classical and quantum. On the classical side, the chosen periodic trajectories are such that all spins instantaneously point in the same direction, which evolves as a function of time.
We find that the largest Lyapunov exponents characterising the stabillity of these trajectories have surprisingly strong and nontrivial dependencies on the interaction constants and chain lengths. In particular, we identify rather long spin chains, where the above periodic trajectories are Lyapunov-stable on many-body energy shells overwhelmingly dominated by chaotic motion. We show that the above phenomenology can be quantitatively described by connecting Lyapunov instabilities of translationally invariant periodic trajectories to irreducible representations of the translational symmetry group with well-defined wave vectors. We also find that instabilities around periodic trajectories in modestly large spin chains develop into a transient nearly quasiperiodic non-ergodic regime. In some cases, the lifetime of this regime is extremely long, which we interpret as  a manifestation of Arnold diffusion in the vicinity of integrable dynamics. On the quantum side, we numerically investigate the dynamics of quantum states starting with all spins initially pointing in the same direction: these are the quantum counterparts of the initial conditions for the above periodic classical trajectories.  Our investigation reveals the existence of quantum many-body scars for numerically accessible finite chains of spins 3/2 and higher.  No evidence of quantum scars was observed for spin-1/2 chains, while spin-1 chains were found to be transitional in this respect. The dynamic thermalization process dominated by quantum scars is shown to exhibit  a slowdown in comparison with generic thermalization at the same energy.
Finally, we identify quantum signatures of the proximity to a classical separatrix of the periodic motion.

\end{abstract}

\maketitle

%\tableofcontents

\section{Introduction}

The role and the properties of chaos in the process of dynamic thermalization of many-body systems remain a subject of intense research that keeps resisting generic conjectures\cite{eisert2015quantum,dAlessio-2016,abanin2019colloquium}. The open issues concern the transition from classical to quantum thermalization\cite{berry1989quantum,Reimann-2013}, as well as  the thermalization in purely classical systems as such\cite{gaspard1998,vulpiani2009chaos}. Central to those issues is the lack of a clear border between systems exhibiting and not exhibiting dynamic thermalization to the Gibbs equilibrium.  
It is often the case that a seemingly exotic non-thermalizing setting, in fact, sets a warning post for a much broader class of physical situations, where the thermalization is absent or poor. In the latter case, the system may be thermalizable in principle but only on a rather long timescale\cite{kinoshita2006quantum,banuls2011strong,gring2012relaxation,neyenhuis2017observation,Luitz-2016,Luitz-2017,Peng-2021,Lando-2023,Pawlowski-2024}. The examples of nonthermalizing/poorly thermalizing behavior are: integrable systems \cite{palzer2009quantum,berman2004irregular,rigol2007relaxation,brandino2015glimmers}, glassy systems \cite{zhang2017observation,rademaker2020slow}, systems possibly exhibiting many-body localisation\cite{Basko-2006,Oganesyan-2007,Pal-2010,Serbyn-2017,morong2021observation}, periodically driven systems\cite{Ketzmerick-PRE-2010,Ji-2011,dAlessio-2014,Lazarides-2015,Bukov-2015,Ji-2018,Pizzi-2020,rylands2020many,Rakcheev-2022}, and, finally, coming to the topic of this article, systems with quantum many-body scars \cite{Shiraishi-2017,bernien2017probing,turner2018weak}. 
%The fragility of systems's dynamics also has an intrinsic potential of being used for designing quantum sensors. 

Even when a system appears sufficiently ``generic", there is still a question\cite{dymarsky2019mechanism,ermakov2020almost}: How does the character and the outcome of the thermalization process depend on the initial conditions?  One should keep in mind here that the process of dynamic thermalization starts from a macroscopic nonequilibrium initial state of an isolated many-body system. Any macroscopic nonequilibrium state is extremely unlikely in comparison with the equilibrium state at the same energy. It is, therefore, intrinsically problematic to discard one nonequilibrium state as being more unlikely than another, as long as both states can be prepared experimentally. 

\begin{figure}
 \includegraphics[width=\columnwidth]{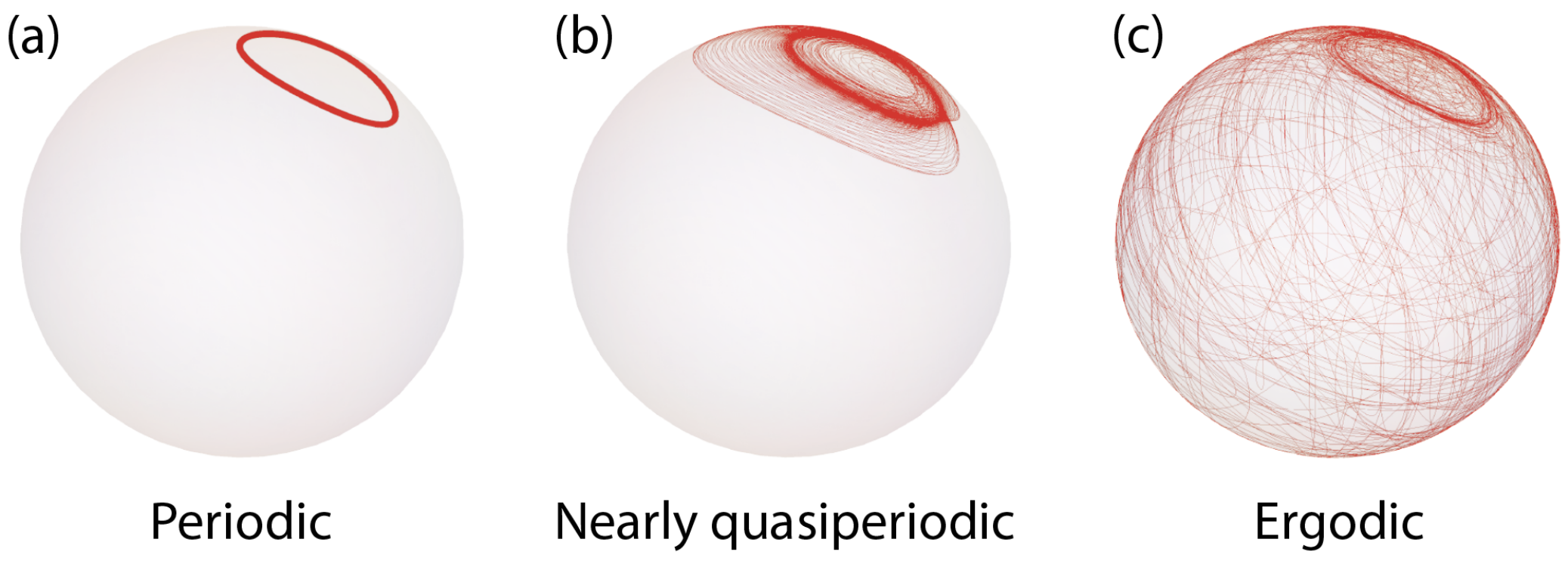}
\caption{Regimes of classical spin dynamics investigated in this work. Pictures represent the trajectory  of the tip of a classical spin on a unit sphere.  The dynamics starts with all spins pointing in the same direction. As a result, each spin has the same periodic classical trajectory shown in (a), which may be Lyapunov-stable or unstable. An unstable periodic trajectory in a finite system decays into a transient nearly quasiperiodic non-ergodic regime illustrated in (b), which then evolves into the ergodic regime pictured in (c). In some case, the lifetime of the nearly quasiperiodic regime is anomalously long, which is attributed to Arnold diffusion. 
\label{scheme_periodic}}
\end{figure} 

The present article is about thermalization and chaos in relatively generic systems of interacting classical and quantum spins for somewhat exotic but experimentally realisable initial conditions where all spins initially point in the same direction. Classically, these initial conditions lead to short periodic trajectories. 
In what follows, we describe the fate of these periodic trajectories in two senses, namely: fate following small deviations from these trajectories in purely classical systems, and also fate in the sense of transitioning from classical to quantum dynamics. 
Our original motivation was to explore whether periodic trajectories in the classical limit entail a quantum-many-body-scar behavior, and, indeed we report here a positive numerical evidence of quantum scars for quantum spins $S\geq 3/2$. However, the investigation also revealed a number of surprising  properties on the purely classical side, such as the non-trivial dependence of the Lyapunov stability  of the periodic orbits on the chain length, including fairly long spin chains with Lyapunov-stable periodic orbits on otherwise strongly chaotic energy shells. 
We also discovered that unstable periodic motion in finite spin systems evolves to exhibit a transient nearly quasiperiodic non-ergodic regime. In several cases, this regime has an anomalously long lifetime, which we interpret as a manifestation of Arnold diffusion\cite{Arnold1964diffusion}. Various classical regimes investigated in this work are schematically summarised in Fig.~\ref{scheme_periodic}.

In order to place these findings in a broader context, we discuss in the next section some of the relevant fundamental issues. Expert readers may skip it and proceed directly to Section~\ref{general}.

\section{Preliminary discussion}
\label{preliminary}

\subsection{General aspects}
\label{aspects}

Classical chaos is an exponentially unstable behavior of  classical phase space trajectories with respect to infinitesimal deviations of initial conditions \cite{lichtenberg2013regular,vulpiani2009chaos}. Classical ergodicity is the property of a phase space trajectory to visit an arbitrarily small region around any point on the energy shell with probability proportional to the ``volume" of that region. 
In principle, a system can be ergodic without exhibiting chaos or, on the contrary,  can exhibit chaos  in a part of a mixed phase space without being fully ergodic\cite{lichtenberg2013regular}. It is, however, a common belief consistent with numerical simulations (e. g., \cite{dewijn2012largest,tarkhov2018estimating}) that a typical nonlinearly interacting many-body classical system is overwhelmingly likely to be both chaotic and ergodic in an ovewhelmingly large part of the phase space. In the classical context, the remaining concerns about ergodicity and chaos can often be traced back to the Kolmogorov-Arnold-Moser (KAM) theorem \cite{kolmogorov1954conservation,moser1962invariant,arnol1963small}, which predicts that, for nearly integrable systems, some invariant tori of quasiperiodic motion associated with the integrable limit survive under a sufficiently small perturbation of integrability.  
Quantum implications of this theorem were explored, e.g., in \cite{brandino2015glimmers,Schmidt-2023,Varma-Cohen-2024}.

The practical relevance of the KAM theorem to the thermalization in many-body systems is sometimes viewed skeptically based on the expectation that the parameter range where the KAM tori survive decreases exponentially with the number of degrees of freedom in the system, and, even when some of them do survive, they do not suppress ergodicity of a typical phase space trajectory. 
Yet, contrary to the latter view, the findings of the present work  indicate the surprising relevance of the KAM-tori even to the dynamics of strongly nonintegrable systems.
These findings include the existence of Lyapunov-stable periodic trajectories (Sections~\ref{classical}) and the existence of a nearly quasiperiodic regime (Sections~\ref{quasiperiodic}) -- both occurring in finite but large systems on energy shells dominated by strong chaotic motion. 

When it comes to the transition from classical to quantum thermalization, one of the primary challenges here  is the fact that the classical description is based on the notions of ergodicity and chaos of phase space trajectories, while quantum-mechanically the very notion of a phase trajectory is not definable -- consequence of the Heisenberg uncertainty principle not allowing one to simultaneously determine the coordinates and momenta of particles with infinite precision. 
On the quantum side, chaotic instabilities and ergodicity become detached  from each other. The common indicators of ``quantum chaos", such as the statistics of energy level spacings \cite{Weidenmueller-2009,Bohigas-1984} or eigenstate thermalization \cite{deutsch1991quantum,srednicki1994chaos,Flambaum-1996,Benenti-2001,d2016quantum,deutsch2018eigenstate}, involve the global properties of many-body energy eigenstates, and therefore, characterise their ergodicity. At the same time, recent research on out-of-time-ordered correlators (OTOCs) indicates\cite{fine2014absence,elsayed2015sensitivity,kukuljan2017weak,craps2020lyapunov,OTOC} that  chaotic instabilities in response to small deviations of initial conditions are, strictly speaking, absent in quantum systems with bounded local operators (e.g. spins 1/2) and finite-range interactions even in the thermodynamic limit. (Here, we sidestep the models with infinite-range interactions, where chaos and thermalization properties may be rather special\cite{Kitaev2015,Maldacena2016,Kolovsky-2017,Schmitt-2019}.) Systems with finite-range interactions, however, may still  exhibit a transient exponential OTOC growth representative of the classical chaotic limit, in particular, for lattices of spins $S > 1/2$ \cite{elsayed2015sensitivity,craps2020lyapunov}.  A similar condition, namely, $S \geq 3/2$, to observe the effect of classical periodic trajectories on quantum dynamics  is reported in Section~\ref{quantum} of the present work. The readers are referred to Refs.\cite{Elsayed-classical-2015,starkov2018hybrid,Schubert-2021,Bilitewski-2021,Lebel-2023,Benet-2023,santos2024quantumclassical} for various complementary aspects of the correspondence between classical and quantum spin dynamics.

Other efforts to identify the indicators and the mechanisms of the proper quantum thermalization targeted the properties of the Hamiltonian matrix elements\cite{Aberg-1990,Santos-2012},   the role of quantum typicality\cite{Gemmer-2009,Goldstein-2006,Popescu-2006,Hahn-2016,Bartsch-2009,Elsayed-2013}, generic features of the relaxation behavior\cite{fine2004long-time,Fine-2005,Lychkovskiy-2010,Khodja-2015,teretenkov2024}, the frequency spectra of the observables\cite{Elsayed-signatures-2014,Parker-2019}, adiabatic eigenstate deformations\cite{Pandey-2020,polkovnikov2024defining}, quantum scrambling\cite{Campisi-2017,Iyoda-2018,von2018operator,khemani2018operator,Chavez-Carlos-2019,Mi-Google-2021,Zhu-2022}  and, more recently, the universal operator growth\cite{Parker-2019,Pappalardi-2024,Uskov-2024,Nandy-2024}.  All the above properties are subject to caution due to possible generic existence of exceptional quantum-scar states associated with the periodic classical orbits of the kind considered in this work.

Another frontier of the thermalization research is the behavior of large but finite systems. The importance of this frontier is that macroscopic systems that exhibit nontrivial thermalization behavior are often those that also exhibit fragmentation into weakly coupled nanoscale subsystems -- e.g. large molecules\cite{Zhang-2014} or clusters of different phases\cite{fine2008phase,Richaud-2018}. Each of such subsystems can in turn exhibit an unusual dynamics associated with its finite size\cite{Ji-2011,Ji-2018,lychkovskiy2018spin,Pizzi-2020}. 
The stationary state of a finite system can realistically deviate from the Gibbs equilibrium, because the number of degrees of freedom and/or the number of quantum levels is insufficiently large. For finite classical systems, the concerns related to the KAM theorem also become more plausible. The present work contributes to the above agenda by showing that the stability of periodic trajectories in finite classical spin chains exhibits a strong non-trivial dependence on the chain length including strongly chaotic and non-chaotic cases, and that unstable periodic motion in finite systems decays into a transient nearly quasiperiodic regime.

\subsection{Quantum scars}
\label{scars}

A recent addition to the zoo of poorly thermalizing settings is the case of the so-called ``quantum many-body scars" associated with anomalously slow relaxation of quantum lattice systems that was first observed experimentally in Ref.\cite{bernien2017probing} and then interpreted theoretically in Ref.\cite{turner2018weak} in terms of the ``PXP" quantum model for a spin-1/2 chain. It was found that the PXP model possesses atypical eigenstates inconsistent with eigenstate thermalization\cite{deutsch1991quantum,srednicki1994chaos}, and, at the same time,  these atypical eigenstates are prominently present in the initial nonequilibrium state of the experimentally observed system of Ref.\cite{bernien2017probing}. 
Eigenstate thermalization hypothesis (ETH)  stipulates that the quantum expectation values of a physical observable in a normally thermalizing system remain almost the same for all energy eigenstates within a given energy window. The atypical eigenstates of the PXP model named ``quantum scars" stood out in this respect. A similar kind of ETH violation was, actually, reported earlier in Ref.~\cite{Shiraishi-2017} without reference to quantum scars. The above findings were followed a large body of research \cite{lin2019exact,serbyn2021quantum,shibata2020onsager,michailidis2020stabilizing,bull2022tuning,ljubotina2022optimal,sinha2020chaos,banerjee2021quantum,Hummel-2023}. 
 
Since the PXP model and much of the follow-up research implemented a constraint-based quantum dynamics, which is somewhat special, the questions remain regarding the occurrence of quantum scars in generic quantum systems.

Let us recall here that the concept of quantum scars as such was initially introduced by Heller\cite{heller1984bound} in the eighties for one particle moving in a chaotic two-dimensional billiard: it referred to atypical quantum eigenstates corresponding to  classical periodic trajectories with short periods. When a classical periodic trajectory is sufficiently short, then the quasiclassical limit of the quantum dynamics implies the onset of quantum interference corrections associated with the trajectory closing onto itself before the chaotic Lyapunov instability suppresses such interference\cite{Berry-1989}, and these corrections, in turn, amplify the probability of the corresponding eigenstate to remain localized around the classical trajectory as opposed to being evenly spread over the entire system which the typical eigenstates do. Numerous examples of low-dimensional quantum scars have been investigated since then -- see, e.g., \cite{Luukko-2016,Keski-Rahkonen_2019}. 
Yet, a missing link between the original low-dimensional quantum scars and the recently discovered quantum many-body scars was that the latter were not connected to periodic classical trajectories.  In other words, the quantum many-body scars as discovered were rather ``distant relatives" to the quantum scars known from the physics of billiards. This issue of missing periodic trajectories was addressed in \cite{ho2019periodic,turner2021correspondence}, although not in the classical limit but rather in the parameter space of quantum matrix product states. 

When completing this article, we discovered that a very similar agenda was pursued for two-spin spin systems in Refs.~\cite{Mondal-PRE2021,Mondal-PRE2022}, for a two-site Bose-Hubbard model in Ref.~\cite{Mondal-PRA2022} and  for a many-site Bose-Hubbard model in Ref.~\cite{Hummel-2023}. The above references and the present work appear to largely agree on their conclusions, while arriving to them by rather complementary routes.  We further note in this regard that most of the findings of the present paper were first reported in the Ph.D. thesis \cite{Ermakov-PhD-2022}.

\subsection{Outline of the article}
\label{preview}

The rest of the article is organised as follows: In Section~\ref{general}, we give a general formulation of our spin chain setting applicable to both classical and quantum problems. Section~\ref{classical} deals with the Lyapunov stability of the classical periodic orbits at hand. In particular, it explains the classification of these orbits in terms of ``librations" and ``rotations" and describes a separatrix between them. It also puts focus on the translationally invariant character of these periodic orbits, which allows one to classify Lyapunov instabilities according to the irreducible representations of the translation symmetry group with well-defined wave numbers. The later classification then plays a crucial role in our semi-phenomenological description of the observed dependencies of the largest Lyapunov exponents on the chain lengths. We also numerically investigate the dependence of the Lyapunov exponents on relative strength of the Hamiltonian parameters, describing in particular, a non-analytical part of this dependence near the separatrix. The section ends with our  proposal of a practical way to experimentally measure the Lyapunov exponent of a periodic trajectory  by observing just one degree of freedom in a many-spin system. 

In Section~\ref{quasiperiodic}, we investigate the dynamical regime emerging in a classical system once a Lyapunov instability around the periodic trajectory develops. This regime is not yet ergodic --- we call it ``nearly quasi-periodic". The Section contains the computed frequency spectra in that regime as well as the time evolutions of the intensities of various spatial Fourier harmonics. We identify three mechanisms limiting the lifetime of the nearly quasi-periodic regime. The slowest and, perhaps, the most interesting of the three is attributed to the phenomenon of Arnold diffusion.  

In Section~\ref{quantum}, we numerically investigate chains of quantum spins $1/2$, 1, $3/2$ and 2 with  varying chain lengths  looking for the signatures of quantum scars corresponding to the above classical periodic orbits. We identify such signatures for spins $3/2$ and 2. The signatures include the slowdown of spin relaxation and the violation of the eigenstate thermalization hypothesis. We also report on the quantum signatures of the proximity to the classical separatrix between librations and rotations.

Section\ref{discussion} contains  detailed summary and concluding discussion. Several technical aspects of our analysis are delegated to Appendices.

\section{Hamiltonian and initial conditions}
\label{general}

%\subsection{Model and initial conditions}

We consider translationally invariant periodic chains of $L$ interacting spins -- classical and quantum -- described by the Hamiltonian:
\begin{align}\label{ham}
\mathcal{H} = -\sum\limits^L_{i=1}&\left(J S_i^x S_{i+1}^x + 2J S_i^y S_{i+1}^y \right)\nonumber\\
    &+\sum\limits^L_{i=1}\left(h S_i^x+h S_i^y \right),
\end{align}
where, $i$ is the lattice index, $\alpha=x,y,z$ is the spin projection index; in the classical case, $S^\alpha_i$ are the components of 3-dimensional vectors $\mathbf{S}_i=(S^x_i,S^y_i,S^z_i)$ of length $|\mathbf{S}_i| = 1$; in quantum case, $S^\alpha_i$ are spin-$S$ operators with $S=\frac{1}{2},1,\frac{3}{2},2$; \   $J$ is the interaction constant, and $h$ is  the local field parameter. We set $\hbar = 1$ where relevant.

In all concrete calculations throughout the paper, we fix $h=1$, while varying $J$. This implies that the unit of time in all plots is $1/h$.  

The particular form of the Hamiltonian~\ref{ham} is motivated by its robust non-integrability in the quantum case. In particular, choosing $h$ on the order of $J$ ensures that the chain with spins 1/2 is far from the integrable limits of either $h=0$ or $J=0$.  In addition, the confinement physics often obscuring thermalization in finite systems \cite{banuls2011strong,lin2017quasiparticle,Peng_2022_Bridging} has been avoided by introducing both interaction terms $S_i^x S_{i+1}^x$ and $S_i^y S_{i+1}^y$. Finally, we have avoided symmetry with respect to the exchange $\{S_i^x\leftrightarrow S_i^y\}$ by having the $yy$-coupling twice that of $xx$.
(See Section~\ref{quantum} for further discussion.) 

We investigate one special type of nonequilibrium initial conditions corresponding to all spins polarised in the same direction. Specifically, we choose them polarised along the $z$-direction, which, according to Hamiltonian (\ref{ham}), implies that the initial energy $E$ (energy expectation value in the quantum case) is equal to zero, which, in turn, corresponds to the infinite temperature in the sense of canonical ensemble. Indeed, since the infinite-temperature  classical probability distribution is uniform in  the classical phase space, while the infinite-temperature quantum density matrix is proportional to the unit matrix in the quantum Hilbert space, the average energy in the both cases can be written as 
\begin{align}\label{E_inf}
E_{\infty} = \langle \mathcal{H} \rangle = -\sum\limits^L_{i=1}&\left(J \langle S_i^x S_{i+1}^x \rangle  + 2J \langle S_i^y S_{i+1}^y \rangle \right)\nonumber\\
    &+\sum\limits^L_{i=1}\left(h \langle S_i^x\rangle +h \langle S_i^y \rangle \right) = 0.
\end{align}
Here $\langle ... \rangle$ denotes the average over all possible values of spin projections, quantum or classical, which implies that $\langle S_i^{\alpha} \rangle = 0$ and $\langle S_i^{\alpha} S_{i+1}^{\alpha} \rangle =0$. Further noting the equivalence between the canonical and the microcanonical ensembles in the limit $L \to \infty$, we conclude that an initial spin state with zero energy belongs to the energy shell of maximum entropy (maximum density of states quantum mechanically).

%by $3L$-dimensional vector:

In the classical case, the above initial conditions have the form 
\begin{align}
\mathbf{S}_i(0)=(0,0,1)
\label{eqclassini}
\end{align}
for each spin.
In the quantum case, the fully polarized ``up''-states are described by the initial wave function: 
\begin{align}\label{eqquantini}
|\Psi(0)\rangle =  |\Psi^\text{up}\rangle\equiv\bigotimes^L_{i=1}|m_i=S\rangle,
\end{align}
where $m_i$ is the $z$-projection of the $i$th spin.

We note that all couplings in the Hamiltonian (\ref{ham}) were chosen to involve only $x$ and $y$ spin projections to ensure that the initial state fully polarized along the $z$-axis has zero energy and hence lies in the middle of the energy spectrum, which is important for reducing the finite-size effects on spin dynamics.

\section{Classical spins: stability of periodic trajectories}
\label{classical}

\subsection{Equations of motion}

Classical dynamics of the Hamiltonian (\ref{ham}) is governed by the system of $3L$ nonlinear differential equations:
\begin{align}\label{claseq}
\frac{dS^\alpha_i}{dt}=\{S^\alpha_i,\mathcal{H}\}
\end{align}
where $\{ ... , ... \}$ are the Poisson brackets, with $\{S^\alpha_i, S^\beta_j \} = \delta_{ij} \epsilon_{\alpha\beta\gamma} S^\gamma_i $. Here  $\delta_{ij}$ is the Kronecker tensor, and $\epsilon_{\alpha\beta\gamma}$ is the antisymmetric Levi-Civita tensor. The system of equations (\ref{claseq}) can be rewritten as 
\begin{align}\label{claseq2}
\frac{d\mathbf{S}_i}{dt}=\mathbf{H}_i \times \mathbf{S}_i,
\end{align}
where
\begin{equation}
  {\mathbf{H}}_i = \left(
  \begin{array}{c} 
  - J S_{i-1}^x - J S_{i+1}^x + h \\[4pt]
  - 2 J S_{i-1}^y - 2 J S_{i+1}^y + h  \\[4pt]
  0 
  \end{array}
    \right)
  \label{Hj}
  \end{equation}
is the local field acting on the $i$th spin. 

Where necessary, the set of equations (\ref{claseq2}, \ref{Hj}) is to be solved numerically using the 4th-order Runge-Kutta algorithm with the discretization time steps $\delta t$ being small enough to make the results insensitive to making $\delta t$ smaller; typically $\delta t =  0.001$.

Given the fixed length of each spin $|\mathbf{{S}_i}|=1$, the phase space of the present system is $2L$-dimensional. It is, however convenient both for discussion and for the calculations of Lyapunov instabilities\cite{dewijn2012largest,dewijn2013lyapunov} to represent each phase space trajectory by  a $3L$-dimensional vector
\begin{equation}
\mathcal{S}(t)
\equiv
[S^x_1(t),S^y_1(t),S^z_1(t),\dots,S^x_L(t),S^y_L(t),S^z_L(t)],
\label{St}
\end{equation}
where the projections are the solutions of Eqs.(\ref{claseq2}, \ref{Hj}).

\subsection{Periodic trajectories}
\label{periodic}

Due to the translational invariance of both the Hamiltonian (\ref{ham}) and the initial conditions (\ref{eqclassini}), the classical spins initially polarised in the same direction exhibit the same time evolutions and hence remain parallel to each other at all subsequent moments of time with the direction of their common polarisation evolving as a function of time. As a result, the calculation of a many-spin trajectory $\mathcal{S}(t)$ is reduced to computing the same one-spin trajectory $\mathbf{S}_i(t) = \mathbf{S}_{\text{p}}(t)$ for each spin on the lattice. Since the one-spin trajectory $\mathbf{S}_{\text{p}}(t)$ is limited to a two-dimensional  unit sphere $|\mathbf{S_{\text{p}}}| = 1$, it normally closes onto itself and thus necessarily becomes periodic. (This property is  explicitly demonstrated for our system in Appendix~\ref{periodicity}. The broader context is that of the Poincaré–Bendixson theorem \cite{katok1995introduction,teschl2012ordinary}.)   The many-spin phase space trajectory $\mathcal{S}_{\text{p}}(t)$ corresponding to $\mathbf{S}_{\text{p}}(t)$ also becomes periodic.

Since all spins during this dynamics point in the same direction, the projections of $\mathbf{S}_{i-1}$ and $\mathbf{S}_{i+1}$ in Eq.(\ref{Hj}) can be replaced by those of $\mathbf{S}_{i}$, and, as a result, the periodic one spin trajectories $\mathbf{S}_{\text{p}}(t)$ can be computed with the help of one-spin Hamiltonian :
\begin{align}\label{ham0}
\mathcal{H}_{\text{p}} = -J(S^x_{\text{p}})^2 -2J(S^y_{\text{p}})^2 + h S^x_{\text{p}} + h S^y_{\text{p}}.
\end{align}

\subsection{Librations and rotations}
\label{librot}

Dependent on the parameters of the Hamiltonian $\mathcal{H}_{\text{p}}$ and on the initial orientation  $\mathbf{S}_{\text{p}}(0)$, there are two possible kinds of generic periodic trajectories $\mathbf{S}_{\text{p}}(t)$, which, by analogy with the description of a pendulum\cite{lichtenberg2013regular}, we call ``librations'' and 
``rotations''. Librations are realised by a single trajectory on the spherical surface $|\mathbf{S_{\text{p}}}| = 1$ connecting all points having the same energy. The rotations are realised by two disconnected trajectories -- both corresponding to the same energy. 
Librations and rotations transition to each other as a function of Hamiltonian parameters and/or initial energy through a separatrix as illustrated in Fig.\ref{SP_solutions_red}. (See Appendix~\ref{periodicity} for an explicit derivation.)

When a periodic trajectory is considered as a function of $J$ for fixed $h=1$ and fixed energy $E=0$ (associated with our initial conditions ``all-spins-up"), the separatrix corresponds to $J=J^*\approx 1.15$. Librations are realised for $J<J^*$,
and rotations for $J>J^*$.

\begin{figure}
\includegraphics[width=\columnwidth]{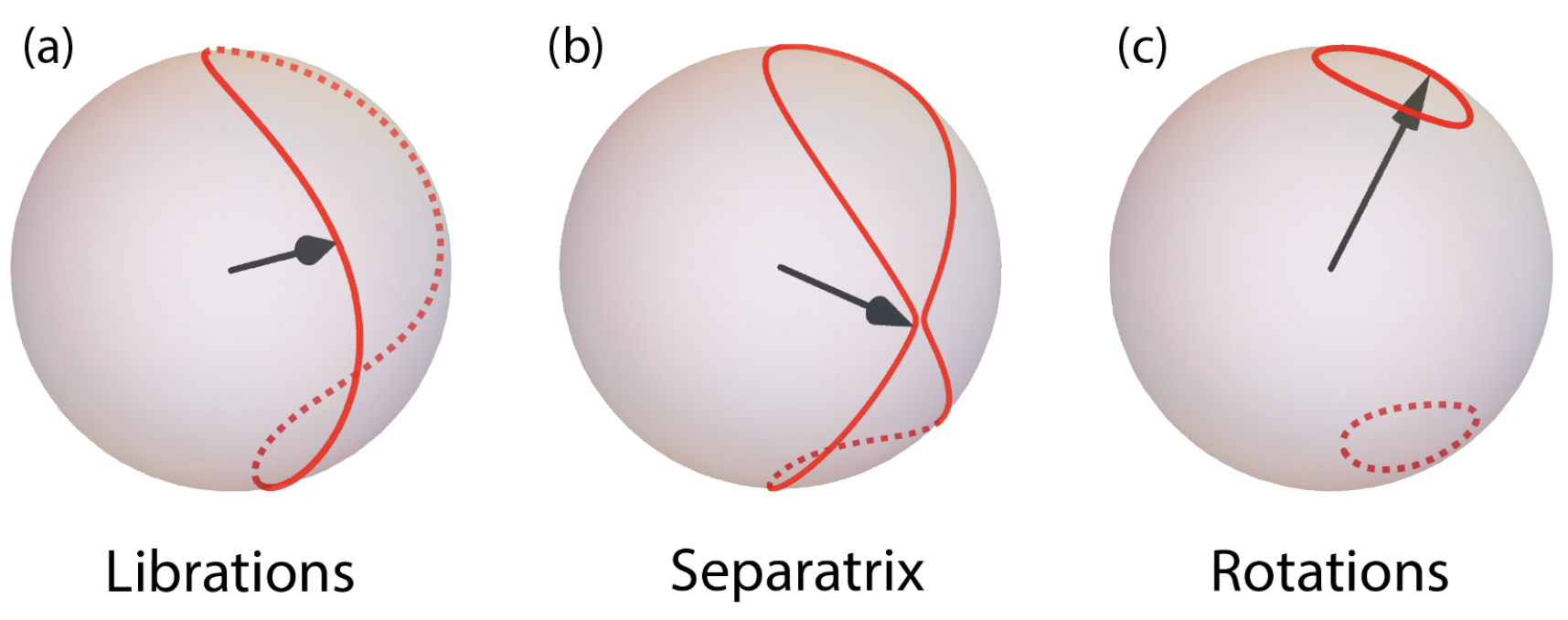}
\caption{
Possible periodic trajectories of one classical spin under conditions that (i)  the spin chain (\ref{ham})  has energy $E=0$ and (ii) all spins point in the same direction:   (a) Libration ($J=0.79$); conditions (i) and (ii) imply a single trajectory. (b) Separatrix ($J = J^* \approx 1.15$) separating librations and rotations. It consists of an unstable fixed point in the middle and two branches approaching the fixed point at $t \to \pm \infty$. (c) Rotations ($J=1.76$): conditions (i) and (ii) are fulfilled by two disconnected trajectories. 
}
\label{SP_solutions_red}
\end{figure}

\subsection{Lyapunov instabilities: general considerations}
\label{Lyapunov1}

Chaos in nonlinearly interacting classical systems is caused by instabilities of phase space trajectories with respect to small deviations of initial conditions. The stability of a phase trajectory in a $2L$-dimensional phase space is characterised by a spectrum of $2L$ Lyapunov exponents $\{ \lambda_k \}$.   The trajectory is called unstable when its largest Lyapunov exponent $\lambda_{\text{max}}$ is positive. 

To compute $\lambda_{\text{max}}$, one chooses a reference phase space trajectory $\mathcal{S}(t)$ and an infinitesimally close one $\mathcal{S}(t) + \delta \mathcal{S}(t)$, then traces the growth of  $\delta \mathcal{S}(t) $, which is eventually overtaken by the largest Lyapunov exponent. This implies that 
\begin{equation}
    \lambda_{\text{max}} =  \lim_{\substack{ t \to \infty, \\ |\delta \mathcal{S}(0) | \to 0}} \frac{1}{t} \log \frac{|\delta \mathcal{S}(t) |}{|\delta \mathcal{S}(0) |}.
    \label{lambdamax-def}
\end{equation}
The finiteness of $\lambda_{\text{max}}$ for the lattices of classical spins with finite number of interacting neighbors is proven in Appendix~\ref{Lambda_Bound} following the general argument of Ref.~\cite{dewijn2013lyapunov}.
Practical numerical computation of $\lambda_{\text{max}}$ requires one to perform a large number of resets (contractions) of $|\delta \mathcal{S}(t)|$ as described in Refs.\cite{benettin1976kolmogorov,Benettin-1980,wimberger2014nonlinear} and in Appendix~\ref{appendix_A}.

Largest Lyapunov exponents and the entire Lyapunov spectra  of classical spin systems have previously been investigated \cite{dewijn2012largest,dewijn2013lyapunov} but only in the vicinity of ergodic trajectories (also called ``stochastic''), which nonperiodically cover the energy shell of the system. If a trajectory is ergodic, it passes in the vicinity of every point on the ergodic part of the energy shell, thereby ``collecting" all possible local growth rates of $|\delta \mathcal{S}(t)|$ and averaging over them. As a result,  the largest ``ergodic" Lyapunov exponent, to be denoted as $\lambda_{\text{e}}$, and the entire Lyapunov spectrum do not depend on the choice of the initial conditions $| \mathcal{S}(0)|$ \cite{benettin1976kolmogorov}. The convergence of the procedure for calculating $\lambda_{\text{e}}$ is, in fact, a good measure of the ergodisation time of the underlying dynamics\cite{tarkhov2018estimating}.  Plots of $\lambda_{\text{e}}$ computed for the ergodic trajectories of our spin chain with different $J$ and $L$ can be found in Appendix~\ref{ergodic}.

The periodic phase space trajectories in many-body systems are rather exceptional as they correspond to subset of initial conditions of measure zero. Periodic trajectories are not ergodic, hence they do not cover the entire energy shell and thus are not supposed to have the same Lyapunov spectrum as the ergodic trajectories. Moreover, it cannot be  excluded {\it a priori} that they are Lyapunov-stable even when their ergodic counterparts are unstable. 

We denote  the largest positive Lyapunov exponent of a spin lattice  in the vicinity of the periodic trajectories $\mathcal{S}_{\text{p}}(t)$ as $\lambda_{\text{p}}$. The numerical algorithm for computing $\lambda_{\text{p}}$ is the same as the one for computing $\lambda_{\text{max}}$ in the ergodic case. 
%The only practical difference is that the reference periodic trajectory is to be computed during only one period and then repeated in time. 

The specific periodic trajectories $\mathcal{S}_{\text{p}}(t)$  considered in this paper have one further defining property, namely, they are translationally invariant in a system where the Hamiltonian is also translationally invariant. To explore the consequences of this property, let us consider  the  operator $\mathcal{L}_t$ known as the fundamental matrix\cite{gaspard1998}, which governs the growth of small deviations in the Lyapunov problem through the relation
\begin{equation}
\delta \mathcal{S}(t) = \mathcal{L}_t[\delta \mathcal{S}(0)]
    \label{L}
\end{equation}
The operator $\mathcal{L}_t[\delta \mathcal{S}(0)]$ is linear with respect to $\delta \mathcal{S}(0)$.
The problem of computing the Lyapunov spectrum is an eigenvalue problem for  $\mathcal{L}_t$. The eigenvalues have the form $e^{\lambda_k t + i \omega_k t}$, where the frequency $\omega_k$ controls the phase factor when present. Each eigenvalue corresponds to an eigenvector $\delta \mathcal{S}_k $.  
Since both the underlying Hamiltonian and the reference trajectory are translationally invariant on the lattice, the operator $\mathcal{L}_t$ must also be translationally invariant. In such a case, the eigenvectors of this operator must realise irreducible representations of the lattice translational symmetry group, which have the form $e^{i \mathbf{q} \mathbf{r}}$, where $\mathbf{r}$ the coordinate of the lattice site and $\mathbf{q}$ is a wave vector. In other words, each Lyapunov exponent $\lambda_k$,   corresponds to an instability around the reference periodic trajectory $\mathcal{S}_{\text{p}}(t)$, which develops with a well defined wave vector $\mathbf{q}_k$. 

Generically, different wave vectors $\mathbf{q}_k$ are supposed to correspond to different values of $\lambda_k$ with 
the exception of symmetry-related or accidental degeneracies. A symmetry-related degeneracy occurs, in particular, when a lattice, such as the one we consider, has inversion symmetry, and, as a result, pairs of wave vectors $\pm \mathbf{q}_k$ correspond to the same $\lambda_k$. In this case, a generic Lyapunov exponent is at least double-degenerate, the possible non-degenerate cases correspond to  $\mathbf{q}_k =0$ and  $\mathbf{q}_k = \pi/L$.

From now on, we focus on our one-dimensional lattice, where symmetry representations are characterised by wave numbers $q$ rather than wave vectors, and so are the Lyapunov instabilities. The above discussion implies that, when the periodic trajectory $\mathcal{S}_{\text{p}}(t)$ is Lyapunov unstable, the instability corresponding to the largest Lyapunov exponent $\lambda_{\text{p}}$ generically develops out of the initially random and hence not translationally invariant deviation $\delta \mathcal{S}(0)$ as a standing wave in real space characterised by a pair of wave numbers $\pm q_{\text{p}}$.

\subsection{Dependence of $\lambda_{\text{p}}$ on $J$ and $L$}

\subsubsection{Overview}
\label{overview}

\begin{figure}
\includegraphics[width=\columnwidth]{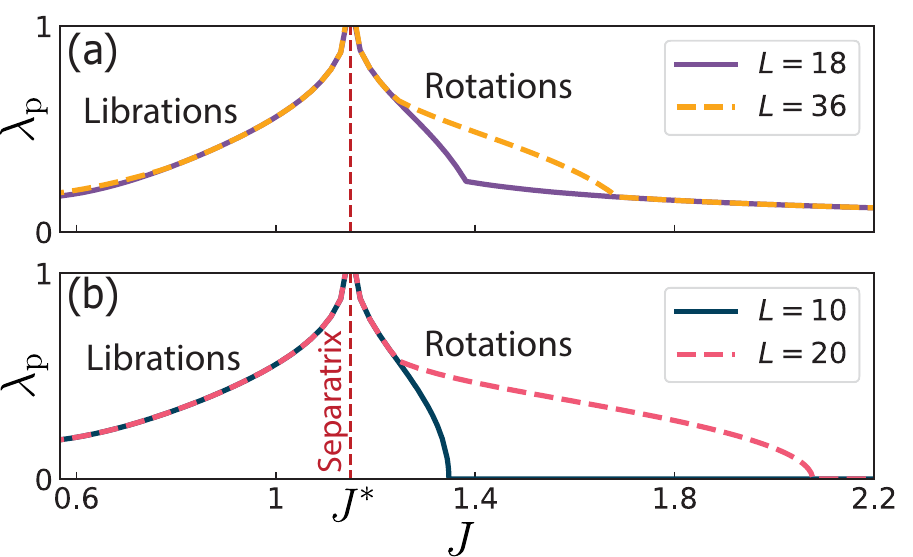}
\caption{
Dependence of periodic Lyapunov exponent $\lambda_{\text{p}}$ on the coupling constant $J$ at constant local field $h=1$  for spin chains of different sizes $L$. The separatrix value $J=J^*$ divides the libration and rotation regimes.  Chains of different lengths $L$ that are multiples of each other often exhibit intervals of equal $\lambda_{\text{p}}$, which, as explained in the text, happens  when the wave numbers characterising the Lyapunov vectors associated with $\lambda_{\text{p}}$ have the same values. On the rotation side, the plots exhibit kinks due to switching of the above wave numbers. A part of the plot coinciding with the horisontal axis implies that $\lambda_{\text{p}} =0$, i.e. the chain is Lyapunov stable.
$\lambda_{\text{p}}(J)$ in the close vicinity of $J=J^*$ is plotted in Fig.~\ref{nearSep}.
\label{gemD}}
\end{figure}

\begin{figure*}
\includegraphics[width=2\columnwidth]{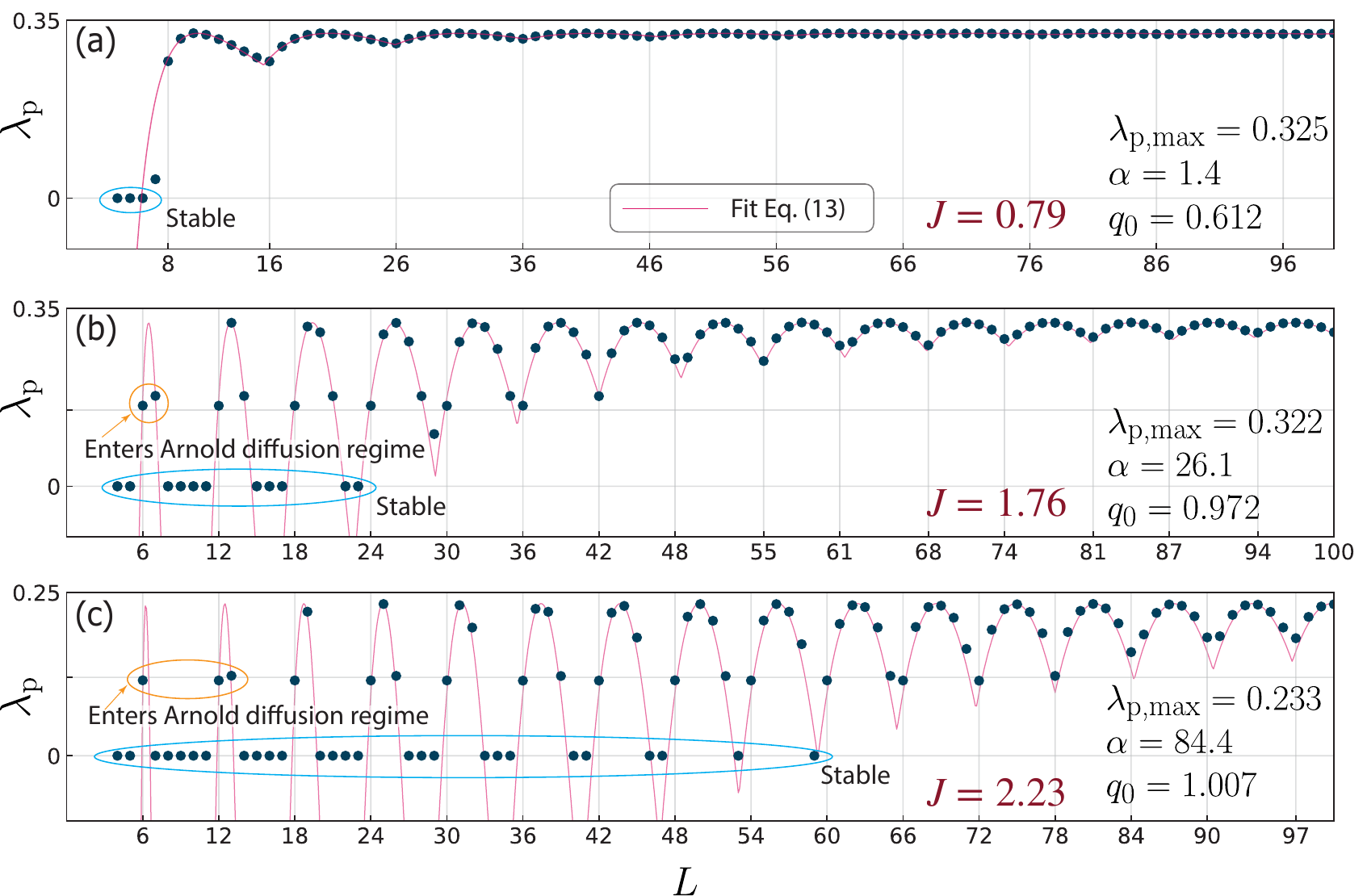}
\caption{
Dependence of periodic Lyapunov exponent $\lambda_{\text{p}}$ on chain length $L$ for three different values of $J$: (a) ~$J=0.79$, (b) $J=1.76$ and  (c) $J=2.23$. Panel (a) corresponds to librations, while panels (b) and (c) correspond to rotations. Points represent numerically computed $\lambda_{\text{p}}$. Red lines are the fits based on Eq.(\ref{lambdap-L}) with parameters indicated in the plot legends. Points  marked as ``Stable" correspond to the $\lambda_{\text{p}} =0$ verified as explained in Appendix~\ref{appendix_A}. In a few cases marked as entering Arnold diffusion regime, Lyapunov instability around a periodic trajectory is followed by anomalously long-living nearly quasiperiodic behavior --- see Section~\ref{quasiperiodic} for further discussion.
\label{LVSL}}
\end{figure*}

The plots of the computed $\lambda_{\text{p}}$ as functions of the interaction strength $J$ and the system size $L$ are presented in Figs.~\ref{gemD} and \ref{LVSL}: Fig.~\ref{gemD} shows $\lambda_{\text{p}}(J)$  for four system sizes $L$, 
while Fig.~\ref{LVSL} displays $\lambda_{\text{p}} (L)$ for three different values of $J$. 
We now describe and discuss these results.

As explained in Section \ref{librot}, the periodic trajectory that starts from the all-spins-up state of our spin chain,   undergoes a transition from the libration to the rotation regime as a function of the interaction parameter $J$  for fixed  $h = 1$ at $J=J^*\approx 1.15$, which corresponds to a separatrix. 

Figures \ref{gemD} and \ref{LVSL}(a) present the results for the libration regime $J<J^*$. The typical dependence of $\lambda_{\text{p}}(J)$ in this case   is a monotonic growth approaching a peculiar singularity at the separatrix value $J=J^*$. (The singularity itself is to be discussed in the subsection~\ref{separatrix} below.)
The dependence $\lambda_{\text{p}}(L)$ in Fig.~\ref{LVSL}(a) starts at 0 for small $L$, followed by a sharp increase and then by a weakly oscillatory dependence as $\lambda_{\text{p}}(L)$ approaches a constant value for $L \to \infty$.

The rotation regime $J>J^*$ has a richer phenomenology evident in Figs. \ref{gemD} and \ref{LVSL}(b,c). 
The function $\lambda_{\text{p}}(J)$ on the rotation side of Fig.~\ref{gemD}, when not equal to zero,  exhibits monotonic decrease interrupted by kinks whose origin is to be explained in subsection~\ref{translational}.  These kinks can be contrasted with the featureless monotonic growth of $\lambda_{\text{p}}(J)$ for librations. Also, as seen in Figs.\ref{LVSL}(b,c), the rotations entail a much stronger oscillatory dependence of $\lambda_{\text{p}}(L)$ than librations.  
We further note that the occurrence of $\lambda_{\text{p}} = 0$ in Fig.~\ref{LVSL} for rotations  extends to significantly larger values of $L$ than for librations. Also, as exemplified in Fig.~\ref{gemD}(b), $\lambda_{\text{p}}(J)$ on the rotation side often becomes equal to 0 above a finite threshold of $J$.

Overall, the strong and nontrivial $J$- and $L$-dependencies of the largest Lyapunov exponent $\lambda_{\text{p}}$ for periodic trajectories (Figs.~\ref{gemD} and \ref{LVSL})  should be compared with the rather featureless $J$- and $L$-dependencies of
the largest Lyapunov exponent $\lambda_{\text{e}}$ for  nonperiodic ergodic trajectories on the same energy shell (Fig.~\ref{lexp_ergpic} in Appendix~\ref{ergodic}). 
Let us also mention that $\lambda_{\text{p}}$ can be either larger or smaller than $\lambda_{\text{e}}$. 
For example, $\lambda_{\text{p}} \approx 0.35$ for a chain with $J = 1.76$ in the limit of large $L$ [see Fig.~\ref{LVSL}(b)], which is significantly smaller than the ergodic exponent $\lambda_{\text{e}} \approx 1.13$ for the same $J$   (see Fig.~\ref{lexp_ergpic}). On the other hand, the periodic exponent near the separatrix $J =J^* $ reaches the value $\lambda_{\text{p}} \simeq 1.43$ (see Fig.~\ref{nearSep}), which is noticeably larger than  the ergodic exponent $\lambda_{\text{e}} \simeq 0.67$ in Fig.~\ref{lexp_ergpic}   for the same $J$.

\subsubsection{ Translational symmetry breaking by Lyapunov vectors and the analytical approximation for $\lambda_{\text{p}}(L)$.}
\label{translational}

The rather diverse phenomenology summarized in the preceding part can be brought into a single perspective on the basis of the considerations of Section \ref{Lyapunov1}, that the Lyapunov vectors lower the full translational symmetry of the system to the one characterized by a particular wave number $q = q_{\text{p}}$. 

\begin{figure}
\includegraphics[width=\columnwidth]{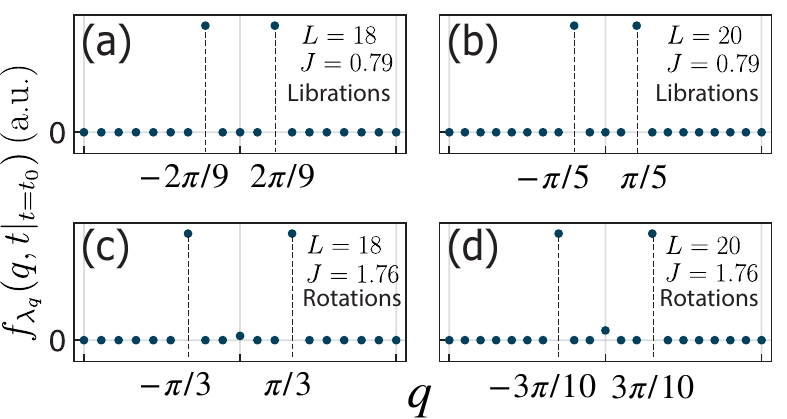}
\caption{Spatial Fourier spectra $f_{\lambda_{\text{p}}} (q, t = t_0)$ of the Lyapunov vectors defined by Eq.~(\ref{flambda}). The chain lengths $L$ and the coupling constants $J$ are indicated in plot legends.   The allowed values of wavenumbers are $q=\frac{2\pi}{L} k$, where $k$ is an integer in the interval $[-L/2, L/2]$, $t_0=750$.  
\label{fourSpace}}
\end{figure}

In Fig.~\ref{fourSpace}, we present four examples of the spatial Fourier spectra
\begin{equation}
 f_{\lambda_{\text{p}}} (q, t) \equiv 
 \frac{1}{L}
\left| 
 \sum_{m=1}^{m=L} \delta \mathbf{S}_m(t) \  e^{-i q m} 
\right|^2
    \label{flambda}
\end{equation}
of the Lyapunov vectors $ \delta {\cal S}(t)$ corresponding to $\lambda_{\text{p}}$ with time $t$ being large enough for the computed $\lambda_{\text{p}}$ to reach the stationary value. Here 
$\delta \mathbf{S}_m \equiv (\delta S_m^x, \delta S_m^y, \delta S_m^z)$ are the components of the high-dimensional vector $ \delta {\cal S}$ entering Eq.(\ref{L}); the Fourier amplitude inside the absolute-value bars $|...|$   is a three-dimensional vector with $x$, $y$ and $z$ complex-valued projections,  $|...|^2$ is the sum of the squares of the moduli of those projections.
The Fourier spectra in  Fig.~\ref{fourSpace} are computed for two chain lengths $L = 18$ and $L=20$ and for two interaction constants $J$ representing the libration and the rotation regimes for each length. In all four cases, one can see two dominant Fourier peaks corresponding to $\pm q_{\text{p}}$.

The underlying translational symmetry breaking helps us to explain the distinct behavior of   $\lambda_{\text{p}}(J)$ plotted in Fig.\ref{gemD}: In both panels of Fig.\ref{gemD}, there are intervals of $J$ where Lyapunov exponents $\lambda_{\text{p}}$ for two chains of different lengths coincide with each other, and also there are intervals where the two Lyapunov exponents suddenly become different, because one of of the two functions $\lambda_{\text{p}}(J)$ exhibits  a kink.  The pair of plots in Fig. \ref{gemD}(a) and another pair in Fig. \ref{gemD}(b) -- both correspond to chain lengths such that one chain is twice the length of the other. Therefore, the wave numbers $q= \frac{2\pi}{L} n $, where $n$ is an integer and $L$ the length of the shorter chain,  are the allowed wave numbers for the both chains,
but then the longer chain has additional allowed wave numbers that are not available for the shorter one. Thus, if the largest Lyapunov exponent for the longer chain corresponds to $q$ available for the shorter chain, then the two Lyapunov exponents $\lambda_{\text{p}}$  are equal to each other. If, however, the above $q$ for the longer chain is not available for the shorter one, then the values of $\lambda_{\text{p}}$ become different, with the larger $\lambda_{\text{p}}$ necessarily corresponding to the longer chain. The fact that the above kind of switching behavior is exhibited in Fig.~\ref{gemD} only in the rotation regime  is, probably, a coincidence; we did not identify any fundamental reason for the librations not to exhibit such a switching.

We now convert the qualitative understanding of the role of the translational symmetry breaking into an analytical formula describing the entire dependence $\lambda_{\text{p}}(L)$ exhibited in Fig.~\ref{LVSL}. 

We conjecture that the oscillatory dependence of $\lambda_{\text{p}}$ on $L$ in Fig.\ref{LVSL}, originates from the fact that, for a chain of a given length $L$, the allowed values of the wave number $q = \frac{2 \pi}{L} n$, where $n = 0, 1, ..., L-1$, cannot exactly match a certain value $q_0$ that is optimal for maximising the Lyapunov exponent $\lambda_{\text{p}}$ in a chain of infinite length. Let us denote that maximal value as $\lambda_{\text{p,max}}$.  As the length of the chain increases, the allowed values of $q$ come increasingly closer to $q_0$, and, as a result, the oscillatory part of $\lambda_{\text{p}}(L)$ becomes gradually suppressed.
We are dealing with a supposedly small difference between $q_0$ and the nearest allowed value  
$q_{\text{p}} = \frac{2 \pi}{L} \  \text{round} \left( \frac{q_0 L}{2 \pi}  
\right)$, where function ``$\text{round}(...)$" rounds its argument to the nearest integer value. It is thus reasonable to parameterise the departure from $\lambda_{\text{p,max}}$ by a quadratic dependence on $(q_0 - q_{\text{p}})$:
\begin{equation}
\lambda_{\text{p}}(L) = \lambda_{\text{p,max}} - \alpha \left[
q_0 - \frac{2 \pi}{L} \ \text{round} \left( \frac{q_0 L}{2 \pi}  
\right) 
\right]^2,
    \label{lambdap-L}
\end{equation}
where $\lambda_{\text{p, max}}$, $q_0$ and $\alpha$ are three adjustable parameters. 

The remarkably good performance of formula (\ref{lambdap-L}) in describing the numerically computed $\lambda_{\text{p}}(L)$ is demostrated in Fig.\ref{LVSL}. One can observe that it gives an excellent quantitative description also in the regime of large differences between $\lambda_{\text{p}}(L)$ and $\lambda_{\text{p,max}}$. In this large-deviations regime, the formula (\ref{lambdap-L})
has to be further supplemented by the condition that, if it predicts $\lambda_{\text{p}}<0$, then $\lambda_{\text{p}} =0$ should be substituted instead, implying that the periodic motion is Lyapunov stable. With such a modification, formula (\ref{lambdap-L}) turns into an excellent predictor of the lengths $L$ corresponding to the stable periodic motion. We, finally, note, that the quality of the approximation (\ref{lambdap-L}) can be improved further by adding there terms with higher-order powers of $(q_0 - q_{\text{p}})$.

\subsubsection{ $\lambda_{\text{p}}(J)$ in the vicinity of a separatrix.}
\label{separatrix}

\begin{figure}
\includegraphics[width=0.6\columnwidth]{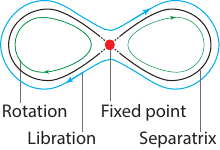}
\caption{
Schematic representation of a fixed point and a separatrix between librations and rotations.}
\label{LRFS}
\end{figure}

\begin{figure}
\includegraphics[width=\columnwidth]{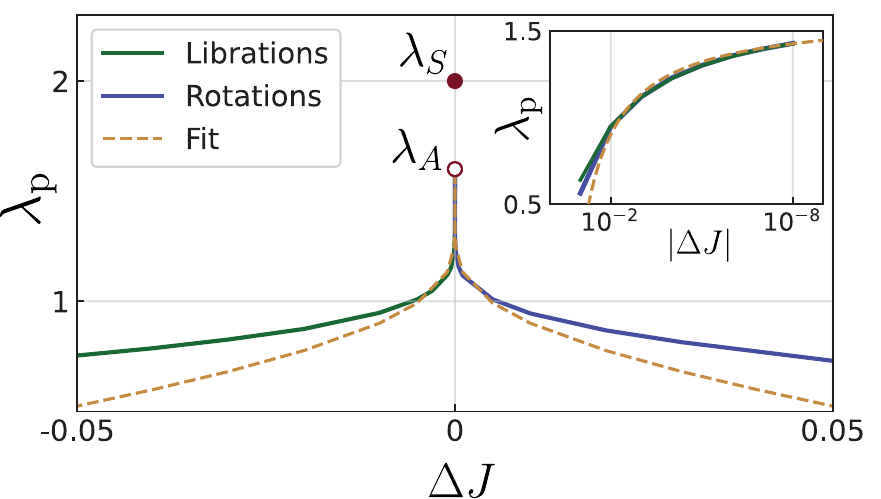}
\caption{
Periodic Lyapunov exponent $\lambda_{\text{p}}(\Delta J)$ computed as a function of the difference $\Delta J \equiv J - J^*$ in the close vicinity of the separatrix corresponding to $\Delta J = 0$. Solid green and blue lines represent the numerically computed $\lambda_{\text{p}}$ on the libration side ($\Delta J <0$) and the rotation side ($\Delta J >0$) respectively; dashed line is the fit given by Eq.(\ref{lmdp-logDJ}); $\lambda_{\text{A}}$ is the asymptotic value on the approach to the separatrix defined by Eq.(\ref{lpJ-pm}); $\lambda_{\text{S}}$ is the value characterising the fixed point.
The inset shows the same data on the semilog plot thereby illustrating their convergence to the fit (\ref{lmdp-logDJ}).
\label{nearSep}}
\end{figure}

The function $\lambda_{\text{p}}(J)$  shown in 
Fig.~\ref{gemD}
exhibits tendency to diverge on the approach of $J$ to the numerically computed separatrix value $J^* = 1.1504059085$. Yet, due to the finiteness of the Hamiltonian parameters $J$ and $h$ and the smoothness of the dynamics, the rate of the divergence between two close phase space trajectories  is limited from above by a number of the order of $J$ or $h$, hence the value of $\lambda_{\text{p}}(J)$ cannot diverge (see Appendix~\ref{Lambda_Bound} for the derivation). 

In this part, we zoom in on the behavior $\lambda_{\text{p}}(J)$ in the close vicinity of the separatrix shown in Fig.~\ref{SP_solutions_red}.  As schematically illustrated in Fig.~{\ref{LRFS}}, the separatrix consists of an unstable fixed point analogous to an upward pointing pendulum and two trajectories starting infinitesimally close to that fixed point at time  $t = -\infty$ and approaching it infinitesimally close  at $t = +\infty$. 
The above fixed point is simultaneously a many-spin trajectory belonging to the $q=0$ subspace of the full phase space. This  trajectory is unique in the sense that, in comparison with the previously considered periodic trajectories, it possesses an additional instability (towards the branches of the separatrix) developing entirely within the $q=0$ subspace, as opposed to involving the subspaces with $q \neq 0$. In the case considered below, that additional instability actually has a Lyapunov exponent which is larger than the largest one for the periodic trajectories.   
The results of our numerical calculation of $\lambda_{\text{p}}(J)$ for a chain of 10 spins  in the close vicinity of the separatrix value $J=J^*$ are presented Fig.~\ref{nearSep}. This calculation becomes increasingly demanding computationally as $J$ approaches $J^*$, because (i) the topological character of the periodic trajectories changes from a connected libration to two disconnected rotations, and (ii) this change is accompanied by the slowing down of the dynamics. Our findings indicate that $\lambda_{\text{p}}(J)$ for $J \to J^*$ has the form of a symmetric cusp approaching the finite value $\lambda_A = 1.6$ with infinite derivative $\lambda'_{\text{p}}(J)$: 
\begin{eqnarray}
\lambda_{\text{p}}(J) & \xrightarrow[J \to J^* \pm 0]{} & \lambda_A ;
\label{lpJ-pm}\\
\lambda'_{\text{p}}(J) & \xrightarrow[J \to J^* \pm 0]{} & \mp \infty .
\label{lppJ-pm}
\end{eqnarray}
Yet, as indicated in Fig.~\ref{nearSep}, for the exact separatrix value $J = J^*$, we obtain another Lyapunov exponent $\lambda_{\text{p}}(J^*) \equiv \lambda_S = 1.99 > \lambda_A$. The latter result is computed by choosing the unstable fixed point on the separatrix as the reference trajectory $\mathcal{S}(t)$ and then extracting $\lambda_S$ as the exponential growth rate of the small deviations $|\delta \mathcal{S}(t)|$ starting from $|\delta \mathcal{S}(0)| \approx 10^{-6} \sqrt{L}$ and letting it grow up to $10^{-1} \sqrt{L}$ without making any resets (see Appendix~\ref{appendix_sep} for further details).  In a related finding, we also observed a switching  of the wave number $q_{\text{p}}(J)$ corresponding to $\lambda_{\text{p}}(J)$, namely: $q_{\text{p}}(J^*) =0$, while $q_{\text{p}}(J) = \pi/5 $ for $J \neq J^*$ in the vicinity of the separatrix. 

We were able to obtain analytically the following symmetric shape  of the cusp function $\lambda_{\text{p}}(J)$ in the vicinity of $J = J^*$: 
\begin{equation}
    \lambda_{\text{p}}(J) \approx  
    \lambda_A 
+ \frac{C}{\log |\Delta J|} ,
    \label{lmdp-logDJ}
\end{equation}
where $\Delta J \equiv J - J^*$,  $\lambda_A$ is the asymptotic value of $\lambda_{\text{p}}$ defined by Eq.(\ref{lpJ-pm}) and  $C$ is a constant. The derivation of Eq.(\ref{lmdp-logDJ}) is relegated to Appendix~\ref{F-form-lJ}. Here we only mention that Fig.~\ref{nearSep} demonstrates that the functional form (\ref{lmdp-logDJ}) is quantitatively consistent with our numerical results.

\subsection{Observability of Lyapunov instability in a many-particle system in the vicinity of a periodic trajectory}
\label{observability}

\begin{figure}
\includegraphics[width=\columnwidth]{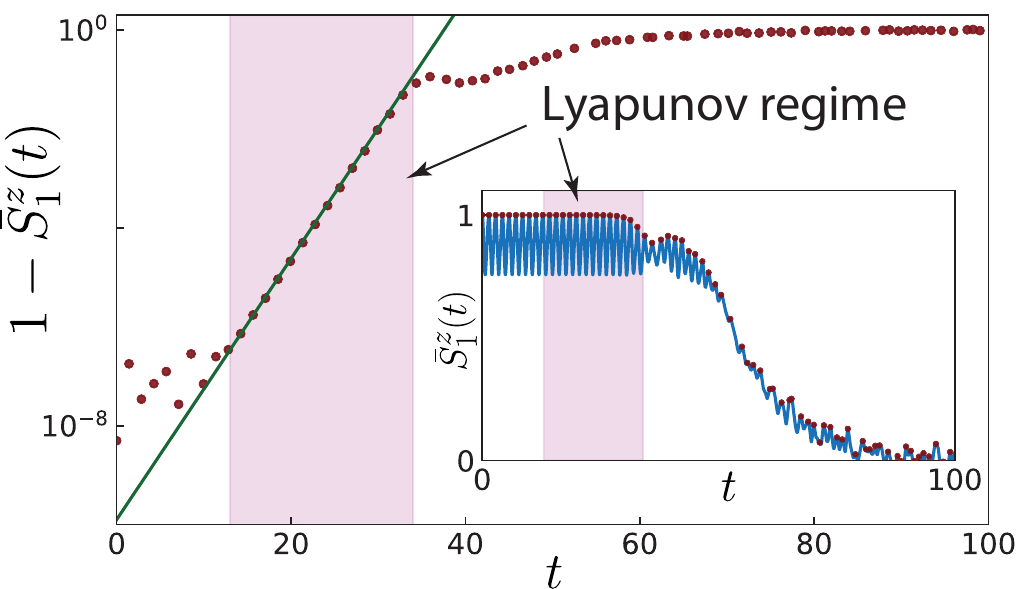}
\caption{Deviation of a slightly perturbed ensemble-averaged dynamics of the $z$-projection of one spin, $\bar{S}^z_1(t)$, from the maximum value 1 corresponding to a perfectly periodic motion. The inset displays $\bar{S}^z_1(t)$ (blue solid line). This plot was computed for a spin chain of length $L=100$ with interaction parameter   ${J=1.76}$. The averaging was performed  over the ensemble of $N=1000$ realisations of $S^z_1(t)$  with  initial conditions ${\mathbf{S}_n(0) = \left(\delta S_{n0}^x, \delta S_{n0}^y, \sqrt{1 - \delta {S_{n0}^x}^{\! \!  ^2} - \delta {S_{n0}^y}^{\! \!  ^2}  } \right)}$,  where $\delta S_{n0}^x$  and $\delta S_{n0}^y$ were randomly sampled from within the circle $ \sqrt{\delta {S_{n0}^x}^{\! \!  ^2} + \delta {S_{n0}^y}^{\! \!  ^2}} \leq 10^{-4}$. The main panel is a semilog plot of the points $1- \bar{S}^z_1(t_m)$, where times $t_m$ correspond to the maxima of $\bar{S}^z_1(t_m)$ plotted in the inset. These points are represented by the red dots in both the main frame and the inset.  The highlighted time interval marks the Lyapunov regime of the exponential growth of $1-\bar{S}^z_1(t_\text{m})$.
The green solid line is the exponential fit
$ \exp(2\lambda_p t_\text{m})$, where $\lambda_p = 0.303$ is the value directly computed in the many-spin phase space. [The initial scatter of the points in the main panel is due to the mismatch between the discretised time points and the true maxima of $\bar{S}^z_1(t)$.]
\label{Ensemble}}
\end{figure}

In this part, we propose a method of experimentally observing the Lyapunov instability around a periodic trajectory in a system of many classical spins. A possible realization may be based on mechanical spin-like simulators, such as those implemented in Ref.~\cite{Nash-2015}. The challenge here is that a direct method based on a very accurate control and monitoring of system's phase trajectory in many-particle phase space  is, likely, impractical. As we show below, it is, in fact, not necessary to monitor the entire phase space trajectory -- rather it is sufficient to monitor just one observable. A similar approach was adopted previously in Refs.~\cite{fine2014absence,elsayed2015sensitivity,tarkhov2017extracting,tarkhov2018estimating}, where the Lyapunov exponents for the ergodic trajectories were obtained through monitoring ensemble-averaged time evolution of a single variable in the course of a slightly imperfect time reversal of the dynamics. 
[In that case, the approach to extracting the Lyapunov exponents involves ``out-of-time-order correlators"(OTOCs)\cite{OTOC}.]

Here we propose to extract the largest Lyapunov exponent of a many-spin periodic trajectory by  monitoring the time evolution of just one projection of a single spin. The Lyapunov exponent is to be extracted from   the growth of the ensemble-averaged deviations of an imperfect periodic behavior of the observed variable from a perfectly periodic one. The deviations are supposed to be unavoidably present in any experiment due to an imperfect preparation of the initial state. When the trajectory is exponentially unstable, the deviations from perfect periodicity in many-particle phase space start growing exponentially after some initial time delay with the rate controlled by $\lambda_{\text{p}}$. This exponential growth is then projected onto the time evolution of the observed variable.  

We implemented the above approach by monitoring the $z$-projection of one of the spins, namely, $S_1^z(t)$. The result is presented in Fig.~\ref{Ensemble}.  Our ideal initial condition (\ref{eqclassini})  is such that  all spins are pointing in the $z$-direction, i.e. $S_1^z(0) = 1$, which is also the maximum possible value of $S_1^z$.  The numerically implemented initial conditions are chosen such that each spin is rotated from the perfect orientation along the $z$-axis by tiny random angles of the order of $10^{-4}$. We then average the function $S_1^z(t)$ over $N=1000$  independently initialised evolutions of the system and denote the result as $\bar{S}_1^z(t)$.  The  subsequent maxima of $\bar{S}_1^z(t)$ at times $t = t_{\text{m}}$ represent the points of the closest return to $\bar{S}_1^z(0)$. Figure~\ref{Ensemble}  shows that there is a regime where 
\begin{equation}
    1 - \bar{S}_1^z(t_{\text{m}}) \ \propto \  \exp(2 \lambda_{\text{p}} t_{\text{m}} ).
    \label{1-S1z}
\end{equation}
The exponential growth rate in Eq.(\ref{1-S1z}) is  $2 \lambda_{\text{p}}$ rather than $\lambda_{\text{p}}$, because the maxima at $t = t_{\text{m}}$ represent the points of the closest return to the orientation ${\mathbf{S}_1(0) = (0, 0, 1 )}$, where $S_1^x(t_{\text{m}})$ and $S_1^y(t_{\text{m}})$ are supposed to be much smaller than 1 in the Lyapunov regime while growing proportionally to $\exp( \lambda_{\text{p}} t_{\text{m}}) $. At the same time, ${S_1^z(t_{\text{m}}) \approx 1 - \frac{1}{2}\left[{S_1^x}^2(t_{\text{m}}) + {S_1^y}^2(t_{\text{m}}) \right]}$, which leads to Eq.(\ref{1-S1z}).

\section{Classical spins: transient nearly quasiperiodic regime}
\label{quasiperiodic}

\begin{figure}
\includegraphics[width=\columnwidth]{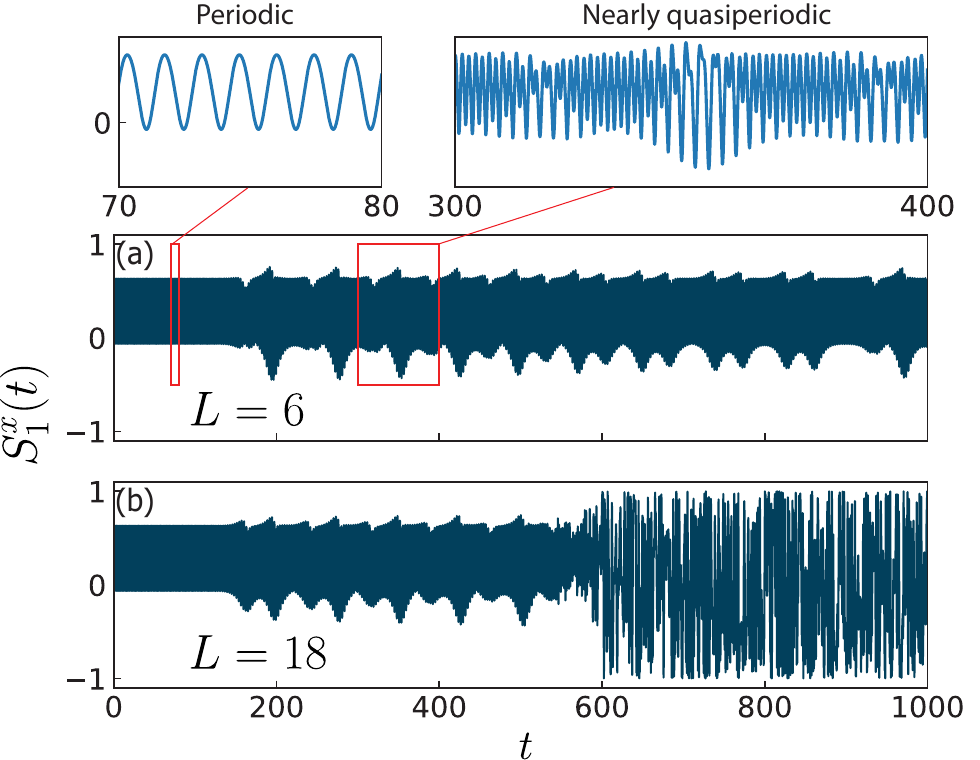}
\caption{Onset of nearly quasiperiodic regime monitored through the time evolution of one projection of one spin, namely, $S^x_1(t)$. The plots are obtained for $J=1.76$ with $L=6$ in panel (a) and $L=18$ in panel (b). The initial dynamics in the both panels is periodic entering the nearly quasiperiodic regime around $t\sim 150$. Two insets above panel (a) magnify the respective rather dense plots.  The nearly quasiperiodic regime in panel (a) has an anomalously long lifetime $\sim 10^6$ further illustrated in Fig.~\ref{QP-Breakdown}.  In panel (b), the nearly quasiperiodic regime terminates by the onset of the ergodic regime around $t \sim 600$. The  periodic, nearly quasiperiodic and ergodic regimes in panel (b) were used to generate the spin trajectories exhibited  in Fig.~\ref{scheme_periodic}. 
\label{transDyn}}
\end{figure}

\begin{figure}
\includegraphics[width=\columnwidth]{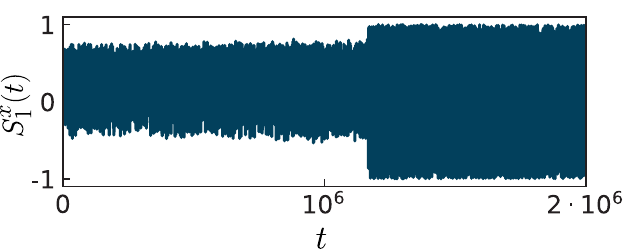}
\caption{
Illustration of the anomalously long-living nearly quasiperiodic regime for the spin chain with $J=1.76$ and  $L=6$. The initial time evolution of this plot for $t < 1000$ is shown in Fig.~\ref{transDyn}. The nearly quasiperiodic regime breaks down at around $t \approx 1.15 \cdot 10^6$.
\label{QP-Breakdown}}
\end{figure}

While computing the Lyapunov exponents for periodic trajectories, we discovered that, as the Lyapunov instability develops, the deviations  from the reference periodic trajectory become large and no longer describable by Lyapunov exponents, and yet the dynamics does not immediately become ergodic --- rather it often enters   a nearly quasiperiodic regime illustrated in Fig.~\ref{transDyn}. The lifetime of this regime exhibits strong variations: it may be relatively short --- a few time units (one unit is $1/h = 1$), or fairly long --- a few hundred time units --- as is the case in Fig.~\ref{transDyn}(b), or anomalously long --- meaning $\sim 10^6$ time units or more; one such example is shown in Figs.~\ref{transDyn}(a) and \ref{QP-Breakdown}. Later in this section we identify the anomalously long-lived quasiperiodic regime with the phenomenon of Arnold diffusion.  In our simulations we came across only five cases of this regime, which are marked as such in Figs.\ref{LVSL}(b,c).

\subsection{Frequency spectra }
\label{Fspectra}

\begin{figure*}
\includegraphics[width=2\columnwidth]{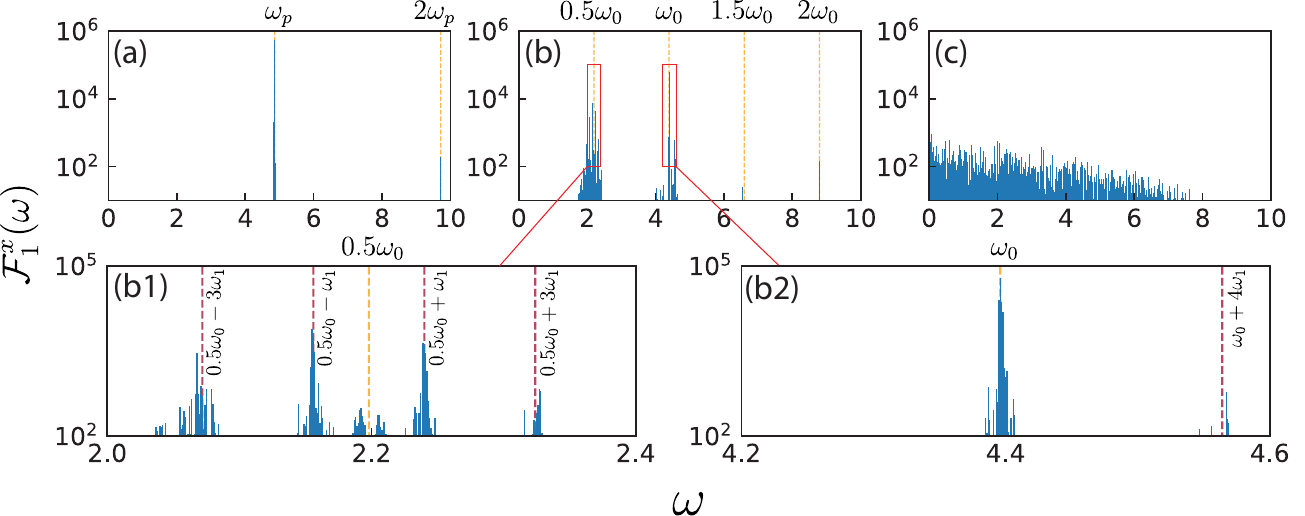}
\caption{
Frequency spectra (\ref{Fomega}) of $S_1^x(t)$ in periodic (a), nearly quasiperiodic (b) and ergodic (c) dynamical regimes of classical spin chains with $J=1.76$.   In all cases, the spectra are extracted from time intervals of width $\Delta t=6000$ with $10\%$ Tukey window\cite{bingham1967modern,harris1978use} applied.  Spectrum (a) is obtained for the original periodic trajectory, which is the same for all chain lengths. Spectrum (b) is obtained for the anomalously long nearly quasiperiodic regime of the chain of length $L=6$ plotted in Figs.\ref{transDyn}(a) and \ref{QP-Breakdown} starting at $t=300$. Spectrum (c) is obtained for the ergodic regime of the chain of length $L=18$ plotted Fig.\ref{transDyn}(b) starting at $t=1000$. Panel (a) exhibits the main peak at the frequency of the periodic motion $\omega_p$ and the higher order peak at $2 \omega_p$. Panel (b) and the magnifying panels (b1) and (b2) exhibit the spectrum dominated by the peaks at $n \cdot 0.5 \omega_0 + m \omega_1$, where $m$, and $n$ are integer numbers; however, not all values of $m$ and $n$ are present and additional small peaks can also be seen. Panel (c) exhibits a typical continuous spectrum of chaotic dynamics.
\label{fourierSpec}}
\end{figure*}

A quasiperiodic regime can be distinguished from the periodic and the chaotic regimes on the basis of  the frequency spectrum of a typical degree of freedom.  When the motion is periodic, its spectrum  is supposed to consist of discrete peaks at frequencies, all of which are the multiples of a fundamental frequency $\omega_{\text{p}}$. For a chaotic motion, the spectrum is expected to be continuous (up to statistical fluctuations). The spectrum of a quasiperiodic motion inherits a discrete peak structure from the periodic motion, but the peaks are determined by the combinations of two or more frequencies, whose ratio is an irrational number. A nearly quasiperiodic regime is expected to exhibit frequency spectra reminiscent of a strictly quasiperiodic regime when the spectra are extracted from the time evolution over not too long time intervals.

In Fig.\ref{fourierSpec}, we show the examples of numerically computed frequency spectra
\begin{equation}
 {\cal F}_1^x (\omega) \equiv 
\left| 
 \int_{- \infty}^{\infty} S_1^x(t) e^{-i \omega t} dt
\right|^2
    \label{Fomega}
\end{equation}
for $S_1^x(t)$ within spin lattices exhibiting periodic, quasiperiodic and ergodic behavior. 
Representative time series behind these spectra can be seen  in Fig.~\ref{transDyn} -- all obtained with $J = 1.76$. The integral in Eq.~(\ref{Fomega}) is computed over a long but finite time interval $\Delta t = 6000$. The spectrum in Fig.~\ref{fourierSpec}(a) is for the original periodic trajectory; the spectrum in Fig.~\ref{fourierSpec}(b) is for a time interval  $\Delta t$ within the anomalously long-lived quasiperiodic regime of a 6-spin chain, and the spectrum in Fig.~\ref{fourierSpec}(c) is for the chaotic regime of the 18-spin chain that follows the quasiperiodic regime in that system.   

The tendency to quasiperiodicity in the spectrum in Fig.~\ref{fourierSpec}(b) [also magnified in frames (b1,b2)] can be  identified through the presence of discrete peaks at frequencies $ m \left(\frac{1}{2} \omega_0\right) + n \omega_1$, where $m$ and $n$ are integer numbers, $\omega_0$ is the primary frequency close to $\omega_{\text{p}}$ for the original periodic orbit  and responsible for the fast oscillations in Fig.~\ref{transDyn}, the halved frequency step $\frac{1}{2} \omega_0$ reflects the period doubling accompanying the instability of the original periodic trajectory, and the smaller frequency $\omega_1$ characterises the slow modulations of the fast oscillations. The numerically fitted frequency ratio for the particular spectrum in Figs.~\ref{fourierSpec}(b, b1, b2) is $\omega_0/\omega_1 \approx 10.5$. The transient character of the supposed quasiperiodicity precludes us from quantifying that number more accurately, thereby making the distinction between a truly irrational ratio and a rational one with a large denominator practically irrelevant.

\subsection{Dynamics in a four-dimensional subspace of a many-dimensional phase space as the origin of the nearly quasiperiodic regime}
\label{Origin}

Given the discussion in the subsection~\ref{translational} of the translational symmetry breaking accompanying Lyapunov instabilities, the explanation of the origin of the nearly quasiperiodic regime can be the following. The initial periodic motion corresponds to  the wave number $q=0$. The motion that emerges as a result of the Lyapunov instability is characterised by wave vector $q=q_{\text{p}} \neq 0$.
As the Lyapunov instability develops, the spectral weight becomes gradually transferred from the component with the wave number $q=0$ to the one with  $q=q_{\text{p}}$. (The latter mode can be treated as a superposition of two complex-valued waves proportional to  $e^{\pm i q_{\text{p}} r}$ and forming a real-valued standing wave.)  The growth of the second mode saturates when the amplitude of that mode becomes comparable with that of the $q=0$ mode, while the modes at other $q$ are still too small.
The phase space of the emerging dynamics becomes effectively four-dimensional. Systems with four-dimensional phase spaces often have finite regions of non-chaotic quasiperiodic dynamics  separated from the regions of stochastic dynamics\cite{arnold1978mathematical,lichtenberg2013regular,feingold1983regular}. This is because a quasiperiodic trajectory in a four-dimensional phase space, when present, covers a two-dimensional torus within a three-dimensional energy shell, thereby dividing the energy shell into two parts with the border impenetrable for stochastic trajectories. When a trajectory ends up in a stochastic regime, it can be strongly or weakly chaotic. A weakly chaotic trajectory spends a long time in the vicinity of a truly quasiperiodic trajectory. We believe that our numerical evidence of the quasiperiodic behavior is of the latter kind. 

The motion departing from an unstable periodic trajectory is expected to end up in the stochastic part  in the above 4-dimensional phase space on the basis of the following argument.  Consider a two-dimensional Poincaré surface of section (PSS) perpendicular to the unstable periodic trajectory within the three-dimensional energy shell. The point where that trajectory crosses the PSS is an unstable fixed point for the corresponding 2D dynamical map. An unstable fixed point in a 2D subspace should be accompanied by a separatrix. The motion in a vicinity of a separatrix is supposed to form a stochastic layer\cite{lichtenberg2013regular}, because the time between the subsequent crossings of the PSS  diverges on the approach to the separatrix, implying that there is an increasingly dense set of orbits satisfying the resonance condition $\mathcal{T}_1 = n \mathcal{T}_0$, where $\mathcal{T}_0$ and $\mathcal{T}_1$ are, respectively, the shorter and the longer periods on the KAM-torus and $n$ is an integer. As a result, a stochastic layer emerges, and our trajectory should find itself  ``stuck" within that layer. Such a regime would mean a suppressed ergodicity with a strong quasiperiodic tendency. In particular, the frequency of the slower periodic motion would exhibit a drift over time that is longer than the already long period of that motion. This regime would persist until the four-dimensional character of the dynamics is destroyed by the mechanisms to be introduced in the next part.

\subsection{Mechanisms limiting the lifetime of the nearly quasiperiodic regime: }
\label{Mechanisms}

\begin{figure*}
\includegraphics[width=2\columnwidth]{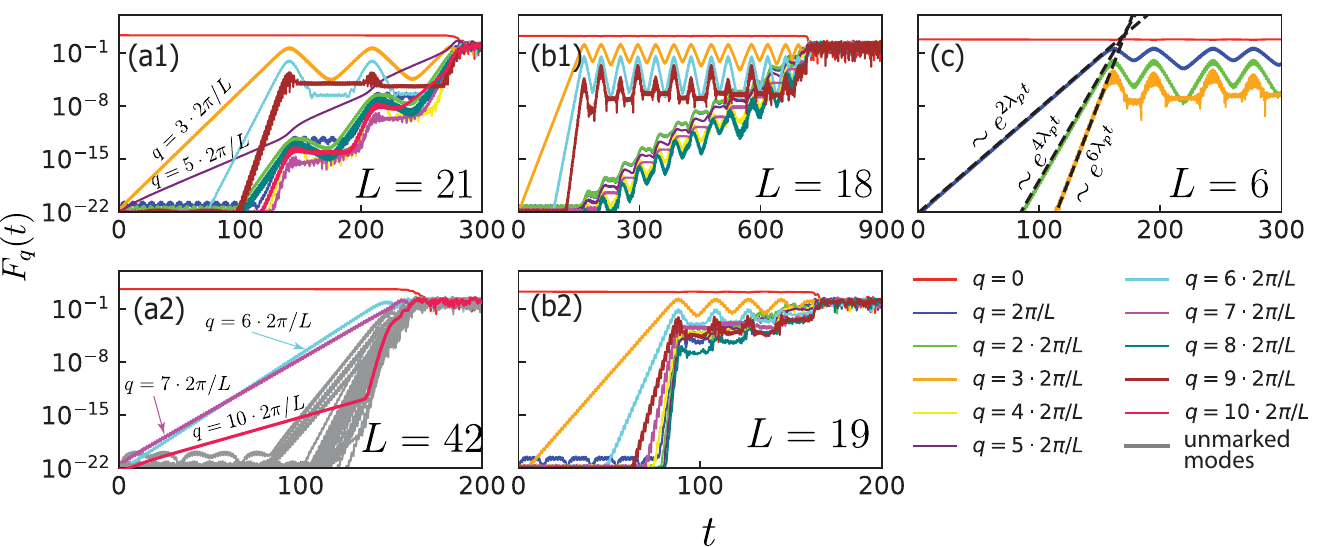}
\caption{Time evolution of  intensities (\ref{Fq}) of  spatial Fourier modes after a tiny random perturbation around a periodic trajectory.  Wave numbers $q$ of Fourier modes are indicated in the plot legend. In all cases, $J=1.76$, while $L$ is given in the plot panels.
The intensity at $q=0$ starts with the value $L$; all other intensities start at $\sim 10^{-22}$. The fastest growing intensity at $q \neq 0$  then destroys the periodic regime and initiates a nearly quasiperiodic one. As indicated in panel (c) but applicable to all panels, the fastest mode growing as $ \exp(2 \lambda_{\text{p}} t) $ ``pulls" with itself higher order harmonics emerging later and growing as $ \exp(4 \lambda_{\text{p}} t) $, $ \exp(6 \lambda_{\text{p}} t) $, etc.  but not signifying additional instabilities --- see subsection~\ref{q-modes}.  Different panels illustrate different mechanisms of destroying the nearly quasiperiodic regime: (a1) and (a2) Mechanism~A: additional Lyapunov instabilities around the original periodic trajectory --- see subsection~\ref{q-criterion}; (b1) and (b2) Mechanism~B: additional Lyapunov instability in the nearly quasiperiodic regime --- subsection~\ref{2nd-mechanism}; (c) Mechanism C: here the absence of additional Lyapunov instabilities required for Mechanisms A and B make the slowest mechanism, namely, the Arnold diffusion operational  --- see subsection~\ref{anomalousQP}.
In the latter case, the nearly quasiperiodic regime is destroyed far beyond the plotted range --- see Fig.~\ref{QP-Breakdown}. 
\label{Fig:Fourier_modes}}
\end{figure*}

We identify three mechanisms that limit the lifetime of the nearly quasiperiodic regime in our system. Each of them will be discussed separately in the following subsections.

Mechanism A is the Lyapunov instability around the original periodic trajectory at yet another wave vector $q = q_2$ that eventually catches up with the modes at $q=0$ and $q = q_{\text{p}}$. The effective phase space then becomes 6-dimensional where the chaotic regions are no longer constrained by 3-dimensional tori of quasiperiodic motion making the occurrence of a stable or nearly stable quasiperiodic motion unlikely. This mechanism is most common for longer chains. When present, it is also the fastest one; we will further characterise it in subsection~\ref{q-criterion}.

Mechanism B becomes relevant, when no additional unstable modes at $q \neq q_{\text{p}}$ exist around the original periodic trajectory. According to Mechanism B, an additional Lyapunov instability still develops around the emergent nearly quasiperiodic trajectory, because that trajectory is noticeably different from the original periodic one, which leads to a different set of Lyapunov exponents. [Compare,  e.g., the frequency $\omega_{\text{p}}$ of the original periodic motion in Fig.~\ref{fourierSpec}~(a)   with the dominant frequency $\omega_0$ of the nearly quasiperiodic motion in Fig.~\ref{fourierSpec}~(b)]. With Mechanism B, as with the previous one, the additional Lyapunov instability destroys the approximate four-dimensional character of the nearly quasiperiodic dynamics. This mechanism is to be discussed further in subsection~\ref{2nd-mechanism}.

Mechanism C is to be identified with the Arnold diffusion. It is the slowest and the most exotic mechanism, becoming operational when no additional Lyapunov instabilities according to the previous two mechanisms arise.  In the process of Arnold diffusion, a phase trajectory stays for a long time in the vicinity of one KAM torus in a high-dimensional phase space, then switches to the vicinity of a different one, continuing these switches  and thereby evolving arbitrarily far from the original KAM torus. While the original context of the Arnold diffusion was the dynamics of nearly integrable systems, our spin system is a strongly non-integrable one with a small part of the phase space, where the dynamics is weakly non-integrable.  The adaption of the Arnold diffusion picture to our case is to be introduced in subsection~\ref{anomalousQP}.

\subsection{Time evolution of Fourier modes with different wave numbers $q$}
\label{q-modes}

In order to substantiate the existence of the three mechanisms identified in the preceding subsection and to further analyse their onset, we numerically computed the time evolutions of the combined intensities $F_{q} (t)$ of the spatial Fourier modes for all three spin projections with a given wave number $q$. These combined intensities are defined as:
\begin{equation}
 F_{q} (t) \equiv 
\frac{1}{L}\left| 
 \sum_{m=1}^{m=L}  \mathbf{S}_m(t) \  e^{-i q m} 
\right|^2.
    \label{Fq}
\end{equation}

Figure~\ref{Fig:Fourier_modes} shows several representative examples of the time evolutions of the Fourier mode intensities $F_{q} (t)$ before, during and after the quasiperiodic regime. Simulations behind all examples in Fig.~\ref{Fig:Fourier_modes} start with an initial condition corresponding to our original translationally invariant periodic trajectory with tiny random deviations of all spins $|\delta \mathbf{S}_m| \sim 10^{-11}$, meaning that the normalised initial intensity of the mode at $q=0$ is $F_0(0)/L \approx 1$, while for all modes with $q \neq 0$,  $F_q(0)/L \sim 10^{-22}$. All plots then show that the Lyapunov instability initiates the exponential growth of a leading unstable mode $F_{q_p}(t) \propto \exp({2 \lambda_{\text{p}} t})$ at wavenumber $q_p\neq 0$, as expected. In all cases, $F_{q_p}(t)$ reaches a maximum value and then starts exhibiting imperfect oscillations as the system enters the nearly quasiperiodic regime.

We further note that the exponential onset of each $q = q_p$ mode in Fig.~\ref{Fig:Fourier_modes} is accompanied by the subsequent onset of the modes growing as $F_q(t) \propto \exp({2 n \lambda_{\text{p}} t})$, where $n$ is a natural number.
These modes are explicitly indicated by the respective exponential fits in Fig.~\ref{Fig:Fourier_modes}(c). The straight lines representing these exponential fits on a semilog plot tend to intersect roughly at the same point located on the red line  representing the mode $F_0(t)$. This phenomenology can be explained as follows. The growth of the leading $q = q_p$ mode  induces a modulation of the chain background, which, on the one hand, constitutes a dominant perturbation to the original translationally invariant periodic trajectory but, on the other hand, still remains very small as long as $F_{q_p}(t) \ll F_0(t)$. As a result, the spatial Fourier harmonics with $q = n q_p$ become coupled to the $q = q_p$ mode in the $n$th order of the perturbation theory, and their amplitudes emerge at later times out of statistical noise exhibiting the Lyapunov growth   with the rate $n \lambda_p$, [$2 n \lambda_p$ for $F_q(t)$]. This phenomenology, however, does not signify independent modes -- rather they are perturbation-induced admixtures to the original unstable $q = q_p$ mode. That perturbation can be described in terms of backfolding of the chain's Brillouin zone, which is to be discussed further below in subsection~\ref{2nd-mechanism} in the context of Mechanism B.

Figures~\ref{Fig:Fourier_modes}(a1, a2) illustrate Mechanism A of destroying the quasiperiodic regime, namely, the instability of a third mode beginning at $t=0$ with a smaller Lyapunov exponent. That mode eventually catches up with the mode at $q= q_p$ thereby inducing the chaotic regime.

Mechanism B, namely, the secondary Lyapunov instability appearing not at $t=0$ but rather  around the onset time of the quasiperiodic regime [understood as the time of the first maximum of $F_{q_p}(t)$] is illustrated Figs.~\ref{Fig:Fourier_modes}(b1,~b2). 

Finally, Fig.~\ref{Fig:Fourier_modes}(c) illustrates the initial evolution of a relatively rare example, where the nearly quasiperiodic regime remains stable during an anomalously long time as compared to the lifetimes in the range from 100 to 1000 exhibited in Figs.~\ref{Fig:Fourier_modes}(a1-b2).  In the particular case of Fig.~\ref{Fig:Fourier_modes}(c), the quasiperiodic regime becomes destroyed only after time  $t \sim 10^6$ as illustrated in Fig.~\ref{QP-Breakdown}. This time exhibits significant fluctuations dependent on the initial randomly chosen tiny deviations from the periodic trajectory.  The destruction of the nearly quasiperiodic regime in this case is governed by the previously mentioned Mechanism C associated with Arnold diffusion. 

We now turn to analysing our numerical findings for each of the three mechanisms in greater detail.

\subsection{Mechanism A: Criterion for the presence of subdominant Lyapunov instabilities around the periodic orbit}
\label{q-criterion}

\begin{figure}
\includegraphics[width=\columnwidth]{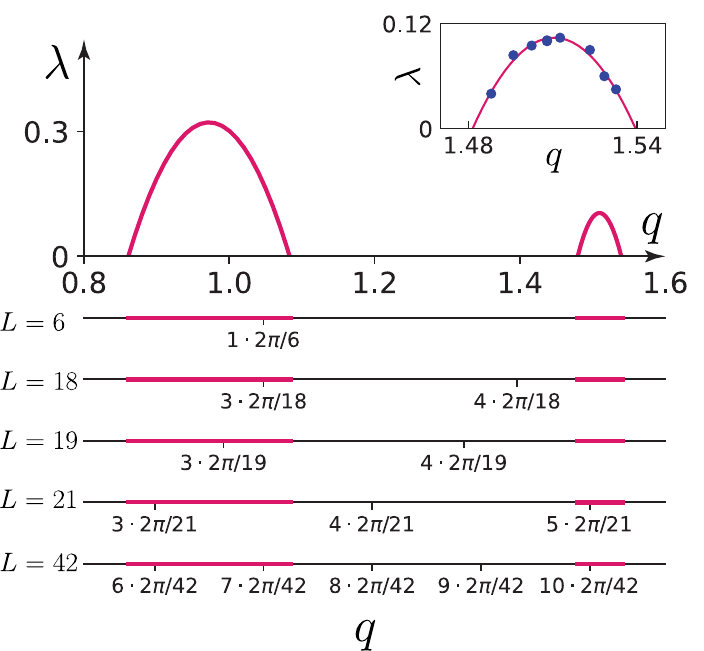}
\caption{
Illustration of criterion for the onset of Mechanism A that destroys the nearly quasiperiodic regime for spin chains with $J=1.76$ and various lengths $L$.
The upper plot $\lambda(q)$ shows parabolic approximations (\ref{lambda-q})  and (\ref{lambda-q1}) for  two ranges of wave numbers $q$ of the Fourier modes~\ref{Fq} possessing positive Lyapunov exponents $\lambda$ that lead to instabilities around the original periodic trajectory. The parameters for the larger range (\ref{lambda-q}) are given in Fig.~\ref{LVSL}(b). The smaller range (\ref{lambda-q1}) is obtained with $q_1 = 1.51$, $\lambda_{\text{1,max}} = 0.104$ and  $\alpha_1 = 1.23 \cdot 10^2$ extracted from the fit to the numerical data shown in the inset. The lengths $L=6, 18, 19, 21, 42$ correspond to chains whose Fourier modes are plotted in Fig.~\ref{Fig:Fourier_modes}. The horisontal lines next to each $L$ represent the $q$-axes with allowed values $q = \frac{2 \pi}{L} n$ marked and the unstable ranges magenta-colored. In each case, there is, at least, one allowed $q$-value falling within an unstable range and responsible for the primary Lyapunov instability around the original periodic trajectory and the subsequent onset of the nearly quasiperiodic regime.  The onset of Mechanism~A requires at least one more allowed $q$ within an unstable range. Such $q$ is present for $L=21$ and $L=42$, hence Mechanism~A is operational --- see Figs.~\ref{Fig:Fourier_modes}(a1,a2). For $L = 6, 18$ and 19, there are no additional values $q = \frac{2 \pi}{L} n$ in the unstable ranges, hence the Mechanism~A is not operational --- see Figs.~\ref{Fig:Fourier_modes}(b1,b2,c).
\label{fig-criterion}}
\end{figure}

Given that not all examples presented in Fig.~\ref{Fig:Fourier_modes} exhibit  Mechanism A, the question arises whether one can identify a relevant criterion here.  

Mechanism A requires the presence of subdominant Lyapunov instabilities around the initial periodic trajectory in addition to the primary one at $q = q_p$. 
In turn, according to the considerations leading to Eq.(\ref{lambdap-L}), the existence of the primary Lyapunov instability requires at least one of the allowed values $q = \frac{2 \pi}{L} n$ to fall within the ``unstable" range defined by condition 
\begin{equation}
    \lambda(q) = \lambda_{\text{p,max}} - \alpha \left(
q_0 - q \right)^2 >0.
\label{lambda-q}
\end{equation} 
After reviewing  the mode-intensity plots of the kind presented in Fig.~\ref{Fig:Fourier_modes} for spin chains with $J= 1.76$ and the lengths of up to $L=85$, we found empirically the following criterion that fairly accurately predicts the presence of the subdominant instabilities and thus the onset of Mechanism A: For a chain of length $L$, at least one more of the  allowed values $q = \frac{2 \pi}{L} n$ in addition to $q_p$  should fall  within the same unstable range   that was found for the primary instability on the basis of condition (\ref{lambda-q}) with the same parameters $\lambda_{\text{p,max}}$, $q_0$ and $\alpha$. [For $J=1.76$, these parameters are given in Fig.~\ref{LVSL}(b).] 

The overall excellent agreement of the above criterion  with our numerical tests is, however, subject to the following three caveats: 
First, as illustrated in Fig.~\ref{fig-criterion},  we identified another much narrower window of unstable modes 
\begin{equation}
    \lambda(q) = \lambda_{\text{1,max}} - \alpha_1 \left(
q_1 - q \right)^2 >0,
\label{lambda-q1}
\end{equation}
where $q_1$, $\lambda_{\text{1,max}}$ and  $\alpha_1$ are fitting parameters --- see the inset of Fig.~\ref{fig-criterion}.
Second, the end points of the both ``unstable" ranges of $q$ are slightly shifted with respect to those obtained on the basis of inequalities (\ref{lambda-q}) and (\ref{lambda-q1}) --- likely consequence of the cubic and higher-order terms in the expansion of $\lambda(q)$ as a function of $\left(
q_0 - q \right)$ or $\left(
q_1 - q \right)$. 
Third, there may exist additional smaller $q$-windows for unstable modes undetected by the present investigation. (Somewhat similar parameter dependence with multiple stable and unstable ranges was reported in Ref.~\cite{schweizer1988lyapunov} for classical orbits of the hydrogen atom in magnetic field.)

The subdominant instability originating from the second unstable window can be observed, e.g., for the chain with $L=21$ presented in Fig.~\ref{Fig:Fourier_modes}(a1): it is exhibited by the mode $q =  \frac{10 \pi}{21}$. In this case, it actually determines the lifetime of the quasiperiodic regime. The same unstable mode can also be seen in Fig.~\ref{Fig:Fourier_modes}(a2) [identified with $q =  \frac{20 \pi}{L}$, where $L=42$], but here the lifetime time of the nearly quasiperiodic regime is limited by a stronger subdominant mode at $q =  \frac{14 \pi}{42}$  within the main $q$-window (\ref{lambda-q}).

We note that, since both the maximum of $\lambda(q)$   and the unstable $q$-range for the second window (\ref{lambda-q1}) are smaller than their counterparts for the main window (\ref{lambda-q}), the former did not manifest itself in $\lambda_{\text{p}}(L)$ plotted in Fig.~\ref{LVSL}(b).   

The complete criterion for the onset of Mechanism~A taking into account both unstable $q$-windows is illustrated in Fig.~\ref{fig-criterion}. 

The above criterion implies that, as the chain length $L$ increases, more than one allowed values $q = \frac{2 \pi}{L} n$ must fall within the main unstable $q$-window (\ref{lambda-q}), which means that not only the quasiperiodic regime unavoidably becomes destroyed through Mechanism A as the chain length grows, but also the duration of that regime becomes shorter due to smaller differences between leading and subleading Lyapunov exponents. The comparison between the chain lengths $L=21$ and $L=42$ in Figs.~\ref{Fig:Fourier_modes}(a1) and (a2) illustrates this general trend.

\subsection{Mechanism B and the backfolding of the Brillouin zone }
\label{2nd-mechanism}

According to the preceding discussion, a not too long spin chain can have only one Lyapunov-unstable mode around the initial periodic trajectory -- the mode with $q=q_p$. In such a case, Mechanism A is not operational, so that Mechanism B can, possibly, take over. The latter stipulates that the trajectory in the nearly quasiperiodic regime is sufficiently different from the initial periodic one and thus can exhibit an additional Lyapunov instability that the periodic trajectory did not have.

In order to classify these possible additional instabilities, we note that, since the quasiperiodic motion involves a $q_{\text{p}}\neq 0$ mode, the setting of the respective Lyapunov problem is no longer translationally invariant with respect to the spin chain in the sense that the instantaneous spin pattern does not repeat under the translation $\mathbf{S}_m \to \mathbf{S}_{m+1}$ (periodically closed).  Moreover, the wave number $q_{\text{p}} = 2 \pi n /L $, where $n$ is an integer, corresponds to the spatial wavelength  $\Lambda_{\text{p}} \equiv {2\pi \over q_{\text{p}} } = L/n$, which is, in general, not an integer number and thus cannot be a period with which a spin pattern repeats itself.  That period is rather  $m \Lambda_{\text{p}}$, where $m$ is the minimal natural number that makes $m \Lambda_{\text{p}} = m L/n $ an integer. This implies that $u \equiv n/m$ should be the largest common divider of $L$ and $n$. The spatial period of the $q_{\text{p}}$-mode is thus $L/u$, while the corresponding wave number determining the size of the reduced Brillouin zone  of the two-mode dynamics is $Q_u = 2 \pi u/L$. 
The wave numbers from the original Brillouin zone of the chain  $[ -\pi, \pi]$ should now be backfolded into the new reduced Brillouin zone $[ -Q_u/2, Q_u/2]$. The backfolded modes then form $L/u$ ``bands" of Lyapunov exponents.

We note, in particular, that $u=1$ when $L$ and $n$ have no common dividers, in which case the spatial period is equal to $L$, while $Q_u = 2 \pi /L$. This is, in particular guaranteed to happen when $L$ is a prime number.
The case $u=1$ implies that the reduced Brillouin zone  has just one $q$-point, i.e. all the wavevectors from the original Brilloin Zone become backfolded on the top of each other, which, in particular, implies that the initial Lyapunov instability of the $q = q_p$ mode will eventually become coupled by the higher order perturbations with the modes at all other wave numbers, thereby exciting all $F_q(t)$. Two examples of such a behavior are  Figs.~\ref{Fig:Fourier_modes}(b2) and (c): In the former case, $L=19$ while $n=3$. In the latter case, $L=6$, while $n=1$.

When $u > 1$, the reduced Brillouin zone has $u$ different $q$ points -- each representing a group of the backfolded $q$ values, in which case the modes belonging to the group containing the wave vectors $\pm q_{\text{p}}$ become perturbatively coupled to the $q_{\text{p}}$-mode, while the rest of the modes remain uninvolved. 

As mentioned in subsection~\ref{q-modes}, the  growth rates $2 n \lambda_{\text{p}}$ of $F_q(t)$ for  the modes coupled with $q_{\text{p}}$ only reflect the fact that these modes contribute to the same leading Lyapunov instability during the onset of the quasiperiodic regime. In other words, the presence of the  initial growth rates $2 n \lambda_{\text{p}}$ does not, as such, signify additional instabilities.

The additional Lyapunov instabilities around the quasiperiodic trajectory  may still emerge later in the subspace of modes coupled to the $q_{\text{p}}$-mode, or they can appear  among the modes not coupled to $q_{\text{p}}$.
The latter situation is more straightforward to diagnose numerically because it implies that a previously unexcited Fourier intensity $F_q(t)$ would start exponentially growing from around the initial tiny value. The former situation is somewhat more difficult to identify, as the effect appears on the top of the already excited intensities $F_q(t)$, but still it can be seen as an exponential destruction of the quasiperiodic regime occurring with a rate, which is expected to be significantly smaller than $2 \lambda_{\text{p}}$. 

We have encountered both of the above scenarios. The scenario with the easily observable exponential growth of initially very small Fourier intensities can be seen in  Fig.~\ref{Fig:Fourier_modes}(b1) for $L = 18$. In this case, $q_p = 3 \ \frac{2\pi}{L}$, which implies that the number of the allowed values of $q$ in the reduced Brillouin zone is $u = 3$ (the largest common divider of 3 and 18). One of these values is $q = 0$, which also ``houses" $q = \pm q_p$ after backfolding, while the Lyapunov instability destroying the nearly quasiperiodic regime emerges in the subspace of the $q$-values not backfolded onto $q=0$. 
The alternative scenario, where the already excited Fourier intensities $F_q(t)$  exhibit a slower additional exponential growth is shown in Fig.~\ref{Fig:Fourier_modes}(b2) for $L=19$. That case, as discussed earlier, corresponds to $u=1$.

To summarise, the present subsection has described the initial phenomenology of Mechanism B without proposing any criterion for the onset of it. We only identified two relevant scenarios in terms of the localisation of unstable Fourier modes in the reduced Brillouin zone. Yet the underlying discussion implies that there might be cases where Mechanism B is not activated, because none of the initially stable Fourier modes becomes unstable in the nearly quasiperiodic regime.

\subsection{Mechanism C: Arnold diffusion }
\label{anomalousQP}

As mentioned earlier, we have encountered several examples indicated in Fig.~\ref{LVSL}, where neither Mechanism A nor Mechanism B are operational and hence the lifetime of the nearly quasiperiodic regime becomes anomalously long. We now put forward a heuristic case that the Mechanism C behind the destruction of the nearly quasiperiodic regime in those examples is the Arnold diffusion. It requires that there are no other finite-time Lyapunov instabilities in the vicinity of the nearly quasiperiodic trajectory besides the one that initiated that trajectory. [The term ``finite-time Lyapunov instability'' implies an instability averaged over a time period much shorter than the very long duration of the Arnold diffusion regime -- to be distinguished from the definition~(\ref{lambdamax-def}).]   

The idea of Arnold diffusion goes back to Arnold's article of 1964 \cite{Arnold1964diffusion}, where an example was given, which implied that a stochastic regime  of weakly nonintegrable dynamical systems with phase space dimensions larger than 4 should be unconstrained. The time required to explore the stochastic part of the energy shell was, however, exponentially long - proportional to the exponent of a negative power of the perturbation strength - and thus not accessible by a perturbation theory.  The above proposition was contrasted with the phase spaces of dimensions $\leq 4$, where the KAM tori of the quasiperiodic motion surviving  a finite small perturbation would separate the existing stochastic regions, thereby making the stochastic part of the energy shell disconnected. We note that the term ``Arnold diffusion" does not mean a diffusion in a conventional sense but rather stands for unboundedness of the stochastic motion. Although Arnold diffusion has remained in the focus of active mathematical and physical research~\cite{Lochak-1999,Treschev-2012,Gelfreich-2017,Clarke-2023, CHIRIKOV1979review,lichtenberg2013regular,Gutzwiller-1990,Piro-1988,chirikov1990kam,chirikov1993fastarnold,chirikov1997jetp,shepelyansky2011strongandweak,Schmidt-2023}, the relevance of this phenomenon to realistic many-body physical systems has remained rather enigmatic.

To connect to the present spin dynamics context, let us introduce a variable for the total spin polarisation of a spin chain of length $L$:
\begin{equation}
\mathbf{M} \equiv  \sum_{i=1}^L \mathbf{S}_i,
    \label{M}
\end{equation}
and then  define for that chain an integrable Hamiltonian 
\begin{eqnarray}
&&
{\cal H}_0  \equiv   - \frac{1}{2} (J_0 M_x^2 + 2 J_0 M_y^2) + h M_x + h M_y  
\label{H0}
\\
&&
= - \frac{1}{2} \sum_{i=1}^L\sum_{j=1}^L (J_0 S_i^x S_j^x + 2 J_0 S_i^y S_j^y) + \sum_{i=1}^L (h S_i^x + h S_i^y)
    \nonumber 
\end{eqnarray}
where $J_0 = 2 J/L$. The parameters $J$ and $h$ are the same as in  the Hamiltonian (\ref{ham}). The Hamiltonian (\ref{H0}) is integrable, because its dynamics is reduced to that of one vector $\mathbf{M}$ subject to the Poisson brackets  
$\{M^\alpha, M^\beta \} =  \epsilon_{\alpha\beta\gamma} M^\gamma $, which leaves the length of that vector time-independent and hence limits the dynamics to an integrable line on the spherical surface of length $|\mathbf{M}(0)|$. In fact, the  Hamiltonian (\ref{H0}) is a many-spin counterpart of the one-spin Hamiltonian (\ref{ham0}).

For the initial condition (\ref{eqclassini}) ``all spins up", the integrable Hamiltonian (\ref{H0})  generates the same periodic trajectory for each spin as our original Hamiltonian (\ref{ham}).
We now reexpress the latter as  
\begin{equation}
{\cal H} = {\cal H}_0 + {\cal H}_1,
    \label{H0+DH}
\end{equation}
where 
\begin{eqnarray}
{\cal H}_1 &\equiv& - \sum_i^L (J S_i^x S_{i+1}^x + 2 J S_i^y S_{i+1}^y ) 
\label{H1} \\
&& + \frac{1}{2} \ \sum_{i=1}^L\sum_{j=1}^L (J_0 S_i^x S_j^x + 2 J_0 S_i^y S_j^y)
\nonumber
\end{eqnarray}
The term ${\cal H}_1$ is, in general, not a small perturbation to ${\cal H}_0$. However, in the vicinity of the considered periodic trajectory it is, because the right-hand-side of Eq.(\ref{H1}) becomes equal to zero once  all spins pointing in the same direction and $J_0 = 2J/L$ are substituted into it.

The following perturbative scenario is thus possible. The initial growth of the Lyapunov instability around the periodic trajectory that precedes  the nearly quasiperiodic regime takes place in the part of the phase space, where ${\cal H}_1$ is small. The quasiperiodic regime itself sets in, presumably, because  at least one of the invariant tori of the quasiperiodic motion survives the perturbation in the vicinity of the unstable periodic trajectory and thus prevent the instability from growing further in the 4-dimensional subspace of the $q=0$ and $q=q_p$ modes. The above picture requires that ${\cal H}_1$ remains sufficiently small on the above torus, which implies that the mode intensity $F_{q_p}(t)$ should remains small for the entire duration of the nearly quasiperiodic regime. The latter is, indeed, consistent with the observation that the maximum value of $F_{q_p}(t)$ in Fig.~\ref{Fig:Fourier_modes}(c)  is always less than $0.1 \  F_{0}(t)$. If the dynamics were strictly limited to the 4-dimensional subspace of the modes $q=0$ and $q=q_p$, the  phase space trajectory of the system would have remained forever constrained by the above torus. However, the presence of additional phase space directions make it possible for the trajectory to drift transversely to the above 4-dimensional subspace.  This drift signifies the Arnold diffusion.
In order to substantiate further the connection to the Arnold diffusion, let us note that our spin setting is, in fact remarkably similar to the original example worked out by Arnold. His example was for a five dimensional phase space, where, in the integrable limit, one pair of action-angle variables exhibited finite periodic motion  - counterpart of our original periodic trajectory. The second action-angle pair had no dynamics in the integrable limit but exhibited an instability once a perturbation term was added. This pair is the counterpart of the two additional phase space dimensions added by the growth of the $q=q_p$ mode. Finally, the fifth dimension was coupled to the previous two through yet much smaller perturbation term, which was responsible for the Arnold diffusion. In our context, this fifth dimension corresponds to all  remaining dimensions of the spin phase space. The initial absence of the Lyapunov instability in those dimensions (evidenced by non-operability of Mechanisms A and B) leaves the modes residing there with yet smaller amplitudes, which therefore have very weak coupling to the $q=0$ and $q=q_p$ modes. These very small amplitudes can be estimated from the apparent noise around the minimum values of the subleading Fourier harmonics $F_q(t)$  at $q=n q_p$; for example, the noise intensity is  $ 10^{-6}$ for the harmonic in Fig.\ref{Fig:Fourier_modes}(c) with $n=3$. 

The difference of our setting from the original Arnold diffusion context is that the latter was conceived for weakly nonintegrable Hamiltonians. Our Hamiltonian (\ref{H0+DH}) is strongly nonintegrable, and, therefore, the overwhelmingly large part of its phase space exhibits quite generic chaotic and ergodic properties. The message of the original Arnold's work was that the stochastic parts of higher dimensional phase spaces are connected to each other. We, therefore, conjecture (in agreement with Fig.~\ref{QP-Breakdown}) that, in the present strongly nonperturbative context, a trajectory initiated in the part of the phase space exhibiting Arnold diffusion eventually ends up in the part exhibiting generic ergodic dynamics.

The role of the above weakly excited modes in inducing the Arnold diffusion can be intuitively understood as follows. 
We note first that, in the examples of the nearly quasiperiodic regime presented in Fig.~\ref{Fig:Fourier_modes}(c), the intensity $F_{q_p}(t)$ of the $q = q_p$ mode reaches certain maximum value but then starts decreasing  coming very close to zero -- however, not as close as its initial value at $t=0$.  The reason is that even though other $q$-modes initially exhibit only tiny oscillations around zero, they are affected by the Lyapunov growth of the $q = q_p$ mode, acquire different amplitudes and phases, which, in turn, perturb the $q = q_p$ mode. This backaction may be very small, but the distance to the initial periodic trajectory at the point of the closest return then becomes much larger than the tiny displacement at $t=0$. Hence the time spent around the minimum of $F_{q_p}(t)$ becomes significantly shorter. A similar but less systematic distortion of the time spent around each subsequent minimum then undermines strictly quasiperiodic character of the motion and induces the wandering of the two slowly evolving action variables in the $q = 0$ and $q = q_p$ subspace. The above fluctuations of the time spent around each minimum of $F_{q_p}(t)$  originate from the mathematical fact exploited by Arnold that an outgoing trajectory on a separatrix manifold near an unstable torus in a high-dimensional phase spaces (manifold referred to by Arnold as a ``wisker") twists itself to arrive to a different unstable torus thereby causing the nearby trajectories to exhibit the Arnold diffusion.

Given the stochastic character of the motion around the minima of $F_{q_p}(t)$, the question arises, how is it then possible that the spectra exhibited in Figs.~\ref{fourierSpec}(b,b1,b2) allow one to identify two fairly well-defined frequencies $\omega_0$ and $\omega_1$. The reason is that the time interval 6000 used to compute these spectra, while being much longer than the natural time of the dynamics $1/h, 1/J \sim 1$, is much shorter than the lifetime of the Arnold diffusion regime, which in this particular case is $\sim 10^6$. It is then likely that a trajectory spends most of the time interval 6000 near the same KAM torus of a perfect quasiperiodic motion and thus exhibits well-defined frequencies $\omega_0$ and $\omega_1$.

\begin{figure}
\includegraphics[width=\columnwidth]{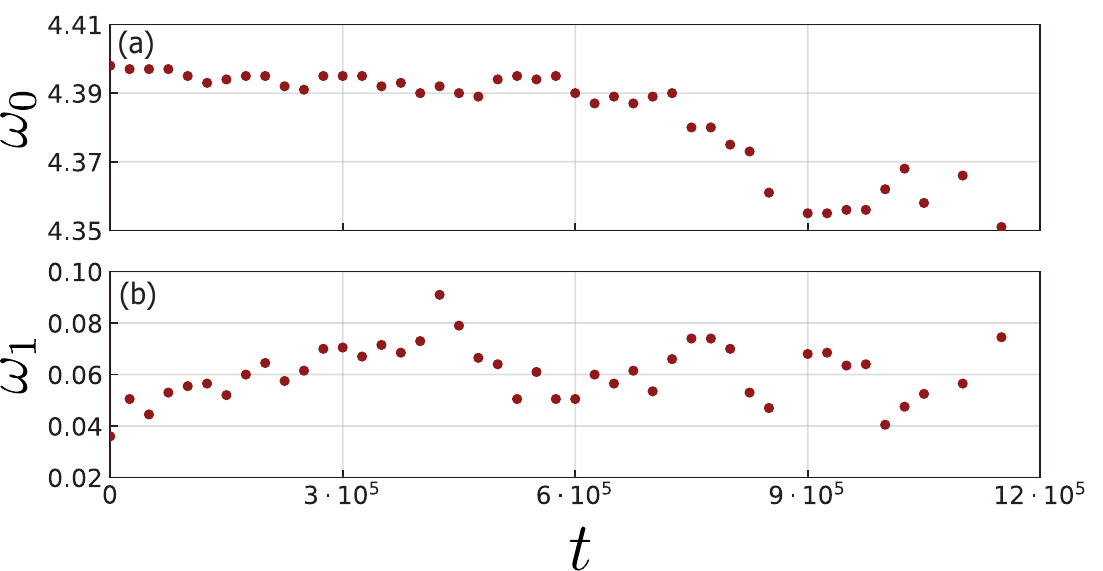}
\caption{Drift of characteristic spectral frequencies during the  anomalously long nearly quasiperiodic regime attributed to Arnold diffusion. Frequences $\omega_0$ and $\omega_1$ are identified as in Figs.~\ref{fourierSpec}(b,b1,b2) for the spectra extracted from different intervals of duration $\Delta t = 6000$ within a much longer  time series of $S_1^x(t)$  shown in Fig.~\ref{QP-Breakdown}. The $10\%$ Tukey window\cite{bingham1967modern,harris1978use} is applied to each interval. The value of $t$ for each point is the beginning of the respective time interval. 
\label{omegas}}
\end{figure}

In order to further corroborate the relevance of the Arnold diffusion to the anomalously long lived nearly quasiperiodic regime, we sampled the frequency spectra of the type presented in Fig.~\ref{fourierSpec}(b1,b2), and thereby extracted the frequencies $\omega_0$ and $\omega_1$ for different time windows of width 6000 during the lifetime $ \sim 10^6$ of the nearly quasiperiodic regime. The results of this sampling are presented in Fig.~\ref{omegas}. They indicate the drift of frequencies $\omega_0$ and $\omega_1$, which is expected to be caused by the Arnold diffusion of the corresponding action variables.

A few remarks about the Arnold diffusion regime are now in  order: 

The transient character of this regime implies that the strict application of the limit $t\to \infty$ to the definition (\ref{lambdamax-def}) of  the Lyapunov exponent would give the same $\lambda_{\text{max}}$ in both Arnold diffusion and ergodic regime. 

On the other hand, the exponentially long time of the Arnold diffusion regime makes the required numerical investigations rather challenging due to the difficult-to-quantify role of the algorithmic and rounding errors. We have checked that, in the case $J=1.76$ and $L=6$, the halving of the discretisation time step does not affect the distribution of the lifetimes of the Arnold diffusion regime for our ensemble of initial conditions.

Finally, we note that the Arnold diffusion regime is similar to a variety of phenomena, where the time-translation symmetry is spontaneously broken: the relevant concepts discussed in the literature are time crystals \cite{shapere2012classical,wilczek2012quantum,sacha2017time,else2020discrete,rovny2018observation,pizzi2019timecryst,yao2020classical,Daviet-2024} and time quasi-crystals \cite{autti2018observation,pizzi2019timecryst,he2024experimental}. The present result adds a new member to this variety. Here we are dealing with the time translational symmetry of the original periodic trajectory that gets spontaneously broken by the Lyapunov instability into a regime, which may be called ``quenched time quasi-crystal", implying that there are no long-range correlations in the symmetry-broken state, and yet locally it looks like a time quasicrystal.
The distinctive feature of our setting is that it involves neither periodic driving nor dissipation. 

\section{Quantum spins}
\label{quantum}

As discussed in the introduction, the present work was initially motivated by our interest to explore whether periodic orbits in classical spin lattices  imply the existence of many-body scar eigenstates in quantum spin lattices. A quantum scar in a semiclassical limit can be associated with self-interference of a quantum wavepacket propagating along a periodic classical trajectory and thereby forming an atypical ``standing wave''\cite{heller1984bound}. In a many-dimensional phase space, the presence of many Lyapunov instabilities along that trajectory is expected\cite{Berry-1989} to make the above wavepacket too broad before it could start interfering with itself, thereby reducing the chances for a distinct scar eigenstate to emerge. From such a perspective, the weaker the Lyapunov instabilities, the better are the chances for observing quantum scars. In this regard, our classical investigation in Section~\ref{classical} indicates that the chances of observing the scar effects in many-spin systems are, actually, somewhat better than what one could expect from the knowledge of the Lyapunov spectra of ergodic trajectories\cite{dewijn2013lyapunov}. 
The reason is that one can choose the parameters of the Hamiltonian such that the largest  Lyapunov exponent for a periodic trajectory  is significantly smaller than that for an ergodic trajectory (cf. Figs.~\ref{gemD} and \ref{lexp_ergpic}), and, moreover, there exist rather long spin chains [see Figs.~\ref{LVSL}(b,c)] for which the Lyapunov instabilities in the vicinity of the periodic trajectories are completely suppressed. 

Below, we explore the quantum dynamics starting with the special initial wave function $|\Psi^\text{up}\rangle$   --``all spins up" defined by Eq.(\ref{eqquantini}), which in the classical limit gives a periodic trajectory. We first check whether the evolution starting from $|\Psi^\text{up}\rangle$ exhibits an anomalously slow thermalization. Then we decompose $|\Psi^\text{up}\rangle$ into the energy eigenstates  and  check whether any of the eigenstates prominently present in the above initial state also violates eigenstate thermalization hypothesis.

This investigation is based on the exact diagonalisation of the Hamiltonian (\ref{ham}). It involves quantum spins with  $S=\frac{1}{2},1,\frac{3}{2},2$. 
They have length   $\sqrt{S (S+1)}$, while in the classical simulations this length was set to 1. Since the terms in the  Hamiltonian (\ref{ham}) proportional to $h$ and to  $J$ scale differently as a function of the spin length,    we compensate this by using the renormalised coupling constant
\begin{equation}
  \tilde{J} \equiv J \sqrt{S (S+1)},
    \label{Jtilde}
\end{equation}
which facilitates the comparison between quantum and classical results, as well as between the results for different quantum spins $S$.

\begin{figure}
\includegraphics[width=\columnwidth]{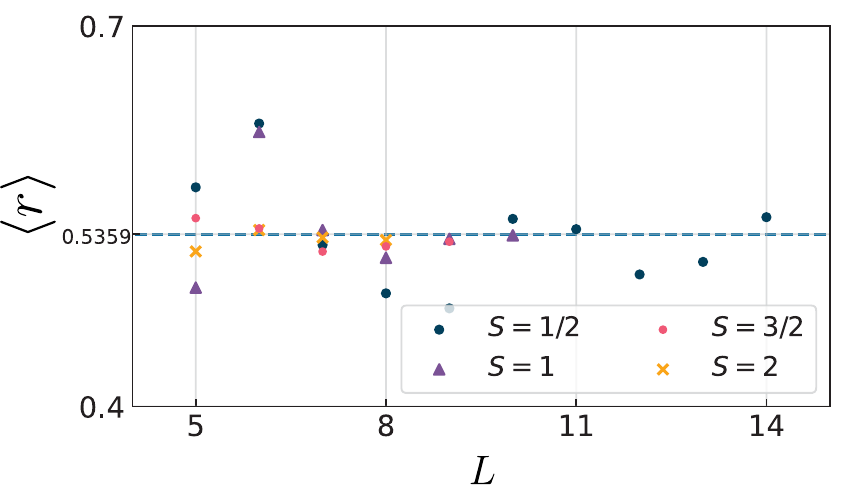}
\caption{Level-spacing parameter $\langle r \rangle$  for chains of quantum spins $S$ with $\tilde{J} =  1.76$ as a function of chain length $L$. The value $\langle r\rangle_\text{GOE} = 0.5359$ indicated by the horisontal dashed line corresponds to the Gaussian Orthogonal Ensemble\cite{Oganesyan-2007,atas2013distribution}. 
\label{rVal}}
\end{figure}

All numerically investigated chains of quantum spins can be classified as ``quantum chaotic", as they exhibit the statistics of energy level spacings consistent with that of the Gaussian orthogonal ensemble (GOE): As an indicator, we used the level-spacing parameter $\langle r\rangle$, defined in \cite{Oganesyan-2007} as the average of 
\begin{align}
r_n=\frac{\min(s_n,s_{n-1})}{\max(s_n,s_{n-1})},
\end{align}
where $s_n=E_{n+1}-E_n$ is the spacing between adjacent eigenenergies $E_{n+1}$ and $E_n$. Examples of $\langle r\rangle$ computed for different quantum spin chains governed by Hamiltonian (\ref{ham}) are shown  in Fig.~\ref{rVal}.

\subsection{Suppressed initial thermalization}
\label{slowdown}

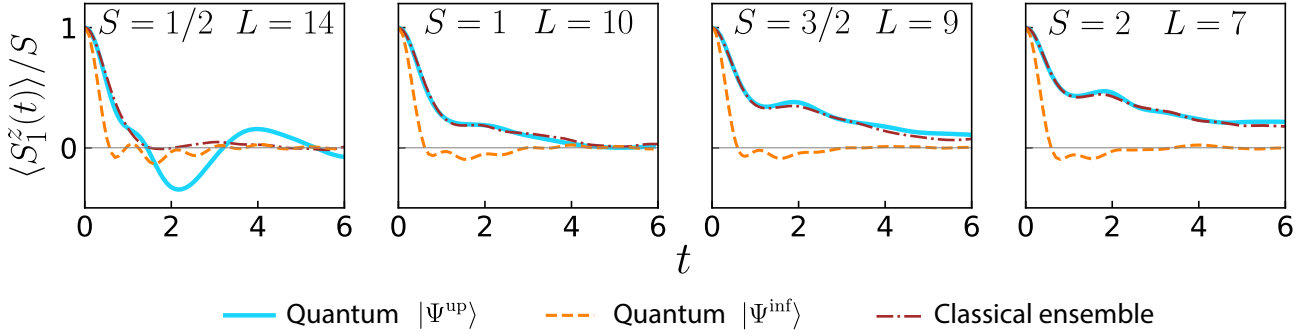
\begin{figure*}
\begin{center}
\setlength{\unitlength}{1mm}
\begin{picture}(228,60) 
  \put(0,0){\includegraphics[width=2\columnwidth]{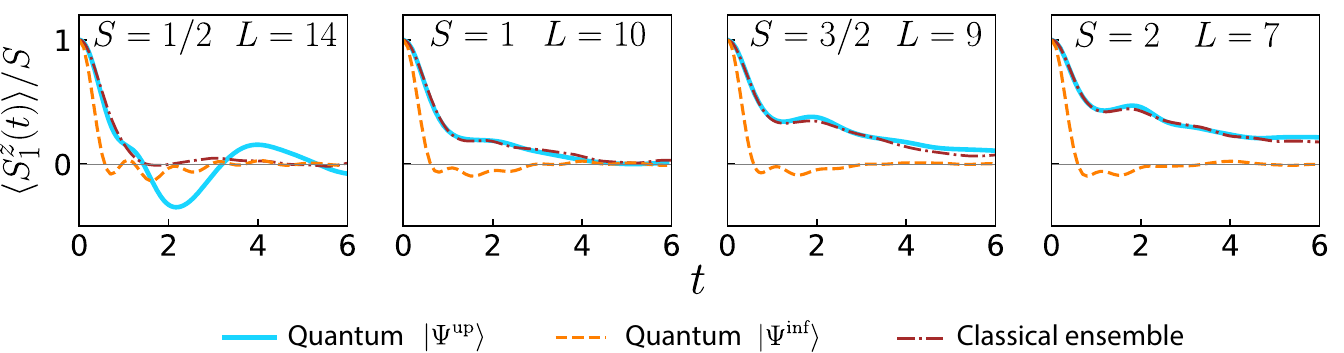}}

  \put(11,22){\makebox(0,0)[lt]{\textbf{(a)}}}
  \put(54,22){\makebox(0,0)[lt]{\textbf{(b)}}}
  \put(95,22){\makebox(0,0)[lt]{\textbf{(c)}}}
  \put(136,22){\makebox(0,0)[lt]{\textbf{(d)}}}
\end{picture}
\caption{Slowdown of thermalization dynamics illustrated by the relaxation of $\langle S^z_1(t)\rangle$ for chains of quantum spins $S$ with lengths $L$; $\tilde{J}=1.76$. Blue solid lines and orange dashed lines represent quantum relaxation starting, respectively, from the initial states $|\Psi^{\text{up}}\rangle$  and $|\Psi^{\text{inf}}\rangle$.  In all cases, the generic relaxation curves starting from $|\Psi^{\text{inf}}\rangle$ quickly approach zero, exhibiting afterwards only small fluctuations. In contrast, the relaxation curves starting from $|\Psi^{\text{up}}\rangle$ exhibit slow decaying positively-valued long-time tails for chains with $S \geq 1$.   (To improve computational efficiency, the relaxation curves for $|\Psi^{\text{inf}}\rangle$ were computed for smaller chain lengths than indicated in the legends, namely, $L =$ 13, 8, 7 and 6 for $S=$ 1/2, 1, 3/2 and 2, respectively. This affected only small fluctuations of $\langle S^z_1(t)\rangle$ around zero.)
Brown dash-dotted lines represent classical simulations averaged over ensembles of $10^4$ initial conditions imitating the quantum state $|\Psi^{\text{up}}\rangle$ as described in the text. 
\label{p2}}
\end{center}
\end{figure*}

\begin{figure}
 \includegraphics[width=\columnwidth]{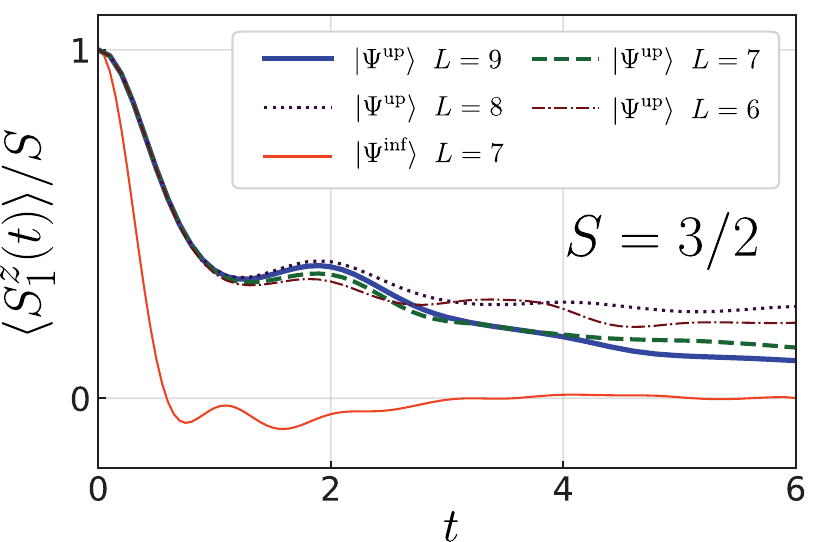}
\caption{Slowdown of thermalization dynamics  for chains of quantum spins 3/2 with different lengths $L$  for the initial state $|\Psi^\text{up}\rangle$. Plotted is the relaxation of $\langle S^z_1(t)\rangle$ with $\tilde{J}=1.76$, same as in Fig.~\ref{p2}. The closeness of the relaxation curves for different $L$ is indicative of a the thermodynamic limit $L \to \infty$. Also included is the plot of generic relaxation starting with $|\Psi^\text{inf}\rangle$.   \label{spin15dynamics}}
\end{figure}

To characterise the initial thermalization, we compute the quantum mechanical expectation value $\langle S^z_1(t)\rangle \equiv \langle \Psi(t)| S^z_1(t) | \Psi(t) \rangle$ starting from the fully polarized state $|\Psi^\text{up}\rangle$ given by Eq.(\ref{eqquantini}) and compare it with the thermalization of the same observable starting from a more generic initial condition where the spin $S^1_z$ is fully polarised but  the rest of the system is at the infinite temperature equilibrium:
\begin{equation}
|\Psi^{\text{inf}}\rangle=|m_1 = S\rangle\otimes|\text{inf}\rangle_{L-1} ,
    \label{Psi-inf}
\end{equation} 
where $|\text{inf}\rangle_{L-1}$ is a pure state randomly sampled in the Hilbert space of spins on sites from 2 to $L$ \footnote{A pure quantum state randomly sampling the infinite temperature equilibrium in a Hilbert space of dimension $N \gg 1$ is obtained as $|\text{inf}\rangle=\frac{1}{\sqrt{2^{L-1}}}\sum^{2^{L-1}}_{n=0}c_n|b_n\rangle$, where $|b_n\rangle$ are basis states, and $c_n=|c_n|e^{i\phi_n}$ are complex amplitudes with phases $\phi_n$ chosen randomly from the interval $[0,2\pi)$ and absolute values $|c_n|$ chosen using the probability distribution $P(|c_n|^2)=Ne^{-N|c_n|^2}$ (see Refs.\cite{fine2009typical,elsayed2013regression})}. Quantum typicality~\cite{Gemmer-2009,fine2009typical,elsayed2013regression} implies that the energy expectation value for $|\Psi^{\text{inf}}\rangle$ is exponentially close to the value for the infinite temperature canonical ensemble, which is, in turn, equal to zero [see Eq.(\ref{E_inf})]. Since the energy expectation value for $|\Psi^{\text{up}}\rangle$ is also zero, both $|\Psi^{\text{up}}\rangle$ and $|\Psi^{\text{inf}}\rangle$ belong to the same ``energy shell" in the sense of the canonical ensemble.

The relaxation plots of  $\langle S^z_1(t)\rangle$ for 
$S =$~1/2, 1, 3/2 and 2 are shown in Fig.~\ref{p2}. There one can observe that the thermalization process for $S \geq 3/2$ starting from the initial condition $|\Psi^{\text{up}}\rangle$ is noticeably slower than the one starting from $|\Psi^{\text{inf}}\rangle$. Such a difference is to be expected when atypical quantum scar eigenstates are prominently present in the expansion of 
$|\Psi^{\text{up}}\rangle$. At the same time, the difference  between the two thermalization curves for $S = 1/2$ is rather small --- suggestive of the absence of quantum scar eigenstates in the expansion of $|\Psi^{\text{up}}\rangle$. The case of $S=1$ appears to be transitional between $S = 1/2$ and $S \geq 3/2$.

Figure \ref{spin15dynamics} further illustrates  the anomalous slowdown of the thermalization for spin-3/2 chains as a function of chain length $L$. The coincidence of the initial slowed down behavior for the chains of different length indicates that the slowdown of the thermalization process starting from the state $|\Psi^{\text{up}}\rangle$ is a feature that is also  present in the thermodynamic limit $L\to \infty$.

We, finally, note that Fig.~\ref{p2} also includes the results of classical simulations for an ensemble of initial conditions devised to imitate the relaxation of quantum spins $S$ starting from the initial state $|\Psi^{\text{up}}\rangle$. In the simulations, each classical spin has length 1. In order to represent a quantum state $|m=S\rangle$, the range $[-1,1] $ of the $z$-projection of a classical spin is divided into $2S + 1$ equal intervals of length $\frac{2}{2S+1}$ and then  the classical projection $S_z$ is randomly sampled from the topmost interval $[1-\frac{2}{2S+1},1] $, while the $x$- and $y$-projections are chosen as $S_x = \sqrt{1 - S_z^2} \cos \varphi$ and $S_y = \sqrt{1 - S_z^2} \sin \varphi$, where  $\varphi$ is a random angle from the interval $[0,2 \pi) $. On the one hand, the above ensemble becomes increasingly close to the considered periodic classical trajectories as the value of quantum spin $S$ increases. On the other hand, as one can see in Fig.~\ref{p2}, the average for that ensemble exhibits an excellent agreement with the quantum simulations starting  already  with $S= 1$. This is yet another indication, in addition to the observations of Ref.\cite{Elsayed-classical-2015}, that classical spin ensembles can quantitatively represent quantum spin dynamics for translationally invariant chains of spins $S \geq 1$ at high temperatures. The classicalization of higher spins  was also recently reported at the operator level in \cite{ermakov2025operator}, further highlighting the difference in dynamical properties between spin-$1/2$ and higher spins. In the present context, the success of the above simulation indicates the proximity of quantum dynamics to the semiclassical limit, which will be put forward below as a condition for the observability of distinct many-body quantum scars.  

\subsection{Quantum scars in finite-size spin clusters}
\label{Quantum-scars}

We searched for the presence of quantum scars by numerically computing the eigenstates of the spin chains, and then computing for each eigenstate the half-chain entanglement entropy $\mathcal{E}$ between the sites from 1 to $L/2$ (rounded down for odd $L$) and the rest of the chain. The entanglement entropy is a popular choice for monitoring eigenstate thermalization; other physical observables could have been used as well.   ETH implies that, in the dense part of the spectrum, $\mathcal{E}$ as a function of eigenstate's energy $E_n$ changes only very little from one state to the next.  A large departure of $\mathcal{E}$ from the value typical of the adjacent-in-energy eigenstates is then an indication of a quantum scar. 

\begin{figure*}
\includegraphics[width=1.75\columnwidth]{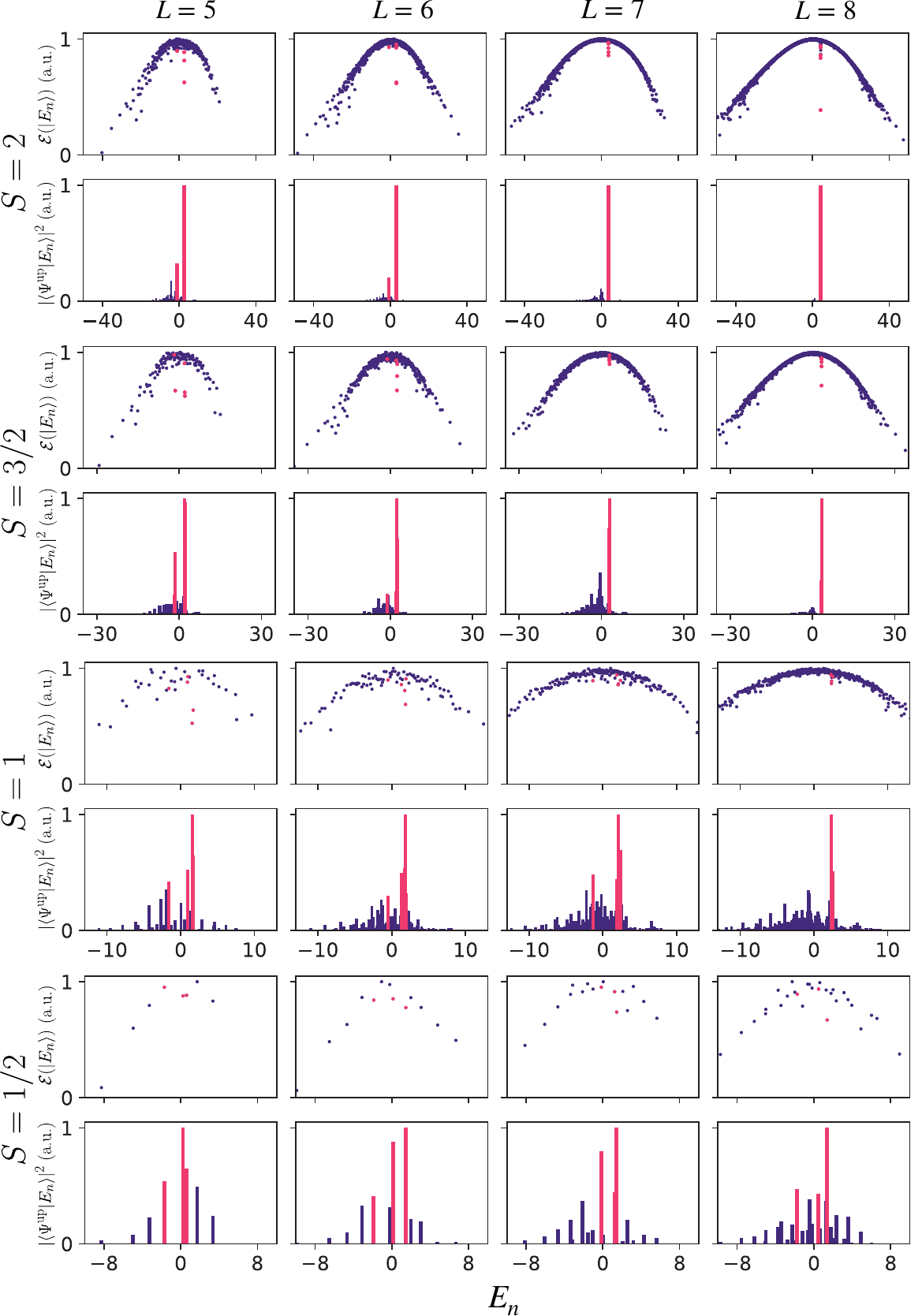}
\caption{Quantum-scar diagnostics for quantum chains of lengths $L$ (indicated above each column of plots) for spins S equal to 1/2, 1, 3/2 and 2 (indicated on the left of the plots). The interaction constant in all cases is $\tilde{J}=1.76$. For each value of $S$, the upper line of plots represents  normalized half-chain entanglement entropy $\mathcal{E}$ for translationally invariant (zero momentum) Hamiltonian eigenstates $|E_n\rangle$ as a function of eigenenergies $E_n$. The outliers of $\mathcal{E}$ are identified with quantum scars.  The bottom line of plots for each $S$ represents the overlaps $|\langle \Psi^{\text{up}}|E_n\rangle |^2 $ of fully polarized ``spins-up'' state  $|\Psi^{\text{up}}\rangle$ with $|E_n\rangle$. Magenta points in the plots $\mathcal{E}(E_n)$ correspond to the top five eigenstates having the largest overlap values $|\langle \Psi^{\text{up}}|E_n\rangle |^2 $ indicated by magenta bars in the respective overlap plots. The vertical axis of each panel is scaled such the maximum projection of the data points on that axis is equal to 1. The trends exhibited in this figure are discussed in Section~\ref{Quantum-scars}.
\label{apentr}}
\end{figure*}

The computed entanglement entropy is plotted  in Fig.~\ref{apentr} for the chains of spins 1/2, 1, 3/2 and 2.
These plots reveal the picture entirely consistent with the expectations  based on observing or not observing   the slowdown of thermalization for the initial state $|\Psi^{\text{up}}\rangle$ --- see Fig.~\ref{p2}. Namely, quantum-scar eigenstates are prominently present for the chains of spins 3/2 and 2 as indicated by the outliers of $\mathcal{E}(E_n)$ around the center of the spectrum, while the chains of spins 1/2 exhibit no signs of quantum scars, and spin-1 chains show rather weak indications thereof. 

Each plot of $\mathcal{E}(E_n)$ in Fig.~\ref{apentr} is accompanied by a plot of the overlap $|\langle \Psi^{\text{up}} |E_n\rangle|^2$ between the initial state $|\Psi^{\text{up}}\rangle$ and the energy eigenstate $|E_n\rangle$ as a function of $E_n$. The overlap plots illustrate that whenever the quantum-scar states are detected in the middle of the spectrum, they are also prominently present in the eigenstate expansion of $|\Psi^{\text{up}}\rangle$. In particular, the largest outlier of $\mathcal{E}(E_n)$ always corresponds to the largest value of $|\langle \Psi^{\text{up}}| E_n \rangle|^2$.   These findings together with the correspondence between quantum (\ref{eqquantini}) and classical (\ref{eqclassini}) initial conditions strongly suggest that the observed quantum scar states are connected to periodic classical trajectories.

\begin{figure}
\includegraphics[width=1.0\columnwidth]{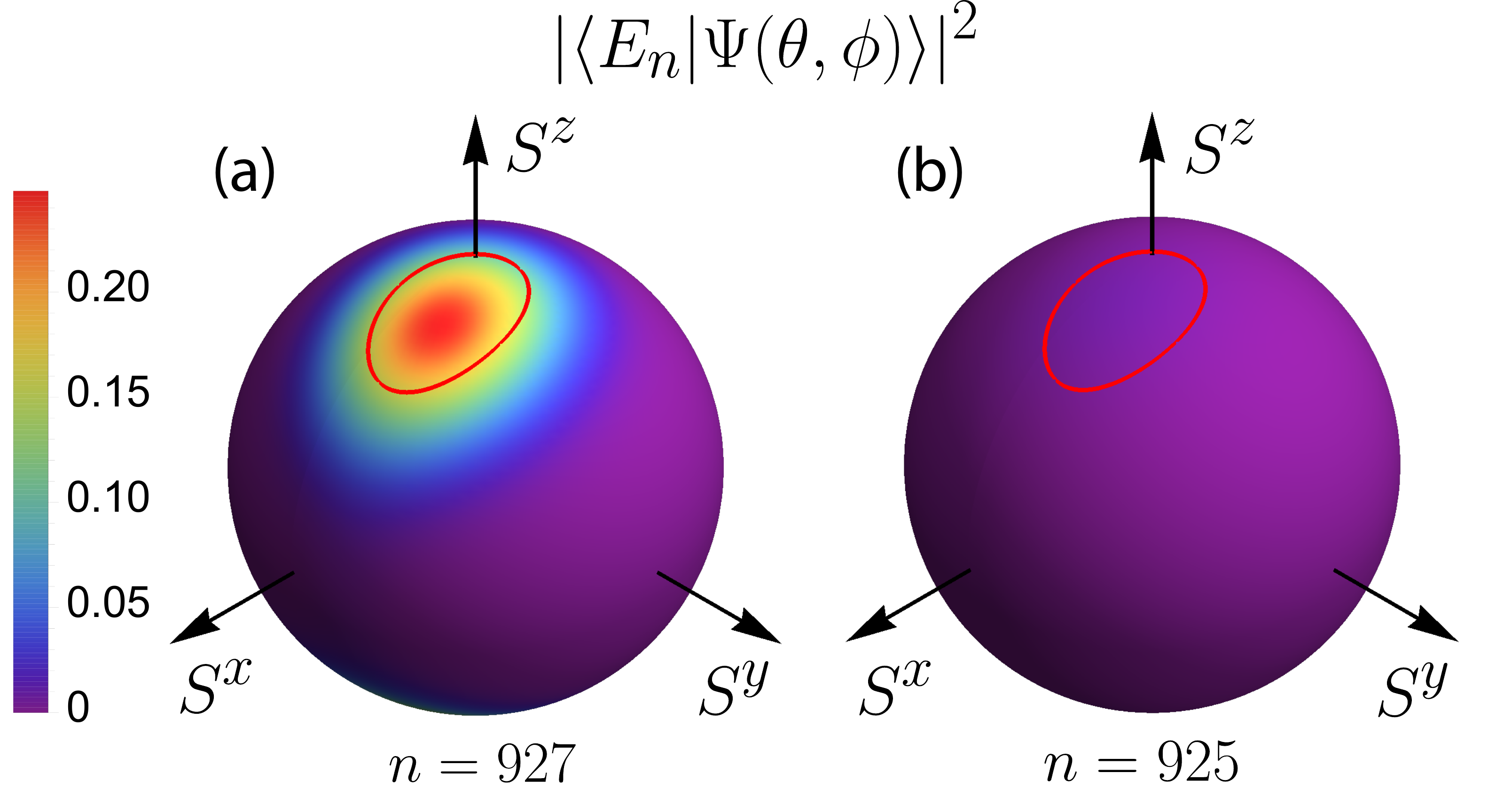}
\caption{[Color online] Spherical map of the overlap $|\langle E_n|\Psi(\theta,\phi)\rangle|^2$ between an energy eigenstate $|E_n\rangle$ and the state   
$|\Psi(\theta,\phi)\rangle$ of all spins maximally polarised along the direction $\mathbf{n} = (\sin{\theta} \cos{\phi}, \sin{\theta} \sin{\phi}, \cos{\theta}) $ for a chain of spins 2 with ${\tilde{J}=1.76}$ and $L=6$. The red line represents the periodic trajectory in the classical limit. (a) Overlap for the most prominent quantum scar eigenstate  with $n=927$ (see the relevant panel in Fig.~\ref{apentr}). (b) Overlap for a nearby generic energy eigenstate  with $n=925$. ($n=926$ corresponds to a weaker quantum scar.)
\label{scar-image}}
\end{figure}

In order to visualize the above connection, we present in Fig.~\ref{scar-image}(a) a spherical map of the overlap $|\langle E_n|\Psi(\theta,\phi)\rangle|^2$ between the most prominent quantum-scar eigenstate $|E_n \rangle$ for a chain of spins 2 and the state 
\begin{equation}
|\Psi(\theta,\phi) \rangle =\prod\limits^L_{i=1}e^{-i\phi S^z_i}e^{-i\theta S^y_i}|\Psi^\text{up}\rangle ,
    \label{Psi-theta-phi}
\end{equation}
where every spin is fully polarized along the direction $\mathbf{n} = (\sin{\theta} \cos{\phi}, \sin{\theta} \sin{\phi}, \cos{\theta}) $ determined by the standard spherical angles $\theta$ and $\phi$. The concentration of this projection around the corresponding classical trajectory is evident there. For comparison, Fig.~\ref{scar-image}(b) shows the overlap with $|\Psi(\theta,\phi) \rangle$ for an adjacent generic energy eigenstate: In the latter case, the overlap  exhibits much smaller values at the level of statistical fluctuations in a high-dimensional Hilbert space.

The tendency of larger quantum spins $S$ to exhibit more pronounced evidence of quantum scarring can be understood as follows.  In order to have a quantum scar distinctively localized around a periodic classical trajectory (as in the case of billiards\cite{heller1984bound}), the quantum dynamics should be close enough to a semiclassical limit in the first place, such that a quantum wave packet initiated around that trajectory does not broaden too much before the trajectory closes upon itself. When this condition is fulfilled, the difference between a quantum scar and a generic eigenstate is as dramatic as the difference between short periodic and ergodic classical trajectories. Without proximity to the semiclassical limit, fluctuations associated with strong quantum interference effects are, presumably, making it difficult to distinguish generic from nongeneric eigenstates.  In our case, Fig.~\ref{p2} illustrates that the accuracy of the semiclassical approximation for quantum spin dynamics indeed improves quickly with the growth of $S$.

The presented evidence of quantum scars is limited only to spin chains of finite size. It remains an intriguing question, whether the quantum-scar eigenstates remain present in the thermodynamic limit $L \to \infty$ for the chains of spins $S \geq 3/2$.
On the one hand, periodic classical trajectories should become more unstable as $L$ grows, because their overall instability is characterised by the sum of all positive Lyapunov exponents (Kolmogorov-Sinai entropy), which should grow proportionally to $L$ \cite{dewijn2013lyapunov}; hence the lifetime of semiclassical quantum wave packets made of coherent spin states\cite{Radcliffe-1971} near periodic trajectories should decrease, thereby suppressing quantum interference behind the formation of quantum scars. On the other hand, the  increasing instability of many-spin classical periodic trajectories does not necessarily exclude the existence of a few quantum scars out of exponentially many, i.e.  $(2S+1)^L$, energy eigenstates. One can further note that the atypical slowdown of thermalization seen in Figs.~\ref{p2}(b-d) and \ref{spin15dynamics} is very likely to remain present in the thermodynamic limit, and this would indicate that, independent of the size of the system, there is an atypical sector of initial conditions in the many-spin Hilbert space.  Although a randomly selected wave function from that sector is not an eigenstate, we conjecture that at least a few quantum-scar eigenstates consisting of the superposition of wave functions predominantly belonging to the above atypical sector do remain present in the thermodynamic limit.

\subsection{Quantum separatrix}

\begin{figure}
\includegraphics[width=1\columnwidth]{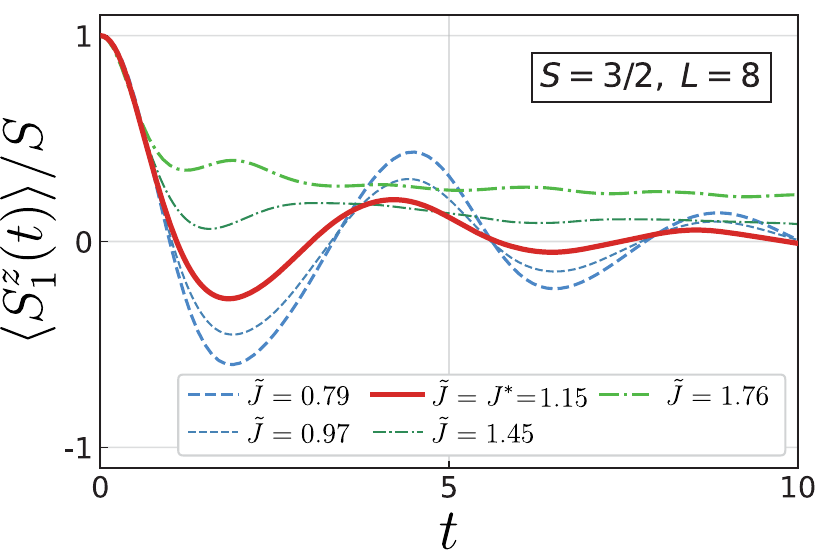}
\caption{Relaxation of quantum mechanical expectation value $\langle S^z_1(t)\rangle$ starting from the initial state $|\Psi^{\text{up}}\rangle$ for a quantum chain of  spins $S=3/2$ with $L=8$ and different $\tilde{J}$ indicated in the plot legend.  The solid red line highlights the classical separatrix value $\tilde{J}=J^*$ separating the regime of oscillations around the zero base line from the oscillations around a positive base line. 
\label{qssq}}
\end{figure}

\begin{figure}
\includegraphics[width=1\columnwidth]{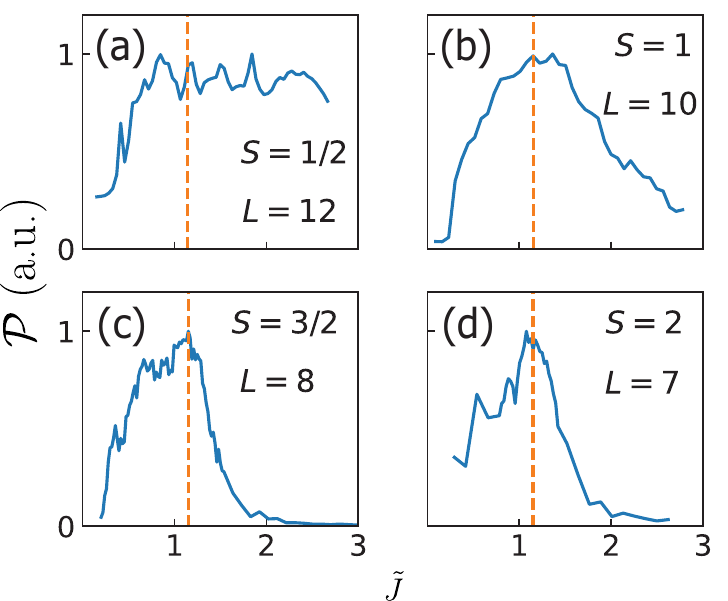}
\caption{Participation ratio $\mathcal{P}$ defined by Eq.(\ref{prpsiup}) as a function of $\tilde{J}$ for chains of different quantum spins  $S=1/2,1,3/2,2$ with lengths $L$ indicated in the plot legends. Vertical dashed lines mark the classical separatrix value $J^*$.
\label{qspic}}
\end{figure}

It was shown in subsection~\ref{librot} that, for initial conditions ``all spins-up" (\ref{eqclassini}), there exist two regimes of the classical periodic motion, librations and rotations, divided by a classical separatrix appearing at the interaction strength $J=J^*$.
It is to be expected that, if the relaxation of, e.g., $\langle S^z_1(t)\rangle$, is monitored for a classical ensemble starting in the vicinity of the  upper rotation trajectory then the $\langle S^z_1(t)\rangle$ would remain positive for a long time, because the spin would tend to stay in the positive-$S^z_1$ hemisphere. In contrast, the proximity to a libration trajectory implies that $\langle S^z_1(t)\rangle$ would immediately exhibit strong oscillations around the zero average value. 

We now show that the quantum dynamics   exhibits a closely related phenomenology.
In Fig.\ref{qssq}, in order to initially expose the two distinct regimes, we present 
$\langle S^z_1(t)\rangle$ for a spin-3/2 chain with the initial state $|\Psi^{\text{up}}\rangle$ and different interaction strengths. 
There one can see that, for $\tilde{J} < J^* $ corresponding to the classical libration regime, the curves $\langle S^z_1(t)\rangle$ display the expected oscillations around zero, while for $\tilde{J} > J^* $ corresponding to the classical rotations, the oscillations, although not disappearing right away, occur around a positively valued base line that very slowly approaches zero. 
Fig.\ref{qssq} also includes the plot of $\langle S^z_1(t)\rangle$ for the separatrix value $\tilde{J} = J^* $. This is indeed approximately the maximum value of $\tilde{J}$ for the oscillations around the zero base line.

The existence of the above two quantum regimes led us to further explore, whether the notion of a separatrix can be defined in the quantum case. 
Indeed, beyond the visual impression of the relaxation crossover, the quantum separatrix value for $\tilde{J}$  can be accurately identified as follows.  We note that, in the classical limit, the separatrix appears at a saddle point of the Hamiltonian function $\mathcal{H}_{\text{p}}(S^x_{\text{p}}, S^y_{\text{p}}, S^z_{\text{p}})$ given by Eq.(\ref{ham0}) and describing the periodic trajectories, where each spin has the same orientation $(S^x_{\text{p}}, S^y_{\text{p}}, S^z_{\text{p}})$. The classical phase-space volume available for the motion around that saddle point has a Van Hove singularity as a function of energy. 
In the quantum case, it is thus reasonable to expect that the number of the {\it relevant} quantum energy eigenstates $|E_n\rangle$ has a maximum around the quantum separatrix for finite systems and, possibly, a singularity in the thermodynamic limit. 
As a measure of that number, we use the participation ratio   $\mathcal{P}$ of eigenstates $\left\{|E_n\rangle \right\}$ in the initial state $|\Psi^{\text{up}}\rangle$:
\begin{align}\label{prpsiup}
\mathcal{P} \equiv \frac{1}{\sum\limits^\mathcal{N}_{n=1}|\langle\Psi^\text{up}|E_n\rangle|^4},
\end{align}
where $\mathcal{N}$ is the number of eigenstates  with zero total quasi-momentum.

In Fig. \ref{qspic}, we plot numerically computed $\mathcal{P}(\tilde{J})$ for different $S$. These plots show that, for $S=3/2$ and $S=2$,  $\mathcal{P}(\tilde{J})$ exhibits well-defined peaks  located with good precision at the classical separatrix value $\tilde{J} = J^*$.  For $S=1/2$, $\mathcal{P}(J)$ does not show any clear peak, while, for $S=1$, the peak around $\tilde{J} = J^* $ does exist, but its maximum is not well defined. The above findings thus consistent with the clear evidence of quantum scars presented in Fig.~\ref{apentr} for the simulated spin-3/2 and spin-2 chains, the weaker evidence for spins 1 and the lack of the evidence for spins-1/2.

\section{Summary and discussion}
\label{discussion}

In this article, we demonstrated that the fate of classical periodic trajectories in spin systems, both in terms of their Lyapunov stability and in terms of transition to quantum dynamics exhibits rather rich phenomenology.  Specifically, we investigated the stability of a class of translationally invariant periodic trajectories in translationally invariant classical spin chains and the corresponding dynamics of their quantum counterparts.

On the classical side, we obtained the following results:

(C1) The largest Lyapunov exponents of the periodic trajectories exhibit nontrivial nonmonotonic dependence on the length of spin chains presented in Fig.~\ref{LVSL}.  We were able to find accurate quantitative approximation (\ref{lambdap-L}) for  the functional form of this dependence on the basis of the insight that the Lyapunov vectors for the Lyapunov instabilities around translationally invariant periodic trajectories constitute irreducible representations of the translational symmetry group with well-defined wave vectors.

The analytical formula (\ref{lambdap-L}), while showing a very good performance, was obtained on the basis of a mathematical conjecture as opposed to the rigorous derivation. The latter, if found, may give a recipe for finding the parameters of that formula without the need to make a fit.

(C2) Our particularly surprising finding is that the periodic trajectories of rather long spin chains -- up to 59 spins in Fig.~\ref{LVSL}(c) -- can be Lyapunov stable. (``Rather long" in the sense of the phase space dimension $2L$ being much larger than 1.) Both our numerical experience and the proposed analytical form (\ref{lambdap-L}) of $\lambda_p(L)$ suggest that the maximum length of chains with stable periodic trajectories is limited for a given value of the coupling constant $J$. However, as the value of $J$ increases, so does the maximum length of the stable periodic motion, i.e. one may, possibly, be able to make the periodic trajectories stable for a spin chain of any given length by choosing large enough $J$.  Another counter-intuitive aspect of the same observation is that  the periodic motion in a system of a large size can be perfectly stable, while it is unstable in a much smaller system with the same coupling constants.

(C3) We have also found a nontrivial dependence of the largest periodic Lyapunov exponent $\lambda_{\text{p}}$ on the value of the coupling constant $J$ presented in Fig.~\ref{gemD}. The periodic trajectories corresponding to the infinite temperature energy $E=0$ are shown to exhibit two topologically distinct regimes, librations and rotations, as a function of the coupling constant $J$. The two regimes are divided by a  separatrix corresponding to the critical value $J = J^*$.
In some cases, $\lambda_{\text{p}}$ exhibits a non-smooth dependence on $J$ associated with the switching of the wave number for the associated Lyapunov eigenvector.
Furthermore, as shown in Fig.~\ref{nearSep}, $\lambda_{\text{p}} (J)$  exhibits a peculiar discontinuous cusp-like behavior  in the vicinity of $J = J^*$, whose functional form (\ref{lmdp-logDJ}) we obtained analytically.

(C4) We have shown in the subsection~\ref{observability} that, in close analogy with the behavior of the out-of-time order correlators, an ensemble of trajectories with very small deviations from the periodic one exhibits a one-spin response characterisied by the exponential growth of the deviation from the perfect periodic behavior with a rate equal to twice the value of the largest periodic Lyapunov exponent. This result implies a possible way of observing the Lyapunov instability of periodic many-spin trajectories experimentally. 

(C5) We have shown in Section~\ref{quasiperiodic} that, once a Lyapunov instability develops around a periodic trajectory, the system generally enters a transient nearly quasiperiodic regime (meaning a stochastic regime close to a quasiperiodic motion). That regime is typically characterised by an interplay of two spatial Fourier harmonics: one  corresponding to the original periodic motion and the other to the leading Lyapunov instability. The nearly quasiperiodic character of the motion arises, because the two Fourier harmonics are described by a four-dimensional phase space, where KAM tori of quasiperiodic motion divide energy shells into impenetrable parts, and, hence, two nearby KAM tori can ``squeeze'' a stochastic trajectory and thus force the nearly quasiperiodic regime.    We have identified three mechanisms that destroy this regime. The first two of them are relatively fast (in comparison with the third one) --- they are associated with additional Lyapunov instabilities involving other spatial Fourier harmonics.  The third mechanism is extremely slow: it is the Arnold diffusion that sets in those rather exceptional cases, where no additional Lyapunov instabilities are available.
This observation is rather remarkable, given that Arnold diffusion is generaly expected for nearly integrable Hamiltonians, while in the present case the Hamiltonian is strongly nonintegrable. Yet Arnold diffusion becomes relevant due to two factors: (i) weak nonintegrability of the dynamics in the close vicinity of the considered periodic trajectories and (ii) the translational invariance of these trajectories that affords the possibility that all spatial Fourier harmonics beside the above mentioned two remain stable.

In general, the observed surprising stability of classical periodic motion increases the chances to detect the signatures of this motion in large quantum systems including, in particular, quantum many-body scar states. 

Our findings on the quantum side were not less rewarding:

(Q1) We discovered that the quantum signatures of the classical periodic motion  become detectable for the values of quantum spin $S \geq 1$.

(Q1.A) The first such signature is the anomalous slowdown of the thermalization process starting from the initial quantum states corresponding to the classical periodic trajectories. This signature is shown to remain present in the thermodynamic limit.

(Q1.B) The second signature is that energy eigenfunctions dominating the expansion of the initial quantum states associated with the classical periodic motion exhibit nonconformance with the eigenstate thermalization hypothesis. While the initial proposal of many-body  quantum scars\cite{turner2018weak,lin2019exact} focused on anomalous energy eigenstates in spin-1/2 systems without a clear classical counterpart, here we demonstrated the viability of the original definition of quantum scars\cite{heller1984bound} as the eigenstates corresponding to periodic trajectories in the classical limit. 

In this work, we numerically identified the quantum-scar states only in the finite-size systems. The survival of these states in the thermodynamic limit remains an interesting mathematical question. Let us, however, remark that the observed anomalous slowdown of thermalization in the thermodynamic limit would constitute, perhaps, the most important practical consequence of the existence of the  quantum scars. At the same time, our result that finite systems possess  quantum-scar eigenstates which, on the one hand, are linked to periodic classical trajectories, and, on the other hand, control the thermalization slowdown, indicates  that the mechanism behind the formation of quantum scars, namely, the quantum interference along short periodic classical trajectories\cite{heller1984bound}, is also the mechanism behind the thermalization slowdown.  Since the slowdown remains present in the thermodynamic limit, the interference effect along the short periodic trajectories should also survive the thermodynamic limit. One can, therefore, argue that the thermalization slowdown together with the presence of quantum scar eigenstates for finite systems can be taken as a primary definition of quantum scars in many-body systems irrespective of whether eigenstates violating ETH actually exist in the thermodynamic limit.

As mentioned in Section~\ref{scars}, our findings are broadly consistent with those of Refs.~\cite{Mondal-PRE2021,Mondal-PRE2022,Mondal-PRA2022,Hummel-2023} and together build the case that translationally invariant quantum many-body systems generically exhibit quantum scars associated with short periodic trajectories realizing irreducible representations of the translational symmetry group with a well-defined wave vector.

We also note that Ref.~\cite{Iyoda-2018} reported that nonintegrable spins-1/2 chains starting with the state $|\Psi^{\text{up}}\rangle$ exhibit atypically poor scrambling properties,  which may imply that, despite not exhibiting quantum scar indicators in our numerical experiments, spin-1/2 chains still inherit nonuniversal behavior from short periodic classical trajectories.

(Q2) We identified and investigated the quantum counterpart of the classical separatrix between librations and rotations. The quantum separatrix is found to be associated with a crossover between oscillatory and monotonic decays of one-spin polarisation. It also manifests itself as a peak of the participation ratio of energy eigenstates in the initial quantum state, when plotted as a function of the coupling constant $\tilde{J}$. This peak  is located at the value of $\tilde{J}$ corresponding to the classical separatrix.

It is an interesting broader question, whether there is a connection between the transition from librations to rotations of the kind considered in this work and the transition between the oscillatory and the monotonic long-time regimes of spin relaxation investigated in Refs.~\cite{fine2003universal,fine2004long-time,starkov2018hybrid}. In the latter case, the quantity of interest is the relaxation of spin polarisation near infinite temperature equilibrium, mainly in the context of nuclear magnetic resonance\cite{morgan2008universal,sorte2011long-time,meier2012eigenmodes}. The presently studied periodic trajectories constitute a part of a nonequilibrium statistical ensemble. If they are more stable than the ergodic trajectories, they could control the long-time relaxation.

Let us now make two final remarks connecting this work to the recent literature: (i) We have not looked for the quantum signatures of the detected Arnold diffusion, but it remains an interesting direction to explore both in terms of quantum localization properties on the Arnold web\cite{Schmidt-2023} and in terms of possible hybridization of eigenstates\cite{Varma-Cohen-2024} between ergodic and non-ergodic regions of the phase space. (ii) According to the proposal of Ref.~\cite{Beringer-2024},  the connection between quantum scars and unstable periodic classical trajectories can be exploited to exercise dynamical control over large clusters of quantum particles.

To conclude, the investigations of periodic classical spin trajectories and their quantum counterparts offer a refreshing perspective on the fundamental aspects of chaotic spin dynamics and may also contribute to the toolbox for quantum engineering of spin clusters.

{\it Note added:} After submitting the present work for publication, we were alerted by the authors of Refs.\cite{pizzi2024spinorcond,pizzi2024arxivjan,Pizzi-NatComm2025} about their closely related publications on the relations between classical periodic orbits in spin systems and quantum many-body scars. We share the optimism of the authors of Ref.~\cite{Pizzi-NatComm2025} that the quantum interference effect associated with the  periodic classical orbits can lead to ``genuine'' many-body quantum scars.

\begin{acknowledgments}
The authors thank D. Treschev for a helpful discussion. This work was supported during years 2019-2021 by the Russian Science Foundation under the grant N$^{\rm o}$ 17-12-01587.
\end{acknowledgments}

\appendix

\section{Periodicity of one-spin classical trajectories $\mathbf{S}_{\text{p}}(t)$ and the transition from librations to rotations}
\label{periodicity}

\begin{figure}
\includegraphics[width=0.6\columnwidth]{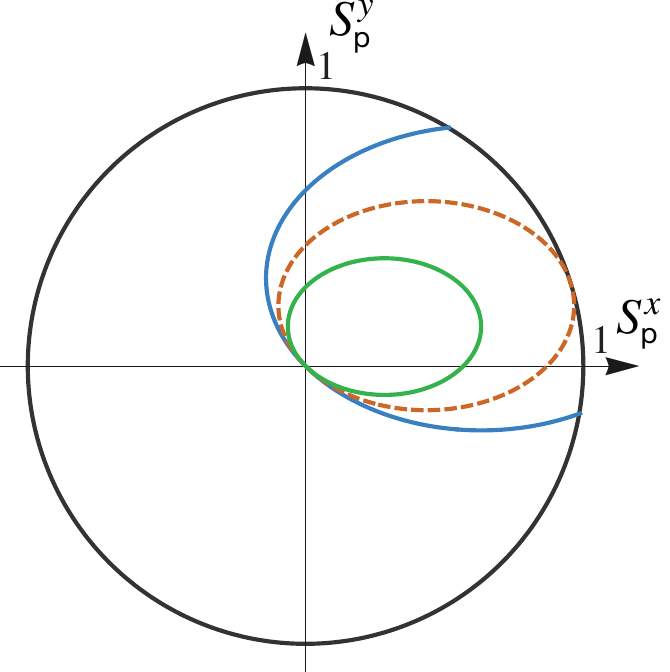}
\caption{
Projections of one-spin trajectories $\mathbf{S}_{\text{p}}(t)$  onto the $(S^x_{\text{p}}, S^y_{\text{p}})$-plane for $h = 1$ and different values of $J$ in the Hamiltonian (\ref{ham0}). Colored lines are the ellipses (or a part thereof) defined by Eq.~(\ref{app0eq3}) - among them: blue solid elliptic segment is a projection of libration ($J = 0.76$), green solid ellipse is a projection of rotation ($J = 1.76$), and brown dashed ellipse is a projection of the separatrix ($J = J^* \approx 1.15$). Black line is the unit circle defined by Eq.~(\ref{app0eq4}). The three plotted projections correspond to the three examples shown in Fig.~\ref{SP_solutions_red}.
}
\label{LRE}
\end{figure}

In order to follow the transition from librations to rotations analytically and simultaneously demonstrate the generic periodicity of the one-spin orbits $\mathbf{S}_{\text{p}}(t)$,  we note that energy $E = 0$ corresponding to our initial conditions ``all-spins-up", implies a constraint   on the time evolution of $\mathbf{S}_{\text{p}}(t)$,  which, given the Hamiltonian (\ref{ham0}), can be rewritten as
\begin{align}
    \label{app0eq3}
    \left(S^x_{\text{p}}-\frac{1}{2}\frac{h}{J}\right)^2+2\left(S^y_{\text{p}}-\frac{1}{4}\frac{h}{J}\right)^2=\frac{3}{8}\left(\frac{h}{J}\right)^2.
\end{align}
This equation defines an ellipse in the $(S^x_{\text{p}}, S^y_{\text{p}})$-plane.
At the same time, the spin normalization condition implies:
\begin{align}
    \label{app0eq4}
    (S^x_{\text{p}})^2+(S^y_{\text{p}})^2=1-(S^z_{\text{p}})^2\leq 1
\end{align}
i.e., the projection of the spin on the $(S^x_{\text{p}}, S^y_{\text{p}})$ plane is bounded by a unit circle. As illustrated in Fig.~\ref{LRE}, the intersection of the ellipse (\ref{app0eq3}) and the unit circle (\ref{app0eq4}) determines the type of motion of the spin on the unit sphere as follows:

(i) A libration corresponds to the case where the ellipse (\ref{app0eq3}) intersects the unit circle at two points associated with the spin transitioning back and forth between the hemispheres $S^z_{\text{p}} > 0$ and  $S^z_{\text{p}} < 0$. The elliptic segment truncated by the unit circle is a projection by the libration trajectory onto the $(S^x_{\text{p}}, S^y_{\text{p}})$-plane. Each point on this projection is associated with two points on the libration trajectory and thus is being passed twice during the libration period.

(ii) A rotation is the case where the ellipse lies entirely within the unit circle without intersecting it. In this case, the motion is limited to either hemispheres $S^z_{\text{p}} > 0$ or  $S^z_{\text{p}} < 0$. Correspondingly, there are two disconnected rotation trajectories having the ellipse as their $(S^x_{\text{p}}, S^y_{\text{p}})$-projection. 

(iii) A separatrix emerges, when the ellipse touches the unit circle at one point, which then becomes the fixed point.

In order to establish the correspondence between the above regimes and the value of $J/h$, we note that the ellipse (\ref{app0eq3}) must pass through the point $(S^x_{\text{p}}, S^y_{\text{p}}) = (0,0) $ for any  $J/h$. Small values of $J/h$ imply large linear size of the ellipse, thereby guaranteeing two crossing points with the unit circle and hence the libration regime, while large values of $J/h$ imply that the ellipse is too small and thus cannot cross the unit circle, which means the rotation regime.

We, finally, note that, for a broader class of spin Hamiltonians, one can  imagine a larger variety of situations, such as, e.g., the ellipse crossing the circle at four points and still implying a pair of rotation trajectories, or the ellipse touching the unit circle at two points and implying heteroclinic  non-periodic trajectories. In any case, the  explicit formula (\ref{app0eq3}) for the $(S^x_{\text{p}}, S^y_{\text{p}})$-projection together with $S^z_{\text{p}} = \pm \sqrt{1- (S^x_{\text{p}})^2 -  (S^y_{\text{p}})^2}$   implies that periodic trajectories $\mathbf{S}_{\text{p}}(t)$ represent a generic case, while the non-periodic ones -- homoclinic or heteroclinic - are also possible (as anticipated by the Poincaré-Bendixson theorem\cite{katok1995introduction,teschl2012ordinary}) but only as an exception.

\section{Upper bound on the largest Lyapunov exponent for classical spin dynamics}
\label{Lambda_Bound}

Here we derive the upper bound on the largest Lyapunov exponent for the classical spin chain considered in this work. The derivation is trivially generalizable to any spin lattice with finite number of interacting neighbors per spin: it amounts to an explicit implementation of the steps of the general argument given in Ref.\cite{dewijn2013lyapunov} that the maximum Lyapunov exponent for classical spin dynamics must be finite.

The largest Lyapunov exponent defined by Eq.(\ref{lambdamax-def}) for any trajectory, periodic or ergodic, can be computed as the time-average of the instantaneous ``local stretching rates'' along the trajectory:
\begin{equation}
  \lambda_{\text{max}} = \frac{1}{T}  \int_0^T \lambda_{\text{inst}}(t) dt ,
    \label{integral-lmd-inst}
\end{equation}
where
\begin{equation}
\lambda_{\text{inst}}(t) \equiv  \frac{|\delta \dot{\mathcal{S}}(t)|}{|\delta \mathcal{S}(t)|} 
    \label{lambda_inst}
\end{equation} 
and $T$ is a sufficiently long integration time for an ergodic trajectory or a period for a periodic trajectory. Therefore, 
\begin{equation}
\lambda_{\text{max}} \leq \text{max}[\lambda_{\text{inst}}] .
    \label{lambda_max-lambda_inst}
\end{equation} 
Below we obtain a constraint from above on $\text{max}[\lambda_{\text{inst}}]$.

The linearization of the spin dynamics based on Eqs.(\ref{claseq2},\ref{Hj}) relatively to the small deviations of spin projections 
\begin{equation}
\delta \mathcal{S}(t)
\equiv
[\delta \mathbf{S}_1(t),\dots,\delta \mathbf{S}_L(t)]
\label{deltaSt}
\end{equation}
gives
\begin{align}\label{claseq2-delta}
\delta \dot{\mathbf{S}}_i= \delta \mathbf{H}_i \times \mathbf{S}_i +  \mathbf{H}_i \times \delta \mathbf{S}_i,
\end{align}
where $\mathbf{H}_i$ is given by Eq.(\ref{Hj}), and 
\begin{equation}
  {\delta \mathbf{H}}_i = \left(
  \begin{array}{c} 
  - J \ \delta S_{i-1}^x - J \  \delta S_{i+1}^x  \\[4pt]
  - 2 J \  \delta S_{i-1}^y - 2 J \  \delta S_{i+1}^y   \\[4pt]
  0 
  \end{array}
    \right)
  \label{deltaHj}
  \end{equation}
After Eq.(\ref{deltaHj}) is substituted into Eq.(\ref{claseq2-delta}), the result can be represented as 
\begin{align}\label{claseq2-delta-A}
\delta \dot{S}_i^{\alpha}= \sum_{j}^{\text{NN}(i)} A_{ij}^{\alpha \beta} \delta S_i^{\beta},
\end{align}
where $\text{NN}(i)$ means that the sum is taken over the nearest neighbors of spin $\mathbf{S}_i$, and $A_{ij}^{\alpha \beta}$ is the tensor determined by the coupling constants and the instantaneous projections of $\mathbf{S}_i$ and its two neighbors. Since each projection of each spin has the maximum value 1, the maximum value of an element in this tensor can be constrained for our system by examining the coefficients in the explicit expressions for the individual projections of $\delta \dot{\mathbf{S}}_i$, which, in the case of Eq.(\ref{claseq2-delta}), gives
\begin{equation}
      A_{\text{max}} \equiv \text{max}[ |A_{ij}^{\alpha \beta} |] \leq 4 |J| + |h| . 
    \label{Amax}
\end{equation}

After we collect all spin deviations into one $3L$-dimensional vector \mbox{$
\delta \mathcal{S}
\equiv
[\delta S^x_1, \delta S^y_1, \delta S^z_1,\dots, \delta S^x_L, \delta S^y_L, \delta S^z_L]
$}, the counterpart of Eq.(\ref{claseq2-delta-A}) reads
\begin{equation}
\delta \dot{\mathcal{S}} = \mathcal{A} \cdot \delta \mathcal{S},
\label{dSdot-A}
\end{equation}
where  $\mathcal{A}$ is an $3L \times 3L$ tensor  collecting all the entries of tensors $A_{ij}^{\alpha \beta}$. Below we also use variable $D = 3L$ for the dimension of the extended phase space, and calligraphic variables $\delta \mathcal{S}_n$ and  $ \mathcal{A}_{mn}$ for the components of $\delta \mathcal{S}$ and $\mathcal{A}$.
We note that, for any spin chain with nearest neighbor spin-spin interaction, each line or column of the tensor $\mathcal{A}$ cannot contain more than  $K=9$  non-zero entries $\delta \mathcal{A}_{mn}$ originating from the three projections of each spin plus six projections of the two neighbors coupled to that spin. 

The expression (\ref{lambda_inst}) for the local stretching rate can now be rewritten as
\begin{equation}
\lambda_{\text{inst}} =  \frac{| \mathcal{A} \delta \mathcal{S}|}{|\delta \mathcal{S}|} ,
    \label{lambda_inst-A}
\end{equation}
which implies that
\begin{widetext}

\begin{align}
    \lambda_{\text{inst}}^2 = \frac{|\mathcal{A} \, \delta \mathcal{S}|^2}{|\delta \mathcal{S}|^2}  
    = \frac{(\mathcal{A}_{11} \delta \mathcal{S}_1 + \dots + \mathcal{A}_{1D} \delta \mathcal{S}_D)^2 + \dots + (\mathcal{A}_{D1} \delta \mathcal{S}_1 + \dots + \mathcal{A}_{DD} \delta \mathcal{S}_D)^2}{\delta \mathcal{S}_1^2 + \dots + \delta \mathcal{S}_D^2}. 
    \label{eq:lambda_bound_1}
\end{align}
We now proceed with the chain of inequalities for each term in the numerator of the right-hand-side (RHS) of Eq.(\ref{eq:lambda_bound_1})
\begin{align}
(\mathcal{A}_{n1} \delta \mathcal{S}_1 + \dots + \mathcal{A}_{nD} \delta \mathcal{S}_D)^2 
\leq
\left( |\mathcal{A}_{n1}| \ |\delta \mathcal{S}_1| + \dots + |\mathcal{A}_{nD}| \  |\delta \mathcal{S}_D|
\right)^2
\leq
A_{\text{max}}^2 \left( \sum_{m(n)}^K |\delta \mathcal{S}_m| \right)^2 \leq
A_{\text{max}}^2 K  \sum_{m(n)}^K |\delta \mathcal{S}_m|^2 ,
\label{eq:lambda_bound_2}
\end{align}
\end{widetext}
where the notation $m(n)$ implies that the sum is taken only over the $K$ projections $|\delta \mathcal{S}_m|$ that were multiplied by those tensor elements $|\mathcal{A}_{nm}|$ that are allowed to have non-zero values. (Once our notations are traced back, a fixed value of index $n$ implies a certain projection $x$, $y$, or $z$ for one of our $L$ spins; the above sum then extends over $|\delta \mathcal{S}_m|$ corresponding to the three projections of that spin plus the six projections of its two coupled neighbors.)  Once the inequality (\ref{eq:lambda_bound_2}) is substituted for each term in the numerator of the RHS of Eq.(\ref{eq:lambda_bound_1}), we further note that each squared projection $|\delta \mathcal{S}_m|^2$ of the vector $\delta \mathcal{S}$ can appear in the numerator maximum $K$ times, which, therefore, gives
\begin{align}
    \lambda_{\text{inst}}^2 \leq   \frac{A_{\max}^2 \ K^2 \   (\delta S_1^2 + \dots + \delta S_M^2)}{\delta S_1^2 + \dots + \delta S_M^2} = A_{\max}^2 K^2.
    \label{eq:lambda_bound_3}
\end{align}

The inequalities (\ref{lambda_max-lambda_inst}) and (\ref{eq:lambda_bound_3}) imply the  upper bound 
\begin{align}
    \lambda_{\text{max}} \leq    A_{\max} K
    \label{eq:lambda_bound_4}
\end{align}
for the largest Lyapunov exponent of both ergodic and periodic phase-space trajectories.

While the final inequality (\ref{eq:lambda_bound_4}) firmly establishes the existence of the upper bound on $\lambda_{\text{max}}$, it is likely that the chain of inequalities leading to it can be made more restrictive, which would reduce that bound. We cannot provide any example where the value of $ \lambda_{\text{max}}$ would actually reach $A_{\max} K$.

\section{Numerical calculation of Lyapunov exponents}
\label{appendix_A}

\begin{figure}
\includegraphics[width=\columnwidth]{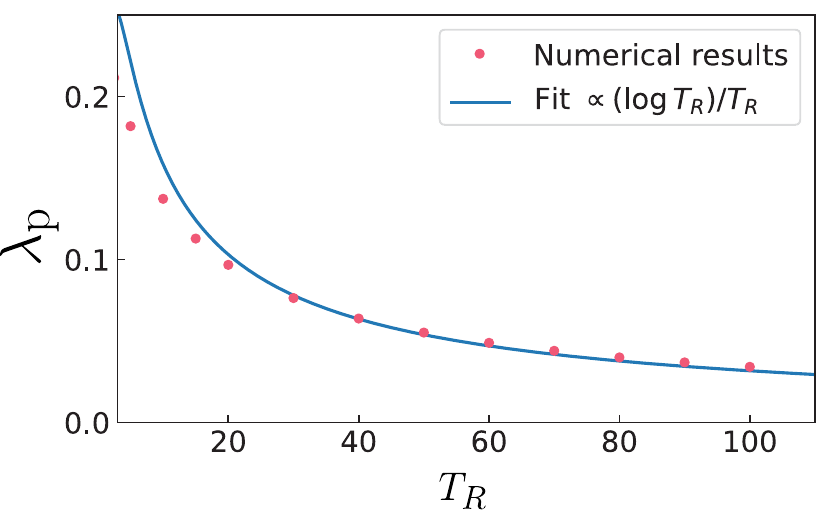}
\caption{
Periodic Lyapunov exponent $\lambda_\text{p}$ as a function of  the reset time $T_R$ for stable periodic trajectory of classical spin chain of length $L=23$ with $J=1.76$. Red points represent the numerically computed $\lambda_\text{p}$ with $M=1000$ resets behind each point. Blue solid line is a fit of the form (\ref{lambdaTR}).
\label{resets}}
\end{figure}

In this Appendix, we describe the algorithm for computing the largest Lyapunov exponent of a many-spin system and then explain how we distinguish between truly zero Lyapunov exponents and finite but small ones. The latter is necessary to justify the zero values of some Lyapunov exponents in Figs.~\ref{gemD} and \ref{LVSL}.  

Our procedure for  computing the largest Lyapunov exponent implements the standard general method of Ref.~\cite{Benettin-1980} for classical spin systems as described, e.g., in Refs.~\cite{dewijn2012largest,Elsayed-signatures-2014}. It tracks two initially close phase trajectories defined by Eq.~(\ref{St}),  $\mathcal{S}_1(t)$ and $\mathcal{S}_2(t)=\mathcal{S}_1(t)+\delta \mathcal{S}(t)$, where $|\delta \mathcal{S}(0)|=d_0$ is very small. The deviation $|\delta \mathcal{S}(t)|$ grows during the time interval $T_R$ (reset time); after that, $\delta \mathcal{S}(t)$ is renormalised such that its direction is preserved and its length is reset back to $d_0$ again. This procedure is repeated until the following quantity converges:
\begin{align}
\lambda=\frac{1}{MT_R}\sum\limits^M_{m=1} \log \left[\frac{|\delta \mathcal{S}(mT_R)|}{d_0} \right],
\label{lexpGenFormula}
\end{align} 
where $M$ is the number of resets. When the reference trajectory $\mathcal{S}_1(t)$ is ergodic, then (\ref{lexpGenFormula}) gives the ergodic Lyapunov exponent $\lambda_\text{e}$. If the reference trajectory $\mathcal{S}_1(t)=\mathcal{S}_p(t)$ is periodic, then (\ref{lexpGenFormula}) yields the periodic Lyapunov exponent $\lambda_\text{p}$. 

The right-hand-side (RHS) of Eq.(\ref{lexpGenFormula}) converges to that of Eq.(\ref{integral-lmd-inst}) in the limit $d_0 \to 0$, $T_R \to 0$ with $M = T/T_R$. However, in numerical simulations, $T_R$ and $d_0$ are held finite while $M$ is increasing until the RHS of Eq. (\ref{lexpGenFormula}) approaches a stationary value. Such a procedure normally results in a finite value of $\lambda$ even when the true Lyapunov exponent is zero. To distinguish zero Lyapunov exponents from finite but small ones, we look at the dependence of the right-hand-side of Eq.(\ref{lexpGenFormula}) on the reset time $T_R$. In the case of Lyapunov-unstable trajectories, the values of $\lambda$ do not depend on $T_R$ (as long as $|\delta \mathcal{S}(T_R)|$ is small enough to apply the Lyapunov stability theory). When a trajectory is Lyapunov-stable, the value of $\lambda$ decreases with $T_R$ as 
\begin{equation}
\lambda_\text{p}\propto \frac{\log T_R}{T_R},
    \label{lambdaTR}
\end{equation}
which, in turn, means that it approaches zero for ${T_R \to \infty}$. In Fig.\ref{resets}, we plot $\lambda_\text{p}$ as a function of $T_R$ in the case of a Lyapunov-stable periodic trajectory for the spin chain of length $L=23$ with $J=1.76$. The plot shows that $\lambda_\text{p}$ decreases with $T_R$ consistent with the asymptotic dependence (\ref{lambdaTR}). Hence the corresponding point in Fig.~\ref{LVSL}(b) indicates that $\lambda_\text{p} = 0$.

\section{Lyapunov exponents for ergodic phase space trajectories of classical spin chains}
\label{ergodic}

\begin{figure}
\includegraphics[width=\columnwidth]{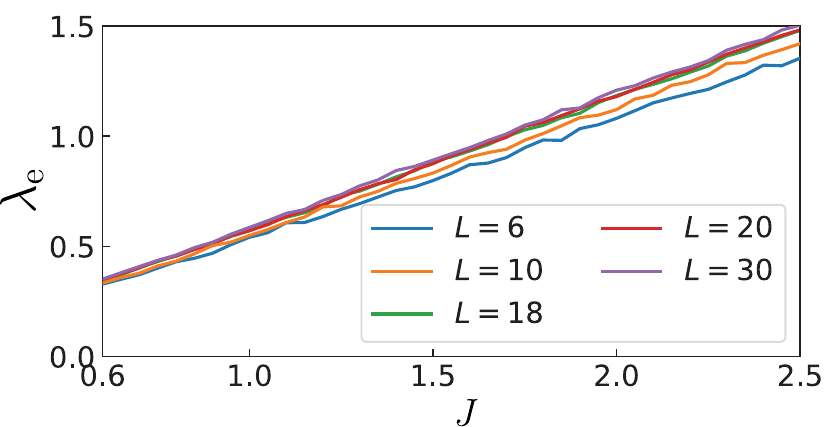}
\caption{Lyapunov exponents $\lambda_\text{e}$ for ergodic trajectories as a function of $J$ for classical spin chains of lengths $L$. The initial conditions $\mathcal{S}_1(0)$ are chosen randomly on the energy shell $E=0$; the reset time is $T_R=5.0$, the number of resets $M=1000$.
\label{lexp_ergpic}}
\end{figure}

We computed the largest Lyapunov exponents $\lambda_\text{e}$ for ergodic trajectories of classical spin chains governed by the Hamiltonian (\ref{ham}) on the infinite temperature energy shell $E=0$. The results for different interaction constants $J$ and chain lengths $L$ are summarised  in Fig.~\ref{lexp_ergpic}. 
One can observe there that $\lambda_\text{e}$ exhibits a rather featureless nearly linear dependence on $J$ and very little dependence on $L$.  Such a behavior stands in a clear contrast with strong and nontrivial dependencies of the periodic Lyapunov exponent on $J$ and $L$ illustrated in the main text by Figs.~\ref{gemD},  \ref{LVSL} and \ref{nearSep}.

\section{Procedure for computing Lyapunov exponent $\lambda_\text{S}$ around the unstable fixed point of the separatrix}
\label{appendix_sep}

\begin{figure}
\includegraphics[width=\columnwidth]{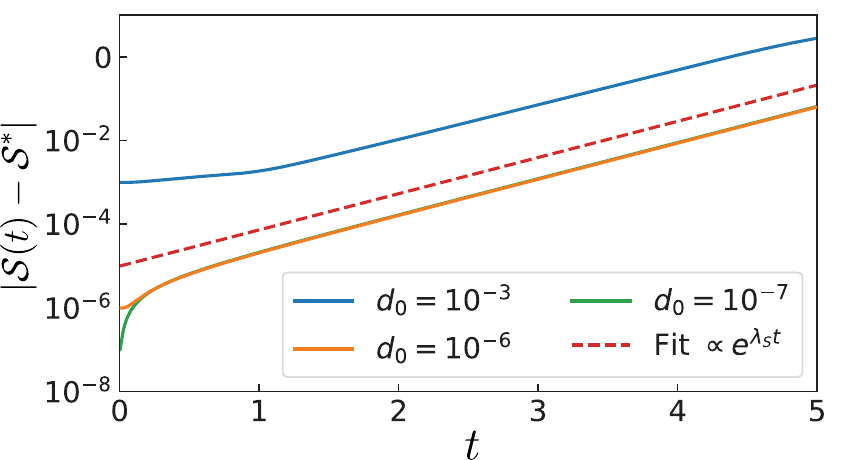}
\caption{Illustration to the procedure of computing the largest Lyapunov exponent $\lambda_\text{S}$ around the unstable fixed point $\mathcal{S}^*$ on the separatrix for a classical spin chain of length $L$ with $J = J^*$. The plot shows the
time dependence of the distance between $\mathcal{S}^*$ and a point on a phase space trajectory $\mathcal{S}(t)$ initiated in the close vicinity of $\mathcal{S}^*$ such that $|\mathcal{S}(0) - \mathcal{S}^*| = d_0 $. Different solid lines represent different values of $d_0$ given in the plot legend. All of them exhibit asymptotic exponential growth determined by $\lambda_\text{S}$. The dashed line is the exponential fit proportional to $e^{\lambda_\text{S} t} $, where $\lambda_\text{S} = 1.99$.  
\label{lamS}}
\end{figure}

As explained in Section~\ref{separatrix}, the separatrix for the class of considered periodic  trajectories consists of an unstable fixed point and two branches asymptotically departing and approaching the fixed point for $t \to - \infty$ and $t \to + \infty$ respectfully (see Figs.~\ref{SP_solutions_red} and \ref{LRFS}). As a result, each of the two branches exhibits in the vicinity of the fixed point an atypical additional Lyapunov instability along the time flow direction. The corresponding Lyapunov exponent $\lambda_\text{S}$ happens to be larger than the largest Lyapunov exponents $\lambda_\text{p}$ for the periodic trajectories in the vicinity of the fixed point.  Having a fixed point as a reference trajectory implies that no multiple resets of $|\delta \mathcal{S}(t)|$ envisioned by Eq.(\ref{lexpGenFormula})  are necessary to fairly sample the Lyapunov growth.  The value of $\lambda_\text{S}$ is, instead, computed as described below. 

We choose $J=J^*$ and then follow the time evolution of a phase trajectory $\mathcal{S}(t)$ starting from the initial conditions $\mathcal{S}(0)=\mathcal{S}^* + \delta \mathcal{S}_0$, where $\delta \mathcal{S}_0$ is a small deviation from the fixed point $\mathcal{S}^*$, meaning that $|\delta \mathcal{S}_0|\equiv d_0 \ll 1$. The growth of $|\mathcal{S}(t) - \mathcal{S}^*|$ is expected to become asymptotically proportional to $e^{\lambda_\text{S} t} $, and, indeed, such a behavior can be observed in Fig.~\ref{lamS}, where the same exponential growth  is exhibited at later times by $|\mathcal{S}(t) - \mathcal{S}^*|$ independent of the initial deviation $\delta \mathcal{S}_0$. The value of $\lambda_\text{S}$ is then extracted as the rate of that growth. 

\section{Functional form of $\lambda_{\text{p}} (J)$ near the separatrix $J = J^*$  }
\label{F-form-lJ}

In this Appendix, we obtain an analytical approximation that gives the functional form (\ref{lmdp-logDJ}) of the cusp  of $\lambda_{\text{p}}(J)$ near the separatrix value $J = J^*$. The cusp itself is shown in Fig.(\ref{nearSep}).

Let us start by recalling, that according to Eq.(\ref{integral-lmd-inst}), the largest Lyapunov exponent for a periodic trajectory can be expressed as the time-average of the instantaneous ``local stretching rates'' $\lambda_{\text{inst}}(t)$ along the trajectory:
\begin{equation}
  \lambda_{\text{p}} = \frac{1}{T}  \int_0^T \lambda_{\text{inst}}(t) dt ,
    \label{integral-lmd-inst1}
\end{equation}
where $T$ is the period of the trajectory, and $\lambda_{\text{inst}}(t) =  \lim_{\Delta t \to 0} \frac{1}{\Delta t} \log \frac{|\Delta \mathcal{S}(t + \Delta t)|}{|\Delta \mathcal{S}(t)|} $, which is the same as  Eq.(\ref{lambda_inst}) but expressed as a limit of finite differences similar to those entering Eq.(\ref{lexpGenFormula}). We further note that the separatrix has a symmetric 8-shaped form with the unstable fixed point in the middle (see Appendix~\ref{periodicity}). The spin orientation corresponding to the fixed point is to be denoted as $\mathbf{S}^*$. (Note: $\mathbf{S}^*$ is a three-dimensional vector for the one-spin representation of a periodic trajectory, while  $\mathcal{S}^*$  is the corresponding point in the many-spin phase space.)  The rotations near the separatrix cover half of that 8-shape, while the librations cover the entire 8-shape. Since the 8-shape is symmetric, it is sufficient in the case of librations to average the stretching rates in Eq.(\ref{integral-lmd-inst1}) only over the half of the libration period. Furthermore, since both the half of the libration trajectory and the rotation trajectory nearly coincide in the vicinity of the separatrix, the function $\lambda_{\text{p}}(J)$  tends to be largely symmetric with respect to small deviations from the separatrix  $\Delta J \equiv J - J^*$  in agreement with the visual impression from Figs.~\ref{nearSep}(a) and with the analysis below. Yet, this analysis also reveals a subdominant asymmetry of $\lambda_{\text{p}}(J)$. 

\begin{figure}
\includegraphics[width=\columnwidth]{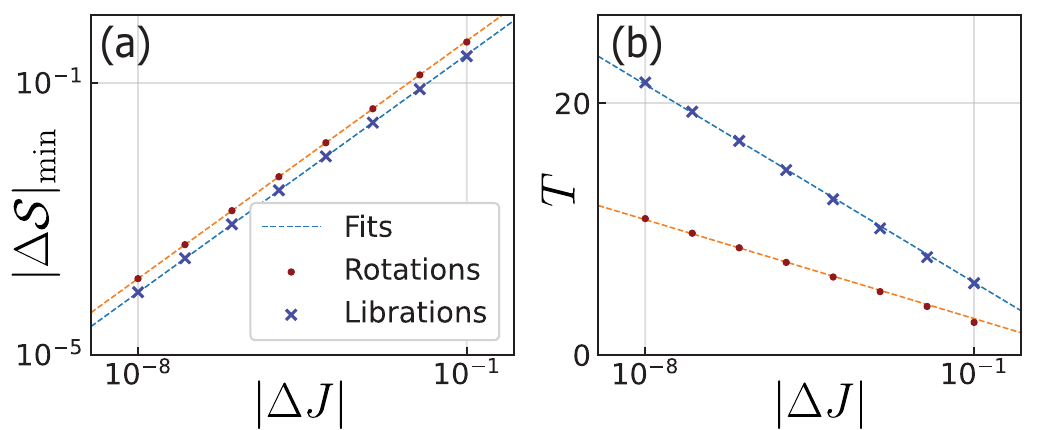}
\caption{Characteristics of periodic trajectory $\mathbf{S}(t) $ in the vicinity of the separatrix as a function of $ |\Delta J| \equiv |J - J^*|  $
(a) Closest distance $\Delta |\mathbf{S}_\text{min}|$ to the fixed point $|\mathbf{S}^*|$. Red and blue lines represent fits of the form (\ref{DS-DJ}) for rotations and librations respectively. The offset between two lines is due to different values of parameter $a$ for the two regimes predicted on the basis of Eq.(\ref{hyperbola}).  (b) Period $T$ of trajectories near the separatrix. The orange line is the fit of the form (\ref{T-T0-TS}) for rotation with the slope $1/\lambda_S$. The blue line is the analogous fit for librations with the slope $2/\lambda_S$. 
\label{Smin_and_T}}
\end{figure}

Since the separatrix has an unstable fixed point $\mathbf{S}^*$ characterized by the pair of Lyapunov exponents $\pm \lambda_S$, the period of a periodic trajectory in the vicinity of the separatrix logarithmically diverges as $J$ approaches $J^*$. Here the Lyapunov exponents $\pm \lambda_S$ are relevant, not because we investigate the instability of the fixed point, but because the reference periodic trajectory itself has a stretch  that exhibits nearly exponential slowdown followed by the nearly exponential acceleration in the vicinity of the fixed point. 

The time the phase trajectory stays near the fixed point is controlled by the smallest distance  $|\Delta \mathbf{S}|_{\text{min}}$  to the fixed point $\mathbf{S}^*$, where $\Delta\mathbf{S}(t) \equiv \mathbf{S}(t) - \mathbf{S}^*$: the smaller this distance, the closer $|\Delta\dot{\mathbf{S}}|$ approaches zero. We now show that this distance has the form
\begin{equation}
|\Delta \mathbf{S}|_{\text{min}} \approx a \sqrt{|\Delta J|},
    \label{DS-DJ}
\end{equation}
where $a$ is a constant that, in general, has different values for librations and rotations. The square-root dependence in Eq.(\ref{DS-DJ}) arises, because the trajectories in the vicinity of the fixed point have a hyperbolic character governed by the equation 
\begin{equation}
- f \Delta s_1^2 + g \Delta s_2^2 = \Delta J,
    \label{hyperbola}
\end{equation}
where $\Delta s_1$ and $\Delta s_2$ are projections of vector $\Delta\mathbf{S}$ on two spin axes orthogonal to $\mathbf{S}^*$ and chosen such that the quadratic form in the left-hand-side of Eq.(\ref{hyperbola}) becomes diagonal; $f$ and $g$ are two positive constants, whose values are not of principal importance here. Equation (\ref{hyperbola}) implies that,  on the rotation side ($\Delta J > 0 $), the minimum distance $|\Delta \mathbf{}{S}|_{\text{min}} = \sqrt{\Delta J /g}$ is realised when $\Delta s_1 = 0$ and $\Delta s_2 = \pm \sqrt{\Delta J /g} $. On the libration side ($\Delta J < 0 $),  $|\Delta \mathbf{S}|_{\text{min}} = \sqrt{|\Delta J| /f}$ corresponds to $\Delta s_1 = \pm \sqrt{|\Delta J| /f}$ and $\Delta s_2 = 0 $. Thus the value of constant $a$ in Eq.(\ref{DS-DJ}) is equal to  $1/\sqrt{g}$ for the rotations and $1/\sqrt{f}$ for the librations. The relation (\ref{DS-DJ}) is verified numerically in Fig.~\ref{Smin_and_T}(a).

Let us now concentrate on the rotation side, i.e. $T$ in Eq.(\ref{integral-lmd-inst1}) becomes the rotation period.
For small enough $\Delta J$, the approach and then the departure of the phase trajectory to the fixed point can be described as follows.

Let us start from the point of the closest approach corresponding to $|\Delta \mathbf{S}(0)| = |\Delta \mathbf{S}|_{\text{min}}$. The  departure from that point is initially controlled by the both positive and negative Lyapunov exponents. However, after time of the order of $ 1/\lambda_S$, the exponential growth of $|\Delta \mathbf{S}(t)|$  due to the positive Lyapunov exponent takes over and dominates the dynamics as long as $|\Delta \mathbf{S}(t)|$ remains in the Lyapunov regime of small deviations from the fixed point. Let us define the maximum value for the applicability of the Lyapunov regime as $|\Delta \mathbf{S}|_{\lambda}$. (The exact value of this threshold is not important, as it is to be later absorbed into a constant.) The time $T_S$ of the Lyapunov growth of $|\Delta \mathbf{S}(t)|$ for small enough $|\Delta \mathbf{S}|_{\text{min}}$   can be taken to be much larger than $1/\lambda_S$ and then  obtained from the approximate relation 
$|\Delta \mathbf{S}|_{\lambda} \approx |\Delta \mathbf{S}|_{\text{min}} e^{\lambda_S T_S}$. This, together with Eq.(\ref{DS-DJ}), gives  
\begin{eqnarray}
T_S &\approx& \frac{1}{\lambda_S} \log \frac{|\Delta \mathbf{S}|_{\lambda}}{|\Delta \mathbf{S}|_{\text{min}}} 
\nonumber \\
&=& \ C_1 - \frac{1}{2 \lambda_S} \log |\Delta J| 
\ \xrightarrow[|\Delta J| \to 0]{} 
\ 
- \frac{1}{2 \lambda_S} \log |\Delta J|, \ \ \ \ \ 
\label{TS}
\end{eqnarray}
where  $C_1 = \frac{1}{\lambda_S} \log \frac{|\Delta \mathbf{S}|_{\lambda}}{a}$.

The time interval $T_S$ of the Lyapunov growth of $|\Delta \mathbf{S}(t)|$  is then followed by the time interval $T_0$ of large deviations from the fixed point; that interval is approximately independent of small $|\Delta J|$, because \mbox{$T_0(J) \approx T_0(J^*) + \left. \frac{d T_0 }{d J} \right|_{J=J^*}\Delta J \approx T_0(J^*)$}. At the end of $T_0$, $|\Delta \mathbf{S}(t)|$ crosses again the threshold value $|\Delta \mathbf{S}|_{\lambda}$ and starts approaching the fixed point in the Lyapunov regime controlled by the negative Lyapunov exponent $-\lambda_S$. Due to the time-reversal symmetry of the Hamiltonian dynamics, it takes another time $T_S$ until the trajectory arrives to the point of the closest approach to the fixed point, thereby completing the period $T$. 
The period $T$ can now be represented as:
\begin{equation}
    T = T_0 + 2T_S = C_2 - \frac{1}{\lambda_S} \log |\Delta J|,
    \label{T-T0-TS}
\end{equation}
where $C_2$ is a constant. The asymptotic dependence (\ref{T-T0-TS}) is confirmed numerically in Fig.~\ref{Smin_and_T}(b) including the prefactor $\frac{1}{\lambda_S}$.
[The analogous expression for the libration period is $T = 2 T_0 + 4T_S = C'_2 - \frac{2}{\lambda_S} \log |\Delta J|$, where $C'_2$ is another constant.]

According to  Eq.(\ref{T-T0-TS}), the contributions to the time averaging in Eq.(\ref{integral-lmd-inst1}) for a rotation can now be divided into two parts:
\begin{equation}
    \lambda_{\text{p}} = \frac{T_0}{T}\lambda_0 + \frac{2T_S}{T}\lambda_A  ,
    \label{lmdp-lmd0-lmdA}
\end{equation}
where $\lambda_0$ is the average stretching rate during the time interval $T_0$ of large deviations, and $\lambda_A$ is the stretching rate during the time interval $2 T_S$, where the phase trajectory is describable by the Lyapunov regime around the fixed point. Since during that interval the trajectory covers a relatively small patch, and the instability behind $\lambda_{\text{p}}$ occurs in a $q\neq 0$ sector, which, in turn, does not contain an unstable fixed point,  it is reasonable to assume that the instant stretching rate during that patch has nearly constant value $\lambda_A$.

In the vicinity of the separatrix, $\frac{T_0}{2 T_S} \ll 1$. Therefore, Eq.(\ref{lmdp-lmd0-lmdA}) with the substitutions from Eqs.~(\ref{TS}), (\ref{T-T0-TS}) and (\ref{lmdp-lmd0-lmdA}) gives
\begin{equation}
    \lambda_{\text{p}} \approx 
    \lambda_A + \frac{T_0}{2 T_S} (
    \lambda_0 -\lambda_A)  
    \xrightarrow[|\Delta J| \to 0]{} 
    \lambda_A 
+ \frac{C}{\log |\Delta J|} ,
    \label{lmdp-logDJ-lim}
\end{equation}
where $C =  T_0 \lambda_S (\lambda_A - \lambda_0)$. This result appears in the main text as Eq.(\ref{lmdp-logDJ}).

%\section{Quantum scars}

\bibliography{spindynamics1}% Produces the bibliography via BibTeX.

\end{document}